\documentclass[aps,prd,reprint,onecolumn,superscriptaddress,nofootinbib,floatfix]{revtex4-1}
\pdfoutput=1
\usepackage{epsfig, subfigure}
\usepackage{setspace}
\usepackage{booktabs, tabularx} 
 
\usepackage{amsmath,bm,bbm}
\usepackage{hyperref}
\usepackage{graphicx}
\usepackage{enumerate}
\usepackage{dcolumn}
\usepackage{bm}
\usepackage{longtable}
\hypersetup{
    pdfnewwindow=true,      
    colorlinks=true,       
    linkcolor=blue,          
    citecolor=blue,        
    filecolor=blue,      
    urlcolor=blue           
}

\def\lsim{\lesssim}  
\def\gsim{\gtrsim}

\newcommand{\beq}{\begin{equation}}  
\newcommand{\eeq}{\end{equation}}
\newcommand{\bea}{\begin{eqnarray}}  
\newcommand{\eea}{\end{eqnarray}}
\newcommand{\epm}{\ensuremath{e^{\pm}\;}}

\newcommand{\omb}{\ensuremath{\Omega_{\rm B} h^{2}}}
\newcommand{\neff}{\ensuremath{{\rm N}_{\rm eff}}}
\newcommand{\Deln}{\ensuremath{\Delta{\rm N}_\nu}}

\newcommand{\mchi}{\ensuremath{m_\chi}}
\def\3he{$^3$He}
\def\4he{$^4$He}
\def\7li{$^7$Li}
\def\Yp{Y$_{\rm P}$}
\def\yd{$y_{\rm DP}$}
\def\hii{H\thinspace{$\scriptstyle{\rm II}$}}

\newcommand{\ie}{{\it i.e.}}
\newcommand{\eg}{{\it e.g.}}
\newcommand{\etal}{{\it et al.}}

\begin{document}

\title{\mbox{\hspace{-0.5cm}BBN And The CMB Constrain Light, Electromagnetically Coupled WIMPs}}

\author{Kenneth M. Nollett}
 \email{nollett@ohio.edu}
 \affiliation{\mbox{Department of Physics and Astronomy, Ohio University, Athens, OH~~45701, USA}}

\author{Gary Steigman}
\email{steigman.1@osu.edu}
\affiliation{\mbox{Center for Cosmology and AstroParticle Physics, Ohio State University,}}
\affiliation{Department of Physics, Ohio State University, 191 W.~Woodruff Ave., Columbus, OH~~43210, USA}

\date{\today}

\begin{abstract}

In the presence of a light weakly interacting massive particle (WIMP;
$m_{\chi} \lsim 30\,{\rm MeV}$), there are degeneracies among the
nature of the WIMP (fermion or boson), its couplings to the
standard-model particles (electromagnetic or to neutrinos only), the
WIMP mass \mchi, and the number of equivalent neutrinos beyond the
standard model (including possible sterile neutrinos) \Deln.  These
degeneracies cannot be broken by the cosmic microwave background (CMB)
constraint on the effective number of neutrinos, \neff.  However, big
bang nucleosynthesis (BBN) is affected by the presence of a light WIMP
and equivalent neutrinos, so the combination of BBN and CMB
constraints can help to break some of these degeneracies.  Here, the
BBN predictions for the primordial abundances of deuterium and \4he
(along with \3he and \7li) in the presence of a light WIMP and
equivalent neutrinos are explored, and the most recent estimates of
their observationally determined relic abundances are used to limit
the light-WIMP mass, the number of equivalent neutrinos, and the
present Universe baryon density (\omb).  These constraints are
explored here for Majorana and Dirac fermion WIMPs, as well as for real
and complex scalar WIMPs that couple to electrons, positrons, and
photons.  In a separate paper, this analysis is repeated for WIMPs that
couple only to the standard-model neutrinos, and the constraints for
the two cases are contrasted.  In the absence of a light WIMP, but
allowing for \Deln~equivalent neutrinos, the combined BBN and CMB
constraints favor N$_{\rm eff} = 3.46 \pm 0.17$, \omb~$= 0.0224 \pm
0.0003$, and \Deln~$= 0.40 \pm 0.17$ (all at a 68\% C.L.).  In this
case, standard BBN (\Deln~= 0) is disfavored at $\sim 98\,\%$
confidence, and the presence of one sterile neutrino (\Deln~= 1) is
disfavored at $\gtrsim 99\,\%$ confidence.  Allowing for a light WIMP
and \Deln~equivalent neutrinos together, the combined BBN and CMB data
provide lower limits to the WIMP masses (\mchi~$\gsim 0.5 - 5\,{\rm
  MeV}$) that depend on the nature of the WIMP, favor \mchi~$\sim
8\,{\rm MeV}$ (with small variations depending on the WIMP type)
slightly over standard BBN, and loosen the constraints on the allowed
number of equivalent neutrinos, \Deln~$= 0.65^{+0.46}_{-0.35}$.  As a
result, while $\Deln = 0$ is still disfavored at $\sim 95\,\%$
confidence when there is a light WIMP, $\Deln = 1$ is now allowed.

\end{abstract}

\maketitle

\section{Introduction}

Although, the weakly interacting massive particles (WIMPs, $\chi$)
present in some extensions of the standard model (SM) of particle
physics are usually very massive, with $m_{\chi} \gsim$ tens or
hundreds of GeV, there has been a long and continuing interest in
light ($m_{e} \lsim m_{\chi} \lsim$ tens of MeV) or very light
($m_{\chi} \lsim m_{e}$) WIMPs
\cite{ktw,serpico,boehm1,boehm2,boehm3,hooper1,boehm4,hooper2,ahn,fayet,hooper3,feng}.
Recently, one of us (G.S.) has explored the effect on the cosmic
microwave background (CMB) measurement of the effective number of
neutrinos, \neff, of a sufficiently light WIMP that its late-time
annihilation heats either the photons or the SM neutrinos beyond the
usual heating from \epm annihilation \cite{chimera}.  This analysis
had some overlap with the earlier work of Kolb \etal\,\cite{ktw} and
of Serpico and Raffelt \cite{serpico}, and with the more recent
analyses of Ho and Scherrer \cite{hoscherrer1,hoscherrer2} and B\oe hm
\etal~\cite{boehm2012}.  While for the standard models of particle
physics and cosmology, the effective number of neutrinos measured in
the late Universe is $\neff = 3$ (more precisely, 3.046
\cite{dolgov,hannestad,mangano}), for extensions of the SM with ``dark
radiation'' equivalent to \Deln~``equivalent neutrinos'' not found in
the SM, it is generally the case that $\neff > 3$.  In
Ref.~\cite{chimera}, this canonical result was revisited, demonstrating
that in the presence of a light WIMP that annihilates only to SM
neutrinos, a measurement of $\neff > 3$ from observations of the CMB
radiation can be consistent with \Deln~= 0 (``dark radiation without
dark radiation'').  It was also shown in Ref.~\cite{chimera} that in
the presence of a sufficiently light WIMP that couples to photons
and/or \epm pairs (like the millicharged particles of
Ref.~\cite{millicharge} and references therein), a CMB measurement
consistent with $\neff = 3$ is not inconsistent with the presence of
dark radiation ($\Delta{\rm N}_{\nu} > 0$).  This latter possibility
opens the window for one or more equivalent neutrinos, including
``sterile neutrinos,'' to be consistent with $\neff = 3$.  Depending on
its couplings, the late-time annihilation of the light WIMP will heat
either the relic photons or the relic SM neutrinos, and this will
affect the CMB constraint on the sum of the SM and equivalent neutrino
masses \cite{chimera}.  It was emphasized in Ref.~\cite{chimera} that
in the presence of a light WIMP and/or equivalent neutrinos, there are
degeneracies among the light-WIMP mass and its nature (fermion or
boson, as well as its couplings to neutrinos and photons), the number
and nature (fermion or boson) of the equivalent neutrinos, and their
decoupling temperature (the strength of their interactions with the SM
particles).  Constraints from the CMB alone are insufficient to break
these degeneracies.

However, as already shown by Kolb \etal\,\cite{ktw} and Serpico and Raffelt \cite{serpico}, and more recently by B\oe hm \etal~\cite{boehm2013}, the presence of a light WIMP modifies the early Universe energy and entropy densities, affecting the synthesis of the light nuclides during big bang nucleosynthesis (BBN).  As a result, BBN provides an additional constraint on light WIMPs and equivalent neutrinos and, in combination with the information from later epochs in the early Universe provided by the CMB, can help to break some of these degeneracies.  However, none of the previous work on BBN in the presence of a light WIMP allowed for equivalent or sterile neutrinos, thereby eliminating by default one of the potentially
interesting options (\eg, sterile neutrinos).  Here, BBN is explored allowing for electromagnetically coupled light WIMPs (fermions and bosons) and equivalent neutrinos.  The allowed regions (at 68\% and 95\% C.L.'s) in the multidimensional parameter space of the WIMP mass (\mchi), the number of equivalent neutrinos (\Deln), the effective number of neutrinos (\neff), and the baryon density parameter (\omb) are identified and compared with the independent constraints (on \neff~and \omb) from the Planck CMB results \cite{planck}.

It must be emphasized that, for the present considerations, the light WIMP need {\bf not} be a dark matter candidate ($\Omega_{\chi} \leq \Omega_{\rm DM}$).  The light WIMP could be a subdominant contributor to the present Universe dark matter density ($\Omega_{\chi} \ll \Omega_{\rm DM}$).  What is important here is that the light WIMP remain in thermal equilibrium through its transitions in the early Universe from a relativistic to a nonrelativistic particle, finally annihilating and transferring its energy and entropy to the remaining SM particles.

\subsection{Review and overview}
\label{sec:overview}

To set the stage for the analysis presented here, it is worthwhile to establish various definitions and to review previous (and recent) CMB constraints, following the notation of Ref.~\cite{chimera}.  First, consider the definition of the effective number of neutrinos, \neff, and its connection to the number of equivalent neutrinos, \Deln, in the absence of a light WIMP.  At late times in the early Universe,\footnote{E.g., at recombination or, at the epoch of equal matter and radiation densities.} long after the \epm pairs (and any light WIMPs) have annihilated, the only particles contributing to the radiation energy density are the photons ($\gamma$), the SM neutrinos ($\nu$), and \Deln~equivalent neutrinos ($\xi$).  At these late times, when $T_{\gamma} \rightarrow T_{\gamma 0} \ll m_{e}$, the radiation energy density, normalized to the energy density in photons alone, is
\beq \bigg({\rho_{\rm R} \over \rho_{\gamma}}\bigg)_{0} = 1 + {7 \over 8}\bigg[3\bigg({T_{\nu} \over T_{\gamma}}\bigg)^{4}_{0} + \Delta{\rm N}_{\nu}\bigg({T_{\xi} \over T_{\gamma}}\bigg)^{4}_{0}\bigg] = 1 + {7 \over 8}\bigg({T_{\nu} \over T_{\gamma}}\bigg)^{4}_{0}\bigg[3 + \Delta{\rm N}_{\nu}\bigg({T_{\xi} \over T_{\nu}}\bigg)^{4}_{0}\bigg]\,.  
\eeq 
In the discussion here, ``sterile neutrinos" are defined to be those equivalent neutrinos [extremely light ($m_{\xi} \lsim 10\,{\rm eV}$) fermions or bosons] that decouple along with the SM neutrinos ($T_{\xi d} = T_{\nu d}$), so that $T_{\xi 0} = T_{\nu 0}$.  However, if the equivalent neutrinos are more weakly coupled, decoupling before the SM neutrinos ($T_{\xi d} > T_{\nu d}$), the SM neutrinos may be heated when the equivalent neutrinos are already decoupled, resulting in $(T_{\xi}/T_{\nu})_{0} < 1$.  To account for this possibility, it is convenient to introduce $\Delta\mathrm{N}_\nu^{*} \equiv \Delta\mathrm{N}_\nu^{*}(T_{\xi}/T_{\nu})_{0}^{4} \leq $~\Deln. For sterile neutrinos, $\Delta\mathrm{N}_\nu^{*} = $\Deln~is an integer (or, for a boson that decouples along with the SM neutrinos, an integer multiple of 4/7).  However, in general, $\Delta\mathrm{N}_\nu^{*}$ need not be an integer.  The canonical, textbook assumption is that the SM neutrinos decouple instantaneously, when the only thermally populated relativistic particles are the photons and \epm pairs (along with the SM and equivalent neutrinos).  It is further assumed that the \epm pairs are extremely relativistic ($m_{e} \rightarrow 0$) at neutrino decoupling. Under these assumptions, $(T_{\nu}/T_{\gamma})^{3}_{0} = 4/11$, and the above result may be used to define \neff:
\beq \bigg({\rho_{\rm R} \over \rho_{\gamma}}\bigg)_{0} \equiv 1 + {7 \over 8}\bigg({4 \over 11}\bigg)^{4/3}{\rm N}_{\rm eff}\,.  
\eeq 
Note that the effective number of neutrinos is a function of both \Deln~and $T_{\xi d}$:
\beq {\rm N}_{\rm eff} \equiv \bigg[{11 \over 4}\bigg({T_{\nu} \over T_{\gamma}}\bigg)^{3}_{0}\bigg]^{4/3}\bigg[3 + \Delta{\rm N}_{\nu}\bigg({T_{\xi} \over T_{\nu}}\bigg)^{4}_{0}\bigg] = 3\bigg[{11 \over 4}\bigg({T_{\nu} \over T_{\gamma}}\bigg)^{3}_{0}\bigg]^{4/3}\bigg[1 + {\Delta{\rm N}_{\nu}^{*} \over 3}\bigg] \equiv {\rm N}_{\rm eff}^{0}\bigg[1 + {\Delta{\rm N}_{\nu}^{*} \over 3}\bigg]\,, 
\eeq 
where N$^{0}_{\rm  eff} \equiv 3[(11/4)(T_{\nu}/T_{\gamma})^{3}_{0}]^{4/3}$.  It is also important to realize that \neff~is defined to be a ``late-time'' quantity, characterizing the energy density after BBN has ended and the \epm pairs and any light WIMPs have annihilated.  In the following, the variation of the energy density during BBN is not described as an evolution of \neff~with time.

In Ref.~\cite{chimera}, it was shown that since the SM neutrinos
decouple at temperature $T_{\nu d} = 2\,{\rm MeV}$ \cite{enqvist1992},
the approximation of extremely relativistic \epm pairs is not entirely
accurate.  Prior to neutrino decoupling, there has already been some
heating of the neutrinos (along with the photons) by \epm
annihilations, so that $(T_{\nu}/T_{\gamma})^{3}_{0} > 4/11$, leading
to N$^{0}_{\rm eff} \approx 3.02$ \cite{chimera}.  If the
instantaneous decoupling assumption is relaxed, the SM neutrinos are
further heated \cite{dolgov,hannestad,mangano}, resulting in a slight
increase to N$^{0}_{\rm eff} \approx 3.05$ \cite{mangano}.  In the
presence of a light WIMP, these calculations would need to be repeated
for each value of the WIMP mass, N$^{0}_{\rm eff} = {\rm N}^{0}_{\rm
  eff}(m_{\chi})$.  For consistency with the discussion in
Ref.~\cite{chimera}, the analysis and results here are for the
assumption of instantaneous neutrino decoupling.  Any errors in
\neff~or \Deln~($\sim 0.03$) introduced by this assumption are much
smaller than current observational uncertainties from BBN or the CMB.

\begin{figure}[!t]
\begin{center}
\includegraphics[width=0.45\columnwidth]{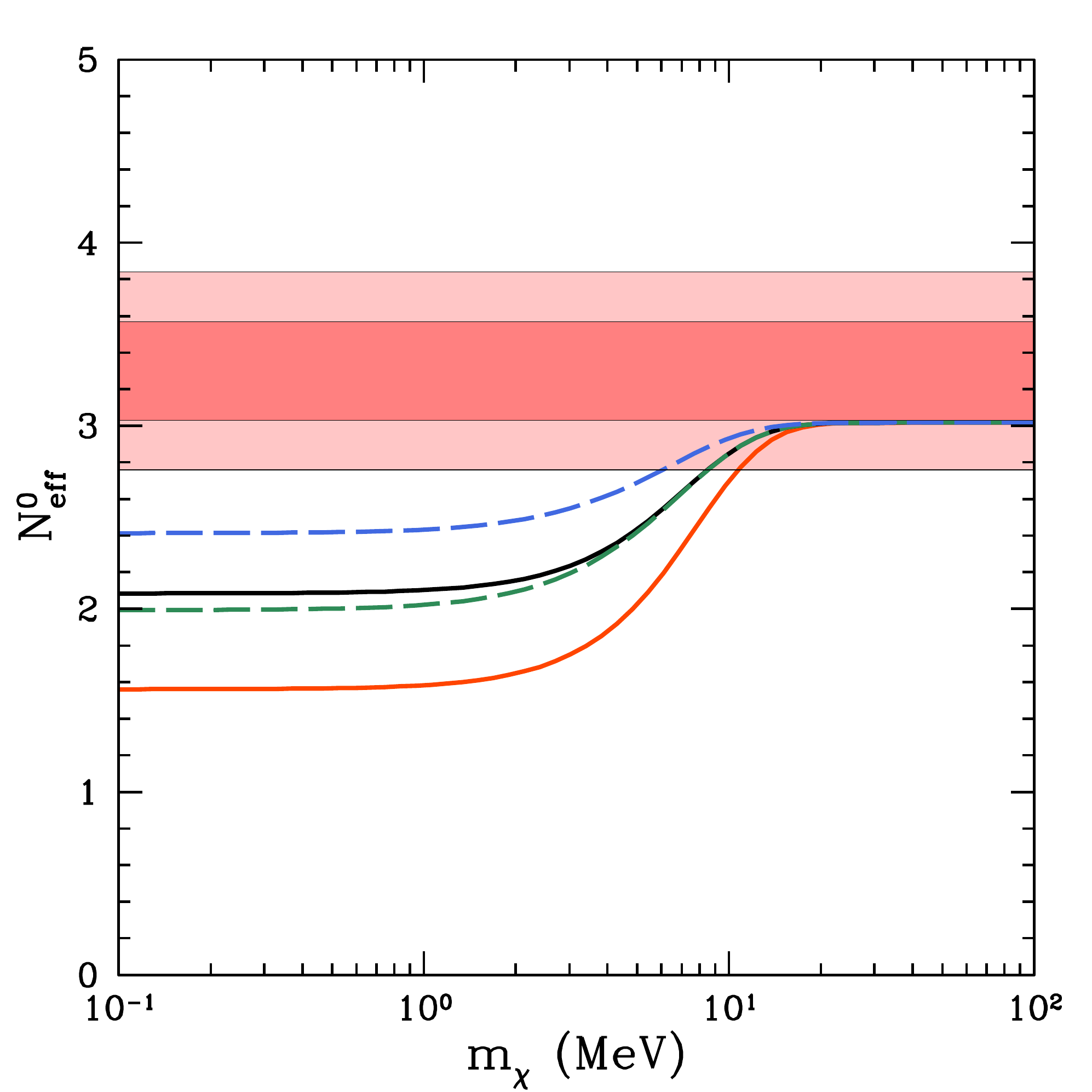}
\hskip .4in
\includegraphics[width=0.45\columnwidth]{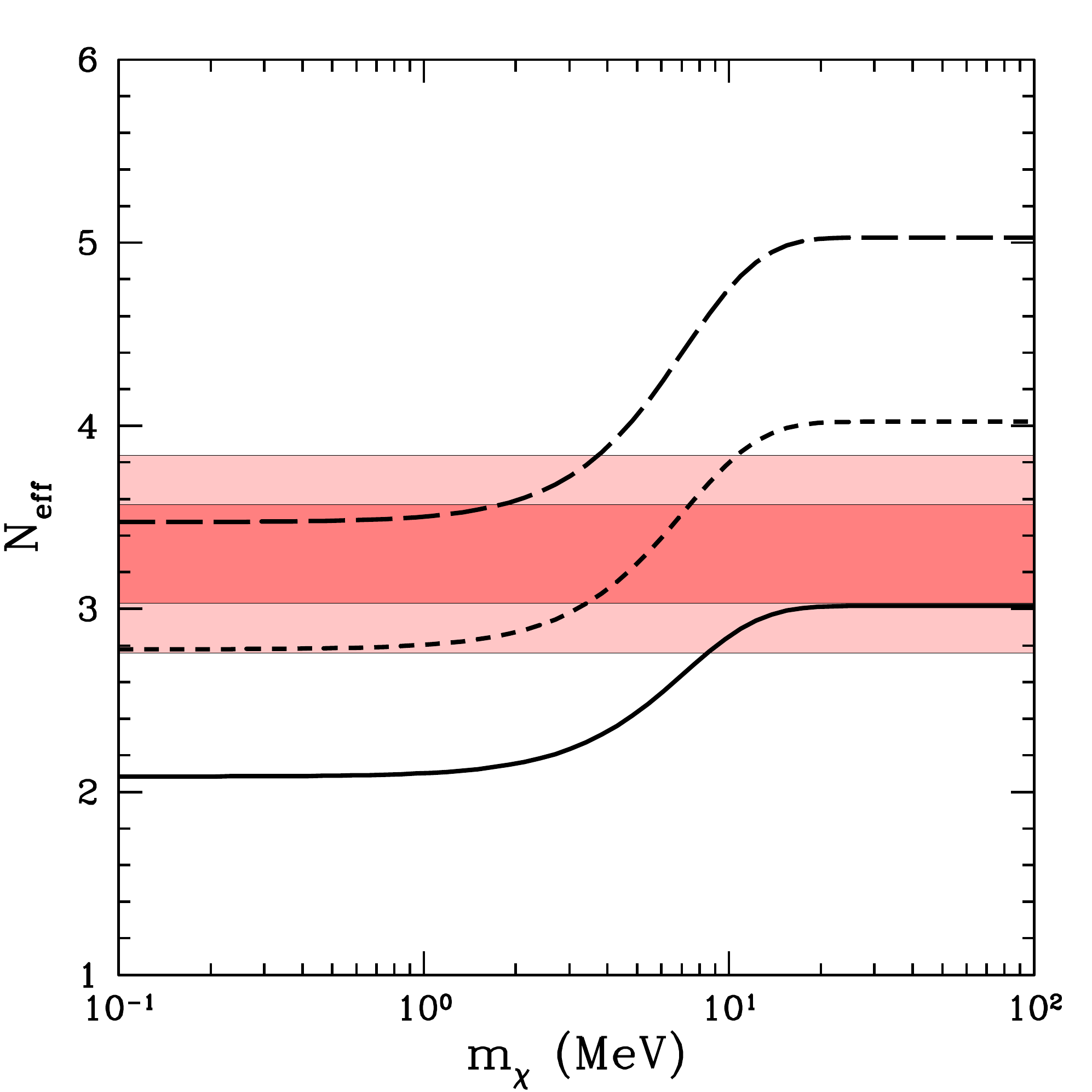}
\\\vskip 0.2in
\caption{(Color online)  The left panel shows N$^{0}_{\rm eff}$ as a function of the WIMP mass for electromagnetically coupled light WIMPs in the absence of equivalent neutrinos.  From bottom to top, the solid red curve is for a Dirac WIMP,  the dashed green curve is for a complex scalar, the solid black curve is for a Majorana fermion, and the dashed blue curve is for a real scalar.  The horizontal, red/pink bands are the Planck CMB 68\% and 95\% allowed ranges for N$_{\rm eff}$.  The right panel specializes to the case of a Majorana fermion WIMP, showing N$_{\rm eff}$ as a function of the WIMP mass for $\Delta\mathrm{N}_\nu^{*}$~equivalent neutrinos.  The solid  curve is for $\Delta {\rm N}_{\nu}^{*} = 0$, the short-dashed curve is for $\Delta {\rm N}_{\nu}^{*} = 1$, and the long-dashed curve is for $\Delta {\rm N}_{\nu}^{*} = 2$.  The horizontal red bands are the Planck CMB 68\% and 95\% allowed ranges for N$_{\rm eff}$, including baryon acoustic oscillations in the CMB constraint.  (After Figs.~7 and 8 of Ref.~\cite{chimera}.)}
\label{fig:neffvsmem}
\end{center}
\end{figure}

A sufficiently light WIMP can only annihilate into photons, \epm
pairs, and the SM neutrinos.  In the presence of a light WIMP $\chi$,
\neff~depends on the nature of the WIMP (fermion or boson) and its
couplings (to photons and \epm pairs, or only to the SM neutrinos), as
well as the WIMP mass.  The annihilation of a WIMP more massive than
$\sim 20\,{\rm MeV}$ occurs prior to the decoupling of the SM
neutrinos.  The strong thermal coupling between the photons and
neutrinos at the higher temperature when a more massive WIMP
annihilates preserves the standard result that $T_{\nu} = T_{\gamma}$
prior to $e^\pm$ annihilation, leading to
$(T_{\nu}/T_{\gamma})^{3}_{0} \approx 4/11$ and N$^{0}_{\rm eff}
\approx 3.02$ at late times.  Now, consider a lighter WIMP that
couples to photons and \epm pairs, but does not couple to the SM
neutrinos.  The late-time annihilation of such a light WIMP ($m_{\chi}
\lsim 20\,{\rm MeV}$) will heat the photons relative to the decoupled
neutrinos, in addition to their usual heating from \epm
annihilation. In this case, at fixed scale factor or fixed photon
temperature, the photons are hotter relative to the SM neutrinos than
in the absence of the light WIMP, so that
$(T_{\nu}/T_{\gamma})^{3}_{0} < 4/11$ and the contribution of the SM
neutrinos to the early Universe energy density is diluted.  This leads
to the possibility that, even in the presence of the three, fully
populated SM neutrinos, N$^{0}_{\rm eff} < 3$, allowing for \Deln~$>
0$ to be consistent with a measurement of \neff~= 3.  For the
contrasting case of a light WIMP that annihilates only to SM
neutrinos, the neutrinos are heated relative to the photons, leading
to $(T_{\nu}/T_{\gamma})^{3}_{0} > 4/11$ and N$^{0}_{\rm eff} > 3$
(``dark radiation without dark radiation").  The amount of heating of
the photons relative to the neutrinos, or vice versa, determines
\neff~as a function of the WIMP mass, depending on the nature of the
WIMP and its couplings to the SM particles.  The analysis here
focuses on the case of a light WIMP coupled to $\gamma$'s and \epm
pairs.  In a subsequent paper \cite{kngs2} the analysis and results
are presented for a light WIMP that couples only to the SM neutrinos.

In the left-hand panel of Fig.~\ref{fig:neffvsmem}, N$^{0}_{\rm eff}$ (\neff~for \Deln~= 0) is shown as a function of the WIMP mass for electromagnetically coupled WIMPs.  These are compared to the Planck CMB result \cite{planck} for \neff~resulting from fitting the six parameters of $\Lambda$CDM plus the additional parameter \neff~and marginalizing over foreground nuisance parameters; the constraint shown includes baryon acoustic oscillations in addition to CMB data.  As may be seen from Fig.\,\ref{fig:neffvsmem}, in the absence of any equivalent neutrinos, the Planck result suggests that \mchi~$\gsim 10\,{\rm MeV}$ (cf.~Ref.~\cite{boehm2013}).  However, for all WIMP masses, the window is open for the presence of equivalent neutrinos. This possibility is illustrated in the right-hand panel of Fig.~\ref{fig:neffvsmem}, which shows \neff~as a function of the WIMP mass for an electromagnetically coupled Majorana fermion WIMP and $\Deln \neq 0$.

\begin{figure}[!t]
\begin{center}
\includegraphics[width=0.45\columnwidth]{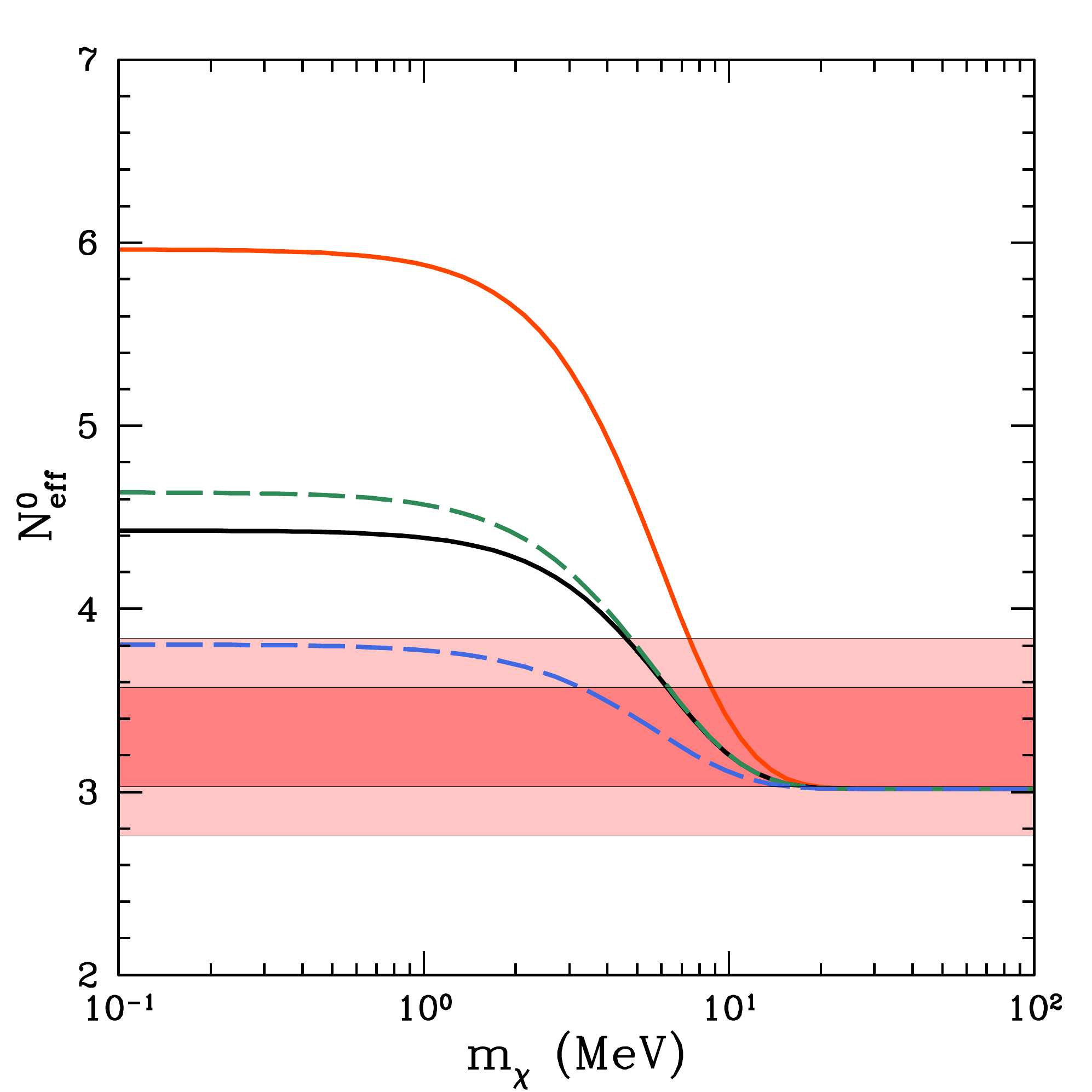}
\hskip .4in
\includegraphics[width=0.45\columnwidth]{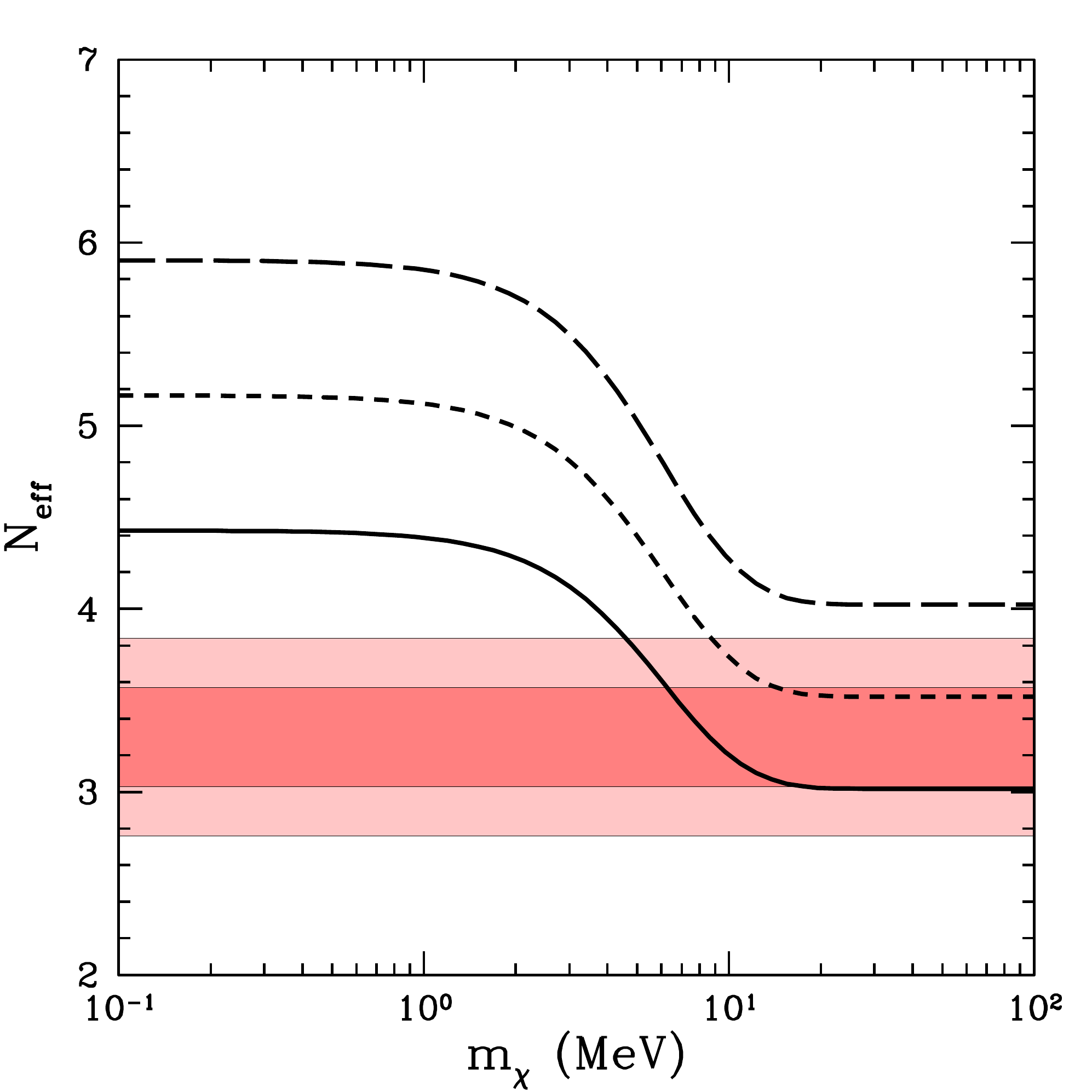}
\\\vskip 0.2in
\caption{(Color online) As for Fig.~\ref{fig:neffvsmem}, but for a WIMP that couples to neutrinos instead of the electromagnetic plasma.   From bottom to top in the left panel, the dashed blue curve is for a real scalar, the solid black curve is for a Majorana WIMP, the dashed green curve is for a complex scalar, and the solid red curve is for a Dirac WIMP.  The horizontal, red/pink bands are the Planck CMB 68\% and 95\% allowed ranges for N$_{\rm eff}$.  The right panel shows N$_{\rm eff}$ as a function of the WIMP mass for a light, Majorana fermion WIMP and $\Delta N_\nu^{*}$~equivalent neutrinos.  The solid curve is for $\Delta {\rm N}_{\nu}^{*} = 0$, the dotted curve is for $\Delta {\rm N}_{\nu}^{*} = 1/2$, and the dashed curve is for $\Delta {\rm N}_{\nu}^{*} = 1$.  The horizontal, red/pink bands are the Planck CMB 68\% and 95\% allowed ranges for N$_{\rm eff}$.  (After Fig.~12 of Ref.~\cite{chimera}.)}
\label{fig:neffvsmnu}
\end{center}
\end{figure}

Figure \ref{fig:neffvsmem} illustrates the degeneracy between the WIMP
mass and the number of equivalent neutrinos.  For example, while the
Planck CMB constraint excludes a very low-mass ($m_{\chi} \lsim
10\,{\rm MeV}$) WIMP when $\Delta {\rm N}_{\nu}^{*} = 0$, it is
consistent with $\Delta\mathrm{N}_\nu^{*} = 2$ in the presence of a
sufficiently low-mass WIMP, and more massive light WIMPs ($m_{\chi}
\gsim 10\,{\rm MeV}$) are excluded for $\Delta\mathrm{N}_\nu^{*}\gsim
1$.  For a Majorana fermion WIMP that couples electromagnetically,
depending on its mass, $-0.2 \lsim \Delta{\rm N}_{\nu} \lsim 2.5$ is
allowed by the CMB.\footnote{In principle, the number of equivalent
  neutrinos should be non-negative, \Deln~$\geq 0$, since it is known
  that the three SM neutrinos mix thoroughly before and after they
  decouple (e.g., Refs.~\cite{mangano,dolgov-osc,pastor}).  In the
  subsequent analysis \Deln~$ < 0$ is allowed and compared to the
  results where a prior is imposed, restricting the number of
  equivalent neutrinos to \Deln~$\geq 0$.  In fact, it is found that
  \Deln~$< 0$ is marginally disfavored when any CMB constraint is
  included.}

The corresponding results for the contrasting case of a light WIMP that couples only to neutrinos are shown in Fig.~\ref{fig:neffvsmnu}.  In the neutrino coupled case, $-0.2 \lsim \Delta{\rm N}_{\nu} \lsim 0.8$ is allowed by the CMB, depending on the WIMP mass.\
It is clear from the discussion here that the CMB -- alone -- is insufficient to break the various degeneracies among \mchi, \Deln, and \neff.  However, since the presence of a light WIMP (and equivalent neutrinos) will also affect the early Universe energy and entropy densities before, during, or immediately after primordial nucleosynthesis, BBN provides an independent probe which may help to break some of the degeneracies.  Here, the changes to standard BBN (SBBN: no light WIMP, \Deln~= 0) in the presence of a light WIMP and \Deln~equivalent neutrinos are investigated.  The BBN and CMB (Planck \cite{planck}) constraints are compared in a joint analysis, leading to lower bounds to \mchi~and, to best fits and 68\% and 95\% ranges for \neff, \Deln, and the baryon density parameter, \omb~$\equiv \eta_{10}/273.9$, where the baryon-to-photon ratio is $\eta \equiv (n_{\rm B}/n_{\gamma})_{0} = 10^{10}\eta_{10}$.

In the analysis here, the key connection among \neff, \Deln, and \mchi~is
\beq
{\rm N}_{\rm eff}(m_{\chi},\,\Delta{\rm N}_{\nu}^{*}) \equiv {\rm N}_{\rm eff}^{0}(m_{\chi})(1 + \Delta{\rm N}_{\nu}^{*}/3)\,,
\eeq
where ${\rm N}_{\rm eff}^{0} \equiv 3[(11/4)(T_{\nu}/T_{\gamma})^{3}_{0}]^{4/3}$ depends on the nature and interactions of the WIMP, along with the WIMP mass.  In our further discussion, the superscript ``$*$" in \Deln~is (usually) suppressed with the understanding that \Deln~need not be an integer or an integer multiple of 4/7.  For the specific case of a sterile neutrino, it is assumed that \Deln~= 1.

\section{BBN In The Presence Of  A Light, Electromagnetically Coupled WIMP}
\label{sec:BBN}

Here, the predictions of BBN are explored in the presence of an electromagnetically coupled light WIMP, allowing for \Deln~equivalent neutrinos.  This analysis extends and updates the earlier BBN calculations of Refs.~\cite{ktw,serpico,boehm2013}, which did not allow for the possibility of equivalent neutrinos, and it compares the BBN predictions with recently updated estimates of the relic abundances of helium and deuterium.

\subsection{Physical effects}
\label{sec:BBN-effects}

Light WIMPs affect BBN in three different ways \cite{ktw,serpico,boehm2013}.  The first of these effects is to modify the expansion rate as a function of the temperature, resulting from the WIMP's contribution to the energy density of the Universe.  Sufficiently light WIMPs ($\lesssim 1$ MeV) will be present as a thermally populated relativistic species during some part of BBN, contributing directly to the energy density.  In addition, a thermally populated WIMP that couples to the electromagnetic plasma and is still at least somewhat relativistic at the time of neutrino decoupling ($m_\chi \lesssim 20$ MeV) heats the photons relative to the neutrinos later, when the temperature is small compared with its rest mass.  The same WIMP species can have both of these effects on the expansion, acting as a relativistic species at the time of weak freeze-out ($T_\gamma\sim 0.8$ MeV) and as a vanished source of photon entropy when the charged-particle nuclear reactions freeze out ($T_\gamma\sim 40$ keV) at the end of BBN.  A $m_{\chi} \sim 1$ MeV WIMP has this property.

\begin{figure}[!t]
\begin{center}
\includegraphics[width=0.45\columnwidth]{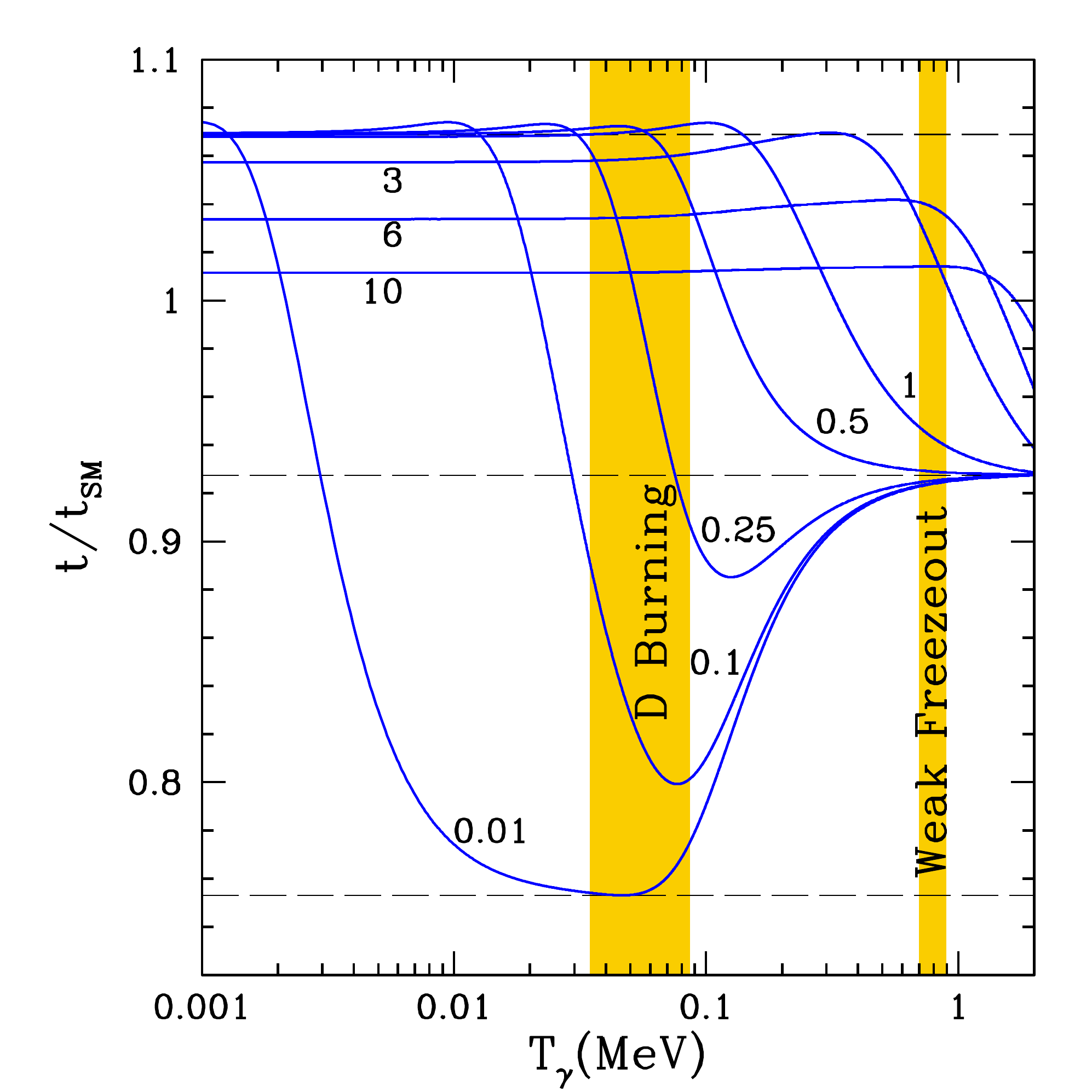}
\includegraphics[width=0.45\columnwidth]{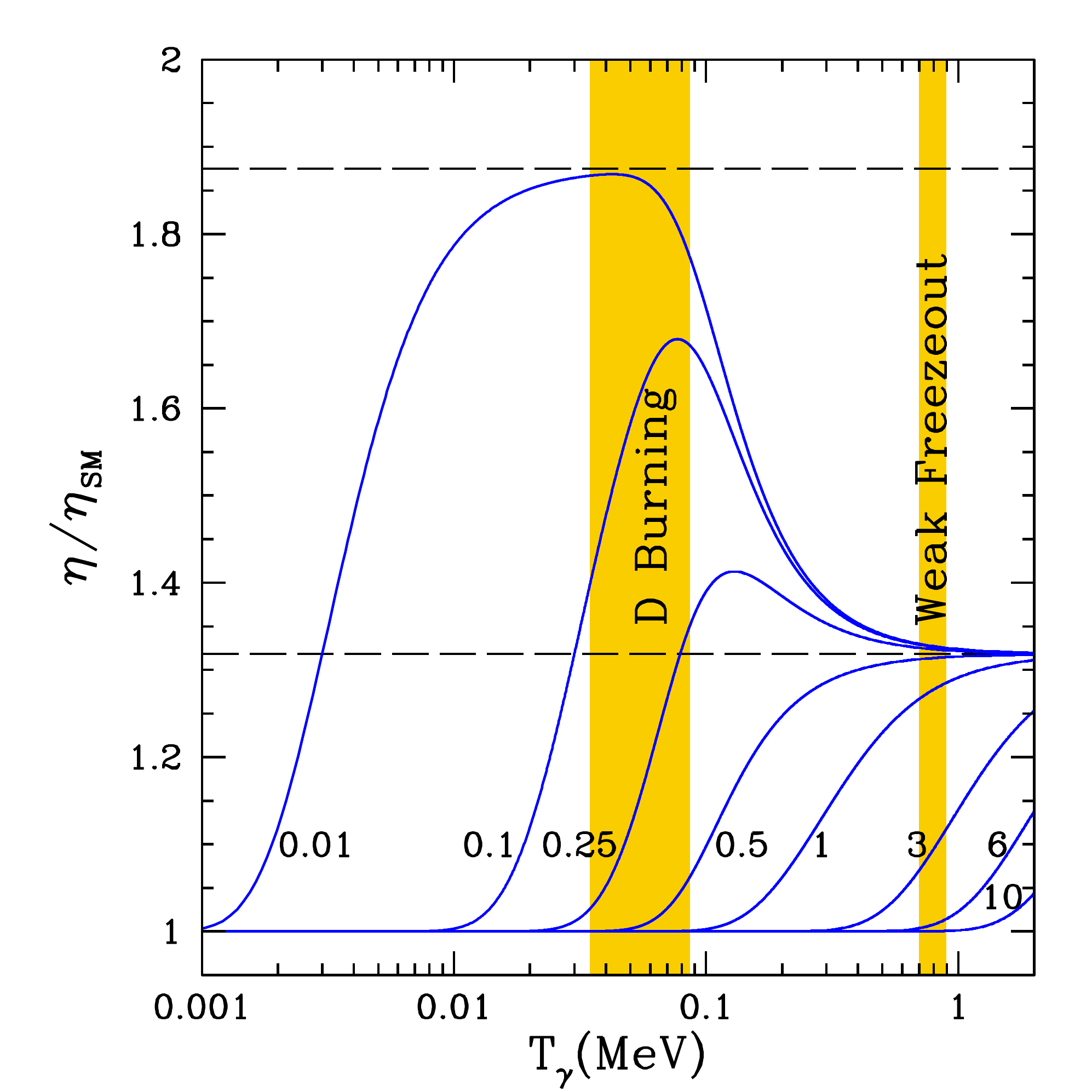}
\caption{(Color online) The left panel shows the elapsed time at fixed photon
  temperature $T_\gamma$ in the presence of a light Majorana WIMP with
  $\Delta N_\nu=0$, relative to elapsed time at $T_\gamma$ in the
  standard model, $t_\mathrm{SM}$.  Each curve is labeled by its value
  of \mchi~in MeV.  Horizontal dashed lines indicate limiting cases in
  which all particles were relativistic at neutrino decoupling, but then
  (top line) all light species ($m < 100$ MeV) have annihilated fully,
  (medium line) all light species including the WIMP remain
  relativistic, and (bottom line) $e^\pm$ annihilation is complete but
  the WIMPs remain relativistic.  The right panel shows the
  baryon-to-photon-number ratio, $\eta$, normalized to this ratio in
  SBBN, as a function of the photon temperature $T_\gamma$, for a
  model including a Majorana WIMP.  The comparison is between models
  that have the same value of $\eta$ today.  For $m_\chi \gtrsim m_e$,
  the largest departure from the standard model is about 30\%, but it
  occurs too early to have a strong influence on the BBN yields.  For
  $m_\chi \lesssim m_e$, $\eta$ during final deuterium burning is
  higher than $\eta$ today.  Here the upper dashed line indicates the
  limit of fully annihilated $e^\pm$ pairs with fully relativistic
  WIMPs, while the lower dashed line indicates the limit with all
  light species fully relativistic.}

\label{fig:timescales}
\end{center}
\end{figure}

These modifications to the expansion rate are shown in the left panel of Fig.~\ref{fig:timescales} for Majorana WIMPs of various masses.  In this figure it is assumed that $\Delta N_\nu=0$.  The vertical axis shows the time elapsed since the initial big bang singularity, which is very nearly equal to the inverse of the expansion rate, divided by the elapsed time at the same photon temperature $T_\gamma$ in the standard model.  Thus, a ratio of unity indicates the same evolution as in SBBN, a ratio greater than unity indicates slower evolution, and a ratio less than unity indicates faster evolution.

By considering separate entropy conservation in the neutrino and electromagnetic fluids after neutrino decoupling, several limiting cases may be explored.  First, if the WIMP mass $m_\chi\gtrsim 20$ MeV, the WIMPs annihilate fully before neutrino decoupling, having no quantitatively significant effect on BBN.  For such large WIMP masses, the time scales are the same as in the standard model.  For lower $m_\chi$, the WIMP behaves at first ($T_\gamma \gtrsim m_\chi/3$) as a relativistic species speeding up the expansion, but later on, it behaves as a vanished source of the enhanced temperature differences between neutrinos and photons, actually slowing the late-time expansion.

Figure \ref{fig:timescales} reveals another important feature of the light-WIMP scenario.  The energy scales discussed so far are set by $m_\chi$, the neutrino decoupling temperature $T_{\nu d}$ (which determines how much WIMP mass/energy enters the electromagnetic plasma), the weak freeze-out temperature $T_{np}\sim 0.8$ MeV, and the charged-particle freeze-out temperature at the end of primordial nucleosynthesis, $T_\mathrm{nuc}\sim 40$ keV.  In Fig.~\ref{fig:timescales}, another important energy scale is evident: the electron mass $m_e$.  A WIMP with $m_\chi< m_e$ remains relativistic even after $e^\pm$ annihilation and can greatly speed up the expansion (relative to that in the standard model) late in BBN.

The expansion-rate effect has a familiar influence on the $^4$He yield of BBN.  Faster expansion than in SBBN (due to the presence of the still-relativistic WIMPs) causes the weak rates interconverting neutrons and protons to fall out of equilibrium at a higher temperature.  Since the equilibrium at higher temperatures is characterized by more neutrons relative to protons, this results in the synthesis of more $^4$He.  Slower expansion at fixed $T_\gamma$ (caused by WIMPs that have annihilated and left $T_\nu<T_\gamma$) has the opposite effect.

However, there is a second effect that more than cancels this first effect if the WIMPs annihilate at least partially before weak freeze-out.  The colder neutrinos (relative to the photons at fixed $T_\gamma$) resulting from WIMP annihilation suppress the rates of neutron-proton interconversion.  Thus, although colder neutrinos tend to cause later freeze-out through slower expansion, they also tend to cause earlier freeze-out through slower weak rates.

A result of the colder neutrinos in this situation that was not
pointed out in previous work (but is implicit in the published
calculations) is that they do {\bf not} slow $n \rightarrow p$ and
$p\rightarrow n$ conversion rates equally.  The first process has a
positive $Q$ value, while the second has a negative $Q$ value.
Consequently, colder neutrinos suppress proton destruction more than
they suppress neutron destruction, disfavoring neutrons by an
additional mechanism.  In SBBN, weak freeze-out is an exit from
thermal equilibrium, but in the light-WIMP scenario the differing
neutrino and photon temperatures keep the neutron/proton ratio out of
equilibrium.  Although it no longer tracks thermal equilibrium, the
$n/p$ ratio reflects a quasi-equilibrium of the interconversion rates
that produces a compromise between the neutrino and photon
temperatures.  The quasi-equilibrium that is exited at $T_{np}$ is one
favoring protons more than in SBBN.\footnote{Even in SBBN there are
  slightly different photon and neutrino temperatures, because $e^\pm$
  annihilation has already begun as noted in Sec. \ref{sec:overview}.
  However, since $T_\gamma > m_e$ at this time, the effect is much
  smaller than for a WIMP with $\mchi \sim T_{\nu d}$.}

The third effect of a light WIMP on BBN emerges when $m_\chi \lesssim
m_e$.  For SBBN, the only parameter is the present-day
baryon-to-photon ratio $\eta$ (strictly speaking, the ratio of baryon
density to entropy density).  During SBBN, this ratio changes
(evolves) only once, due to $e^\pm$ annihilation.  In the light-WIMP
scenario, it also changes when the WIMP annihilates.  This effect is
shown in the right-hand panel of Fig.~\ref{fig:timescales}.  Curves
are shown for the evolution of $\eta$ in the presence of Majorana
WIMPs, for models normalized to the same final value of $\eta$ today.
Values of $\eta$ are shown divided by $\eta$ evaluated in the standard
model at the same value of $T_\gamma$.  Majorana WIMPs that annihilate
before $e^\pm$ annihilation alter $\eta$ at early times by up to $\sim
30\%$.  Since this only affects the $^4$He abundance, which depends
only very weakly on $\eta$, there is no significant change in BBN in
this case.  However, for fixed $\eta$ today, Majorana WIMPs that
annihilate after $e^\pm$ annihilation result in a value of $\eta$
almost a factor of 2 larger than in the standard model during the
later stages of BBN when the D, $^3$He, and $^7$Li abundances are
established.\footnote{For Dirac WIMPs, the effect is even larger; for
  real scalars, the effect is significantly smaller.}

\subsection{BBN calculations}
\label{sec:bbn-calculation}

The BBN yields in the light-WIMP scenario have been computed using a modified version of the Kawano BBN code \cite{kawano88}.  The code's time-step controls have been modified as described in Ref.~\cite{kawano92}, with parameters chosen to give $\lesssim 0.1$\% precision in all yields -- smaller than the precision of several inputs to our analysis below.  The weak rates were computed using the explicit-integration routines in the Kawano code, modified to include Coulomb, radiative, and finite-nucleon-mass corrections as described in Ref.~\cite{lopez99} and references therein.  The weak-rate effects of the altered time-temperature relations in our scenario enter through the Fermi-Dirac distributions of the electron, positron, and neutrino energies in the rate integrals.
The Coulomb and zero-temperature radiative corrections were computed
explicitly in the rate integrations, and it was confirmed that their
effects are consistent with results in the literature; the finite mass
correction was added as a perturbation to the rates using tables
computed in Refs.~\cite{lopez97,lopezxx}.  The latter correction is
presumably somewhat inaccurate for present purposes, since it was
computed assuming the SBBN relation between photon and neutrino
temperatures.  The finite mass effect changes the $^4$He yield by
$\Delta {\rm Y}_{\rm P} = 0.0012$ in SBBN, and corrections to this correction
arising from altered neutrino temperatures are presumably smaller.  
Additional corrections described in Ref.~\cite{lopez99} that result in
$\Delta {\rm Y}_{\rm P} = +0.0006$ in total have been omitted.  About half of this
difference lies in equation-of-state effects rather than in the weak
rates as such.  (Note that the $^4$He abundance is customarily
expressed as a fraction of the baryonic mass -- strictly speaking, as
a fraction of baryon number density -- and is denoted by Y.  The
symbol Y$_{\rm P}$ denotes its \textit{primordial} value.)

To include the effects of light WIMPs, only a few changes to the code were necessary.  First, the expansion rate of the Universe depends on the total energy density according to the Friedmann equation.  Thus, the energy density $\rho_\chi$ of the WIMPs must be included along with the energy densities of the (SM and equivalent) neutrinos, photons, electrons, and positrons when the expansion rate is computed.  Depending on the nature of the light WIMP, the energy density for this contribution is given by a Fermi-Dirac or Bose-Einstein distribution with assumed zero chemical potential (with $T$ in energy units),
\begin{equation}
\label{eq:wimpenergy}
\rho_\chi = \frac{g_\chi}{\pi^2}\int_{m_\chi}^\infty
\frac{\left(E^2-m_\chi^2\right)^{1/2}}{\exp\left(E/T\right)\pm 1}E^2\,dE.
\end{equation}
The choice of positive or negative sign in the denominator corresponds to fermions or bosons, respectively, and $g_\chi$ is the number of ``internal'' degrees of freedom of the WIMP (1 for real scalars, 2 for complex scalars or Majorana fermions, and 4 for Dirac fermions).  This integral was approximated by the first six terms of a well-established and accurate Bessel function expansion of the integral (also used in the Kawano code to compute the $e^\pm$ energy density)~\cite{chandra1939,serpico}.

The second change is to include the WIMP entropy in computing the thermodynamic effect of universal expansion on the particle temperatures.  Computation of this effect in the Kawano code follows that of the original Wagoner code from which it derives and is somewhat convoluted in its mathematical details.  However, it amounts to computing the time derivatives of $\eta$ and $T_\gamma$ from entropy conservation.  The addition of a quantity $\rho_\chi+p_\chi$ (proportional to the WIMP entropy density) into the expression for the temperature derivative of $\eta$ suffices to incorporate the effects of the WIMPs on the thermodynamics.  [The WIMP pressure, $p_\chi$, is computed from an expression similar to Eq.~(\ref{eq:wimpenergy}) with an analogous Bessel-function expansion.]  The full effect of the WIMPs on the weak rates is contained in the temperature evolution, since the rate integrations are carried out numerically at each $T_\gamma$.

In setting up a BBN calculation, it is necessary to specify an initial value of $\eta$.  This is most usefully done by choosing the value, $\eta_0$, of this parameter today and then computing the value at the start of BBN using conservation of comoving entropy.  For the standard-model particles alone, at $T_{\gamma} = T_{\nu d}$,\begin{equation}
\label{eq:initialeta}
  \eta_{\nu d} = \eta_0\left(1+\frac{s_e}{s_\gamma}\right)\,,
\end{equation}
 with $s_e$ the entropy density of electrons and positrons, and $s_\gamma$ that of photons.  For fully relativistic $e^\pm$ pairs ($m_{e} \rightarrow 0$) at the start of BBN, the factor in parentheses is 11/4.  At the start of the BBN calculation, an initial temperature is chosen, the initial $\eta$ is computed from the target $\eta_0$, and then the calculation begins.  As mentioned in Sec.~\ref{sec:overview} above, the $e^\pm$ are not fully relativistic at neutrino decoupling, so that $s_e/s_\gamma \neq 7/4$ (see \eg, \cite{chimera}).  This correction is a small effect and is ignored in the Kawano code, which by default assumes, fictitiously, early neutrino decoupling at very high temperature ($T_{\nu d} \gg m_{e}$) and ignores muons ($T_{\nu d} \ll m_{\mu}$).  The approximation works reasonably well because $m_e \approx T_{\nu d}/4$ and $T_{\nu d} \ll m_\mu$, where $m_\mu$ is the mass of the muon.

In the light-WIMP scenario, it is necessary to deal explicitly with neutrino decoupling.  Ideally, a complete calculation tracking the interaction rates and departures from equilibrium energy distributions would be performed, as was done for SBBN in Ref.~\cite{mangano}.  Here, instantaneous decoupling of all neutrino species from the electromagnetic plasma at $T_{\nu d} = T_\gamma = T_\nu = 2$ MeV is assumed, the value of $\eta$ is computed using Eq.~(\ref{eq:initialeta}), and the BBN calculation is started at $T_{\nu d}$.  This approximation gives $s_e$ a value about 2\% below its relativistic approximation.  With the entropy $s_\chi$ of WIMPs included, Eq.~(\ref{eq:initialeta}) becomes
\begin{equation}
  \label{eq:wimpinitialeta}
  \eta = \eta_0\left(1+\frac{s_e+s_\chi}{s_\gamma}\right).
\end{equation}
In principle, $\eta$ could continue to evolve after the end of BBN, if $m_\chi\ll T_\mathrm{nuc}$.  A calculation analogous to Eq.~(\ref{eq:wimpinitialeta}) was included at the end of the BBN calculation, but the stopping temperature is low enough that this is insignificant for the $m_\chi \geq 10$ keV cases considered here.  All of these effects were incorporated into the BBN code.  It was explicitly confirmed for several cases (including those in Fig.~\ref{fig:timescales}) that the expansion rate and $\eta$ computed by integrating the differential equations followed the trajectories predicted by algebraic entropy-conservation expressions given in Ref.~\cite{chimera}.

Finally, a comment on the nuclear rates used here is in order.  All weak rates in the BBN calculations are normalized to the measured mean lifetime of free neutrons.  Here, the value $\tau_n = 880.1\pm 1.1$ s, recommended in the last full update of the Particle Data Book~\cite{pdg2012}, is adopted.  This value has since been updated to $880.0 \pm 0.9$ s, reflecting a reanalysis of an experiment contributing to the world average.  Since we learned of this minor change late in our work, we have not used it here.  The measurement of the neutron lifetime continues to suffer from systematic discrepancies between experiments, so it should be kept in mind that both of these values are averages that include inconsistent data.  We look forward to the results of new experiments currently underway.

Other nuclear rates have been updated as described in
Ref.~\cite{nollett11}.  The important rates for the present study are
that for the capture of neutrons on protons, $p(n,\gamma)d$, for which the
cross sections of Rupak \cite{rupak} are used; that of the
deuteron-deuteron reactions $d(d,n)^3\mathrm{He}$ and
$d(d,p)^3\mathrm{H}$, fitted to the data discussed in
Ref.~\cite{nollettburles} plus the recent data of
Ref.~\cite{leonard06} (forced to give the same central BBN yields as
the Monte Carlo method of Ref.~\cite{nollettburles}); and that of the
additional deuterium-burning reaction $d(p,\gamma)^3\mathrm{He}$.  As
discussed in Ref.~\cite{nollett11}, there are few data for this last
cross section at BBN energies, and they are in tension with a
high-quality theoretical calculation \cite{viviani00} that agrees with
other data.  The calculations here use the theoretical rate.  It is
estimated that the overall error from all rates amounts to $\sim
2.5$\% for the predicted D/H and to $\Delta{\rm Y}_{\rm P} \sim
0.0005$ for $^4$He (from the error on $\tau_n$ alone there is a
contribution to \Yp~of $\sim 0.0002$).  The systematic difference
between empirical and theoretical cross sections for
$d(p,\gamma)^3\mathrm{He}$ amounts to $\sim 6$\% in D/H, so that D/H
would be larger by this amount if the empirical cross sections had
been used.

\section{BBN Results}
\label{sec:results}

\subsection{General trends}
\label{sec:trends}

\begin{figure}[!t]
\includegraphics[width=0.49\columnwidth,angle=0]{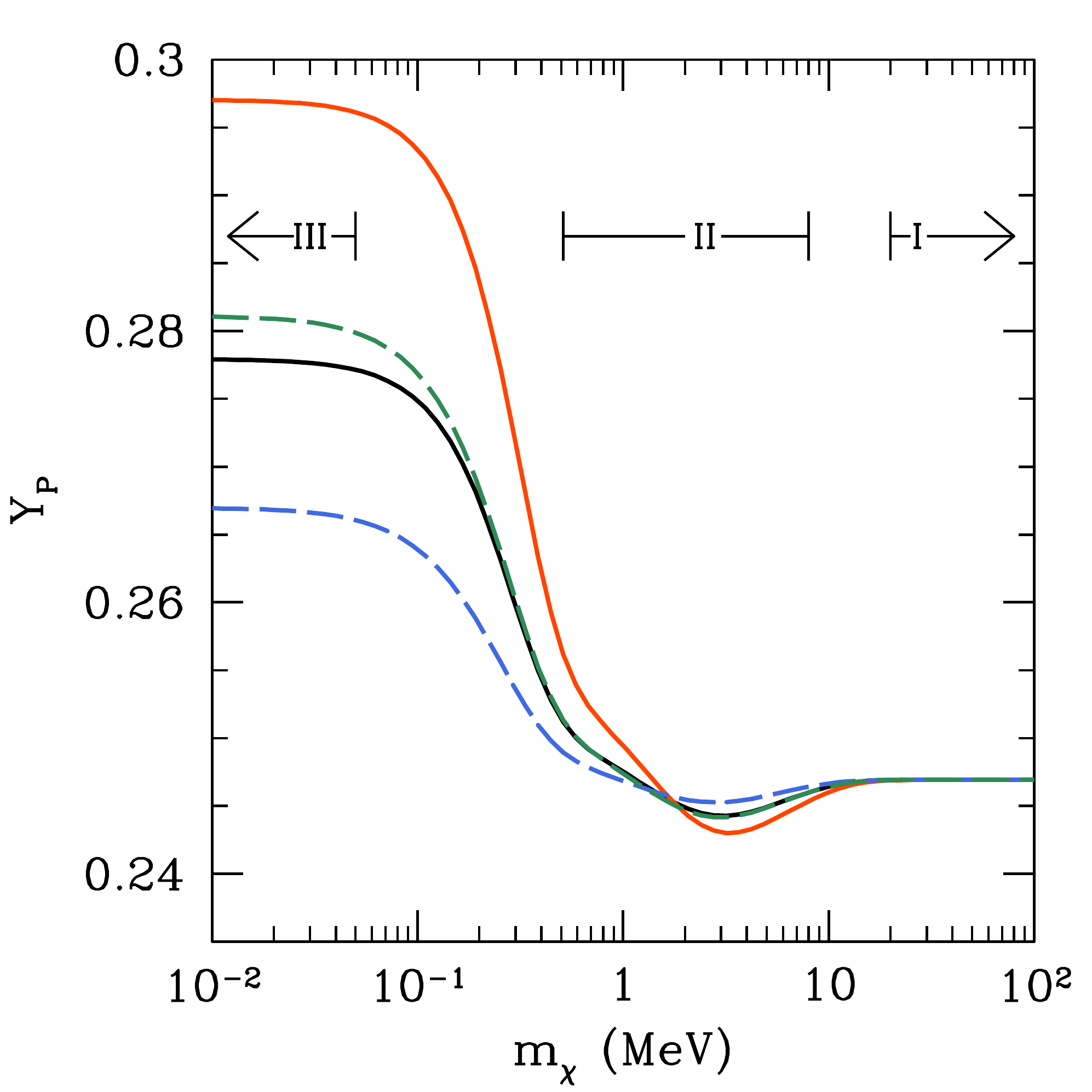}
\includegraphics[width=0.49\columnwidth,angle=0]{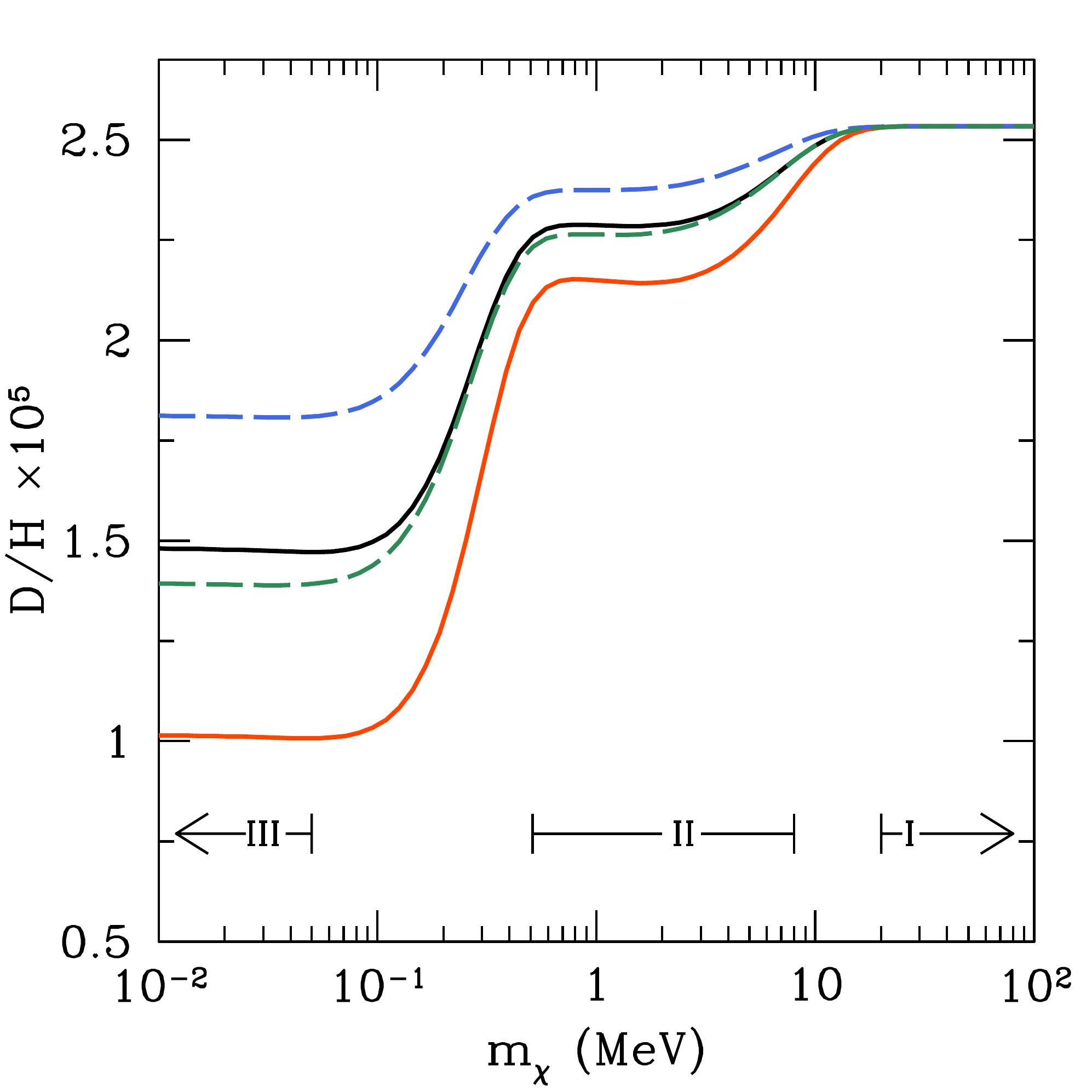}\\
\includegraphics[width=0.49\columnwidth,angle=0]{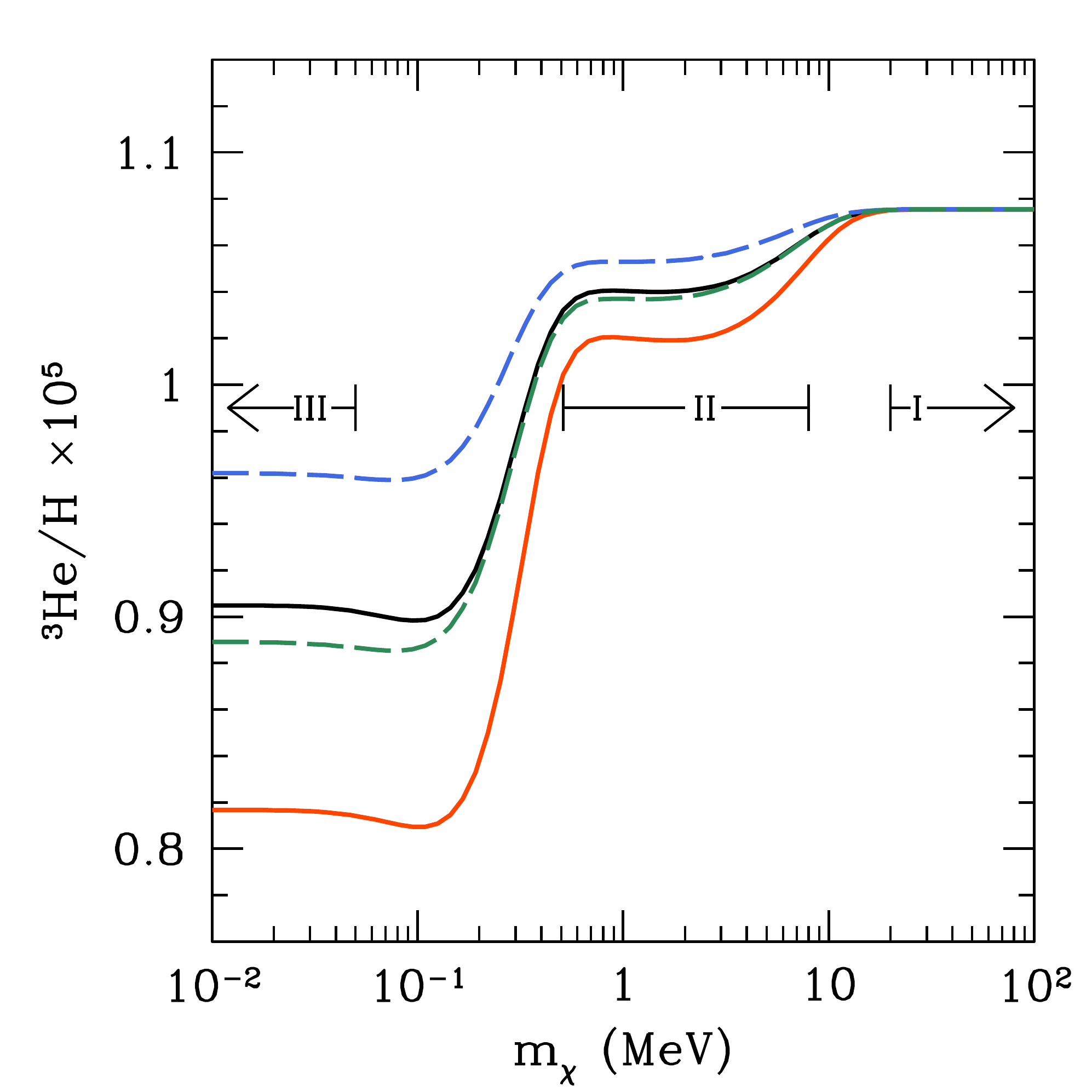}
\includegraphics[width=0.49\columnwidth,angle=0]{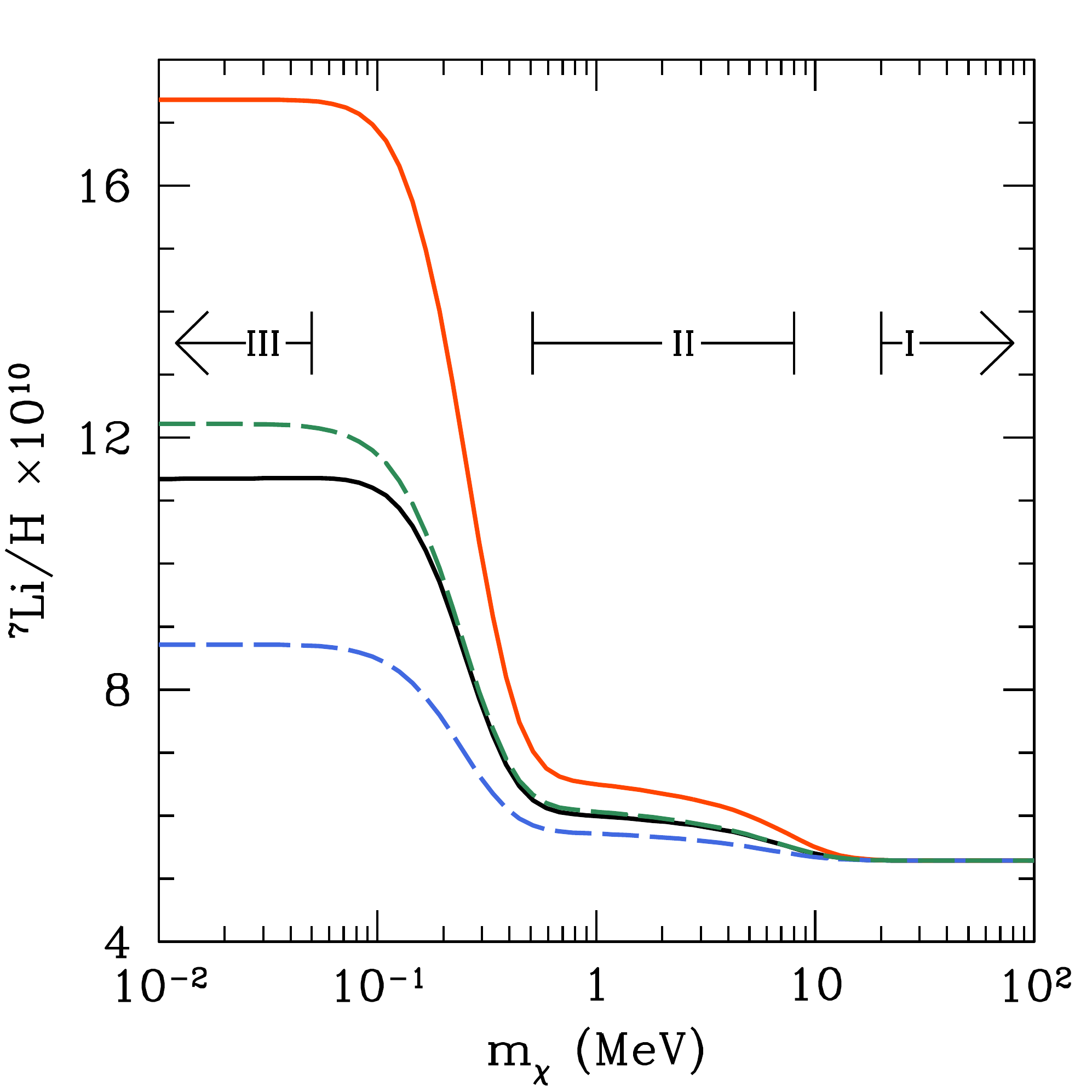}
\caption{(Color online) The four panels show the BBN yields of $^4$He (upper left), D (upper right), $^3$He (lower left), and $^7$Li (lower right) as a function of the WIMP mass, \mchi, for $\Omega_Bh^2 = 0.022$ and $\Deln = 0$.  Solid curves show results for fermionic WIMPs (red for Dirac, black for Majorana) and dashed curves show results for bosonic WIMPs (green for a complex scalar, blue for a real scalar).  In the upper left and lower right panels, the curves in region III are from top to bottom, Dirac fermions, complex scalars, Majorana fermions, real scalars. In the lower left and upper right, the sequence is reversed.  The $^4$He abundance is shown as a mass fraction Y$_{\rm P}$, and the other abundances are shown as ratios by number to hydrogen.}
\label{fig:Deln=0yields}
\end{figure}

As described above, the BBN yields for D and \4he (along with \7li and \3he) are calculated here as functions of $\eta_{10} \equiv 10^{10}\eta_{0}$ (or \omb~$\equiv \eta_{10}/273.9$) and \Deln~for WIMP masses from 10 keV to 40 MeV.  As a starting point for understanding the results and the parameter constraints they provide, the yields for fermionic and bosonic WIMPs are shown as functions of $m_\chi$ for $\Deln = 0$ and the CMB value of \omb~in Fig.~\ref{fig:Deln=0yields}.

Similar results may be found in the prior literature
\cite{ktw,serpico,boehm2013}.  The results here are in excellent
agreement with those presented in Ref.~\cite{boehm2013}.  They are in
fair agreement with those shown in Ref.~\cite{ktw}, the latter having
been computed in 1986 with different rates and a much lower
adopted value of \omb.  There is a small, but real disagreement with
Ref.~\cite{serpico} (and between Refs.~\cite{serpico} and
\cite{ktw}) in the middle mass range of each graph, including the entire region
between the label ``III'' and the label ``I.''  We suspect that this
is most likely the result of a programming error in Ref.~\cite{serpico},
because their results do not show physical effects discussed in the
following paragraphs.

The influence of light WIMPs on BBN may be understood by dividing Fig.~\ref{fig:Deln=0yields} into three regions as in Ref.~\cite{ktw}.  Region I is at the right, on the higher-mass side of the graph: For $m_\chi \gtrsim 20$ MeV, the thermal population of the WIMP is negligible for the entire time after neutrino decoupling and, unable to contribute significantly to the density or to drive a difference between neutrino and photon temperatures, the more massive WIMP has no effect on the BBN yields.  

Region II is in the middle of the graph, for $m_e \lesssim m_\chi \lesssim 3\,T_{\nu d}$.  In this intermediate mass range, most of the WIMPs annihilate after $T_{\nu d}$ but before much BBN has occurred (cf.~Fig.~\ref{fig:timescales}).  In this case, the neutrinos are colder than the photons (which have been heated by WIMP annihilation) by a larger difference than in SBBN.  As a result, the expansion at fixed $T_\gamma$ is slower than in SBBN, so there is more time to destroy deuterium and $^3$He and to assemble $^7$Be (which later becomes $^7$Li).  Region II is characterized in Fig.~\ref{fig:Deln=0yields} by plateaus in the four yield curves, corresponding to WIMPs that are almost fully relativistic at neutrino decoupling but fully annihilated before electrons and positrons annihilate (the top dashed line in the left panel of Fig.~\ref{fig:timescales}).  It is the absence of this plateau effect in region II of Ref.\,\cite{serpico} (but not in Ref.~\cite{ktw} or \cite{boehm2013}) that leads us to suspect a problem in their calculations.  The effect of the WIMPs on the $^4$He abundance in region II is somewhat complicated.  Figure \ref{fig:timescales} shows that a WIMP with $m_\chi \gtrsim 2$ MeV has mostly annihilated before $np$ weak freeze-out ($T_\gamma \sim 0.8$ MeV), leading to the effects discussed early on in Sec.~\ref{sec:BBN-effects}: at fixed $T_\gamma$, expansion is slower than in SBBN, the weak rates overall are slower by a comparable amount, and conversion of protons to neutrons is inhibited even before freeze-out.  These three effects very nearly cancel each other completely, as may be seen in Fig.~\ref{fig:Deln=0yields}.

In region III of very light WIMPs, where $m_\chi < m_e$, the leading effect on the primordial $^4$He abundance is the expansion-rate effect due to the WIMPs acting essentially as an additional neutrino species during weak freeze-out.  There is also an expansion-rate effect on the other abundances, which would tend to raise D/H and lower $^7$Li/H through earlier (compared to SBBN) charged-particle freeze-out.  However, in region III, $\eta$ has not yet reached its final value when BBN is completed.  WIMP annihilation for masses in region III reaches completion (as far as the WIMP contributions to early Universe conditions is concerned) after BBN has ended, so that $\eta_0$ is lower than $\eta$ at the end of BBN.  Since D/H decreases and $^7$Li/H increases with increasing $\eta$, BBN then produces lower D/H and higher $^7$Li/H than expected in SBBN for the present value of $\eta = \eta_{0}$.  Figure \ref{fig:Deln=0yields} shows that this is the dominant effect.  If additional light neutrino species ($\Deln \neq 0$) are allowed along with a light WIMP, the story in all three regions remains unchanged apart from the additional expansion rate effects from the energy density contributed by the equivalent neutrinos.

There are some similarities between the light-WIMP scenario considered
here and cosmologies with low reheating temperatures that were
proposed several years ago \cite{hannestad2004,ichikawa2006}.  For
values of the reheating temperature $T_R$ below $\sim 7$ MeV,
neutrinos do not come into full equilibrium with photons, with the
result that neutrinos can have different temperatures than in the
standard cosmology.  In the light-WIMP scenario,
Fig.~\ref{fig:neffvsmem} and the upper left-hand panel of
Fig.~\ref{fig:Deln=0yields} show that as the WIMP mass decreases,
\neff\ decreases (for fixed \Deln), and the BBN predicted helium
abundance \Yp~increases (assuming fixed baryon density and ignoring
the small dip in \Yp\ around 3 MeV).  If $T_R$ is small and neutrinos
are not produced directly in reheating, $T_R$ plays a similar role to
\mchi: as $T_R$ decreases, \neff\ decreases and \Yp\ increases.  When
a low $T_R$ was most recently proposed, the observationally favored
value of the primordial helium abundance was \Yp~$ \simeq 0.239$,
significantly different from the SBBN-predicted value of \Yp~$\simeq
0.247$.  At face value, this suggested that $\Deln < 0$ and $\neff <
3$.  While a low reheat temperature might account for such a low value of
\neff, it actually exacerbates the helium abundance problem by
producing higher \Yp.  The BBN-related physics of a low $T_R$ is closely related
to that of a light WIMP, in that the interplay between the neutrinos'
roles in the expansion rate and in the weak rates can have
counterintuitive effects on \Yp.  In contrast to the low-$T_R$
scenario, in the light-WIMP scenario considered here, a low value of
$\mathrm{N_{eff}^0}$ can be offset by $\Deln > 0$, so that $\neff \geq 3$.
Extensions of low-$T_R$ models to include sterile neutrinos have also
been considered \cite{gelmini2004}.

\subsection{Observed abundances}
\label{sec:observed-abundances}

For several reasons, among the light nuclides synthesized during BBN, D and \4he have the most value in constraining the cosmological parameters.  First, the relic abundances are inferred observationally long after BBN has ended, so the post-BBN evolution of the elements needs to be accounted for.  (For a more extended discussion and further references, see Ref.~\cite{steig2007}.)  For D and \4he the expected post-BBN evolution is simple and monotonic.  As gas is cycled through stars, any prestellar deuterium is burned away and any deuterium produced in the course of stellar nucleosynthesis is immediately transformed to \3he, \4he, and heavier nuclei~\cite{epstein}.  As a result, the deuterium abundance measured anywhere in the Universe, at any time in its evolution, should provide a lower bound to the primordial value.  In particular, for observations in systems of low metallicity (where very little gas has cycled through stars), the observationally inferred D abundance should provide a good estimate of its primordial value.  Here, \yd~$\equiv 10^{5}({\rm D/H})_{\rm P} = 2.60 \pm 0.12$ is adopted, based on the relatively recent Pettini and Cooke (2012) \cite{pettini} compilation of deuterium abundances inferred from observations of low-metallicity, high-redshift, absorption line systems along 11 lines of sight.

In a similar manner, as gas cycles through stars, hydrogen is burned to \4he, an effect that dominates over the stellar burning of \4he to heavier elements.  The post-BBN \4he abundance is consequently expected to be nondecreasing with time and/or metallicity.  By observing helium in systems (\ie, extragalactic \hii~regions) whose metallicity (\eg, oxygen abundance, O/H) is overall low but spans a sufficiently large range, it is possible to account for the increase in helium mass fraction Y with metallicity and to infer the primordial value.  From a very recent study of more than 1600 \hii~regions, Izotov \etal~\cite{izotov}, extracted a subset of more than 100 low-metallicity systems, measuring the helium abundance to an accuracy better than 3\% and deriving Y$_{\rm P} = 0.254 \pm 0.003$ from a linear regression of their observed Y -- O/H relation.  This value is adopted for the primordial helium abundance used in the analyses presented here.  Clearly, the central values and errors for the parameters (\omb, \Deln, \mchi) inferred from our BBN analysis here are directly tied to the central values and errors adopted for D/H and Y$_\mathrm{P}$.

The other light nuclides produced in significant abundances during
BBN, \3he and \7li, are of less value in constraining the cosmological
parameters because, in part, their post-BBN evolutionary histories are
more complicated, involving the competition between production,
destruction, and survival, some of it occurring in the sites where
they are observed.  Moreover, once \omb~has been fixed by the CMB and
the late-BBN time scale has been fixed by the observed D/H, the predicted
\3he and \7li abundances are very nearly uniquely determined in the
light-WIMP scenario.  This is because their abundances are set at the
same time as that of deuterium (cf. Fig.~\ref{fig:timescales}).

\3he has only been observed in a handful of \hii~regions in the Galaxy
\cite{bania}, and the inferred \3he abundances show no correlation
with metallicity or location in the Galaxy despite the expectation of
significant post-BBN production from the simplest models of stellar
evolution.  This makes any attempt to infer the primordial \3he
abundance from such data model dependent.  Furthermore, the
BBN-predicted \3he abundance depends on \omb~and \Deln~very similarly
to that of deuterium, but it is less sensitive to them than is the BBN
D abundance.  \3he is not used in the analysis here, but its observed
abundance is consistent with the fitted BBN models presented below.
For lithium, in addition to the uncertain, model-dependent post-BBN
evolution, there is the well-known ``lithium problem" (see \eg,
\cite{fields,spite}).  That is, in the absence of a light WIMP, the
BBN-predicted lithium abundance is higher than the abundance inferred
from observations of metal-poor halo stars in the Galaxy by a factor
$\gsim 3$.  As may be seen in Fig.~\ref{fig:Deln=0yields}, a light
WIMP increases the BBN-predicted \7li abundance, exacerbating this
problem.  Increasing \Deln\ reduces the predicted lithium abundance,
helping with the lithium problem, but it does not solve it (even if there
are no WIMPs).  As with \3he, after our best-fit parameters (and their
ranges) are identified, without using lithium to constrain them, the
BBN-predicted \7li abundance is larger than the observationally
determined values, confirming (and exacerbating) the lithium problem.
This is illustrated in Fig.~\ref{fig:lithium} below.

\subsection{Parameter constraints from BBN alone}
\label{sec:constraints-from-bbn}

\begin{figure}
\includegraphics[width=0.5\columnwidth]{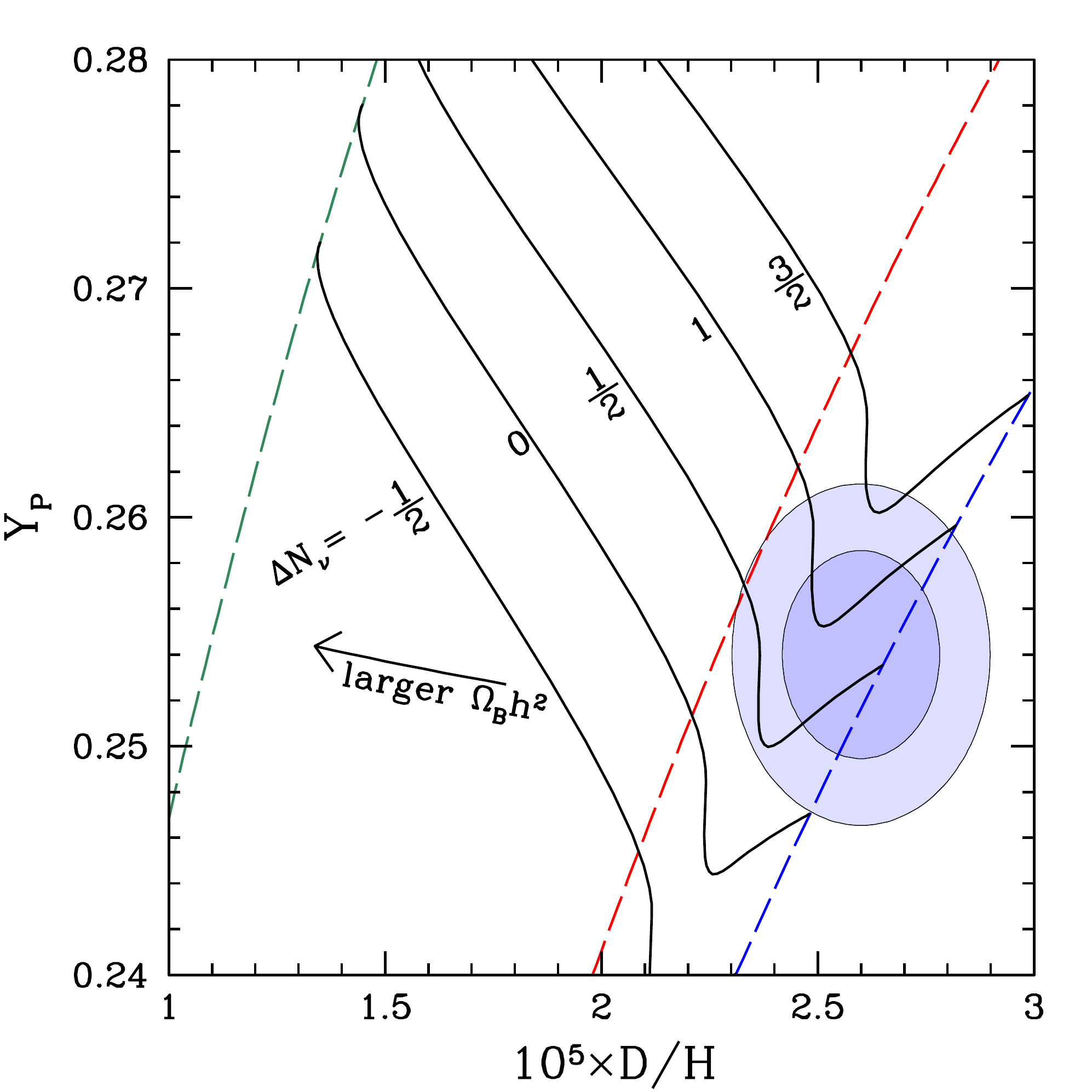}
  \caption{(Color online) BBN yields of D/H and \Yp, computed for Majorana WIMPs at \omb~$= 0.022$ and several values of \Deln.  Values of \Deln~are labeled.  Along each curve of fixed \Deln, $m_\chi$ varies from $\infty$ at the right end to 10 keV at the left end.  The dashed curves show the yields at fixed $m_\chi$ but varying \Deln~with $m_\chi\rightarrow\infty$ (blue, right side), $m_\chi = m_e$ (red, middle), and $m_\chi = 0$ (green, left side).  The observational constraints Y$_{\rm P} = 0.254\pm 0.003$ and $y_\mathrm{D} = 2.60\pm 0.12$ are shown as 68\% (darker) and 95\% (lighter) joint confidence intervals.  The arrow indicates the direction that the BBN yields shift if \omb~is changed.}
  \label{fig:wimpyields-majorana}
\end{figure}

With D/H and Y$_{\rm P}$ identified as the observables for our BBN
model, the behavior of the BBN predictions in the space of these two
variables is now examined.  To illustrate the results, the BBN yields
for a Majorana WIMP are shown in Fig.~\ref{fig:wimpyields-majorana},
along with the 68\% and 95\% contours for the observed \4he and D
abundances.  Each solid curve shows the BBN yields at fixed \omb~$=
0.022$ (equal to its value from Planck \cite{planck} for \Deln~= 0)
and fixed \Deln, with \mchi~varying along each curve from 40 MeV
($m_{\chi} \geq 40\,{\rm MeV}$; lower-right end) to 10 keV ($m_{\chi}
\leq 10\,{\rm keV}$; upper-left end).  The yields along the $\Deln =
0$ curve are those shown in Fig.~\ref{fig:Deln=0yields}.  As
\mchi~decreases from infinity ($m_{\chi} \gg 40\,{\rm MeV}$), no
abundance changes occur until $m_\chi\lesssim 20$ MeV.  Then, for
lower WIMP masses, D/H and Y$_{\rm P}$ both decrease until Y$_{\rm P}$
reaches a minimum at $m_\chi \sim 3$ MeV.  A ``hook'' forms in the
yield curves where, for smaller \mchi, Y$_{\rm P}$ increases at nearly
fixed D/H until $m_\chi \sim m_e$.  At $m_\chi \sim m_e$, both
abundances begin to evolve rapidly with decreasing \mchi, with Y$_{\rm
  P}$ increasing while D/H decreases.  Finally, for $m_\chi\lesssim
20$ keV, a low-$m_\chi$ limit is reached beyond which the BBN yields
cease changing.

In Fig.\,\ref{fig:wimpyields-majorana}, yield curves are also shown for
$\Deln = -0.5,\,0.5,\,1.0$, and\,1.5.  It can be seen that adding
equivalent neutrinos shifts the entire yield curve toward higher
Y$_{\rm P}$ (and slightly higher D/H) while making only small changes
in its shape.  In particular, there is always a minimum of Y$_{\rm P}$
at $m_\chi \sim 3$ MeV and always rapid evolution of the yields toward
very high Y$_{\rm P}$ and very low D/H for $m_\chi < m_e$.  Curves are
also shown in Fig.~\ref{fig:wimpyields-majorana} for continuously
varying \Deln~at fixed values of $m_\chi = 0$, $m_e,$ and $\infty$.
It can be seen that at the CMB-inferred value of \omb, any $\Delta
{\rm N}_{\nu}$ outside the range $0 \leq \Deln < 1.5$ is
disfavored.\footnote{Despite having not restricted our BBN analysis
  to the physical range $\Deln \geq 0$, in fact, the data disfavor
  $\Deln < 0$.}  This conclusion does not change by very much if
$\Omega_Bh^2$ is varied, causing the locations of the yield curves to
shift: the well-known strong dependence of D/H and weak dependence of
Y$_{\rm P}$ on $\Omega_Bh^2$ simply move the yield curves left or
right, but not very far up or down, as indicated by the arrow in
Fig.~\ref{fig:wimpyields-majorana}.  We return to the effect of
decreases in \omb~shortly.

\begin{figure}
\includegraphics[width=0.5\columnwidth]{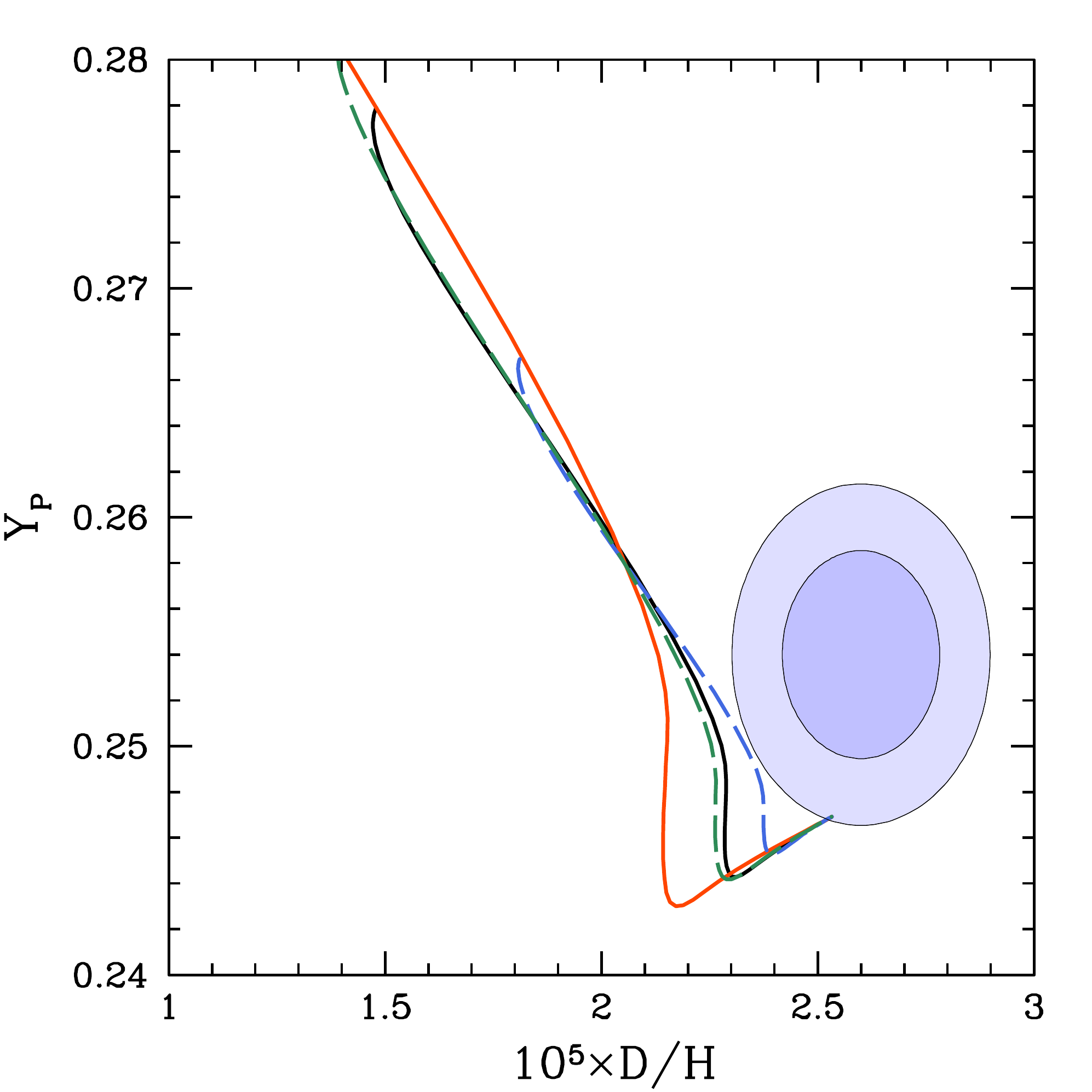}
  \caption{(Color online) Yield curves for \Yp~and $y_{\rm DP}$ as a
    function of \mchi~at constant $\Deln = 0$ and \omb, for the four
    kinds of WIMPs considered here: real and complex bosons ($g_\chi =
    1,2$), and Majorana and Dirac fermions ($g_\chi = 2,4$); colors are
    as in Fig.\,\ref{fig:Deln=0yields}, so that at \Yp$ = 0.25$, the
    curves are from left to right Dirac fermion, complex scalar,
    Majorana fermion, real scalar.  It can be seen that while the
    overall scale of the deviations from SBBN (where the curves end at
    the right) depends on the nature of the WIMP, the general shape of
    the curves -- particularly the hook at $m_\chi \sim 3$ MeV and the
    rapid change at $m_\chi \lesssim m_e$ -- does not.}
  \label{fig:wimpyields-allkinds}
\end{figure}

Although the results here are shown for Majorana WIMPs, the story is
not very different for WIMPs with different spin statistics.  This is
demonstrated in Fig.~\ref{fig:wimpyields-allkinds}, where the yield
curves are shown at fixed $\Deln=0$ and \omb~for the four kinds of WIMPs
considered here.  All these curves show very much the same shape, with
a ``hook'' at $m_\chi \sim 3$ MeV and rapid changes for $m_\chi
\lesssim m_e$.  All four curves converge to a single point in the D/H
vs.~Y$_{\rm P}$ plane, corresponding to $m_\chi \rightarrow \infty$ or
SBBN.  It is evident from this figure that the BBN constraints on a
light WIMP do depend on its spin: \eg, a $g_\chi = 1$ boson has a
smaller effect than a Dirac fermion of the same mass.  For the
remainder of this paper the results in the figures are shown for a
Majorana WIMP.  The detailed quantitative results for all WIMP types
are found in Table~\ref{tab:bounds}.  As shown below, it is only the
constraints on $m_\chi$ that are significantly affected by the nature
of the WIMP.

\begin{figure}[!t]
\begin{center}
\includegraphics[width=0.32\columnwidth]{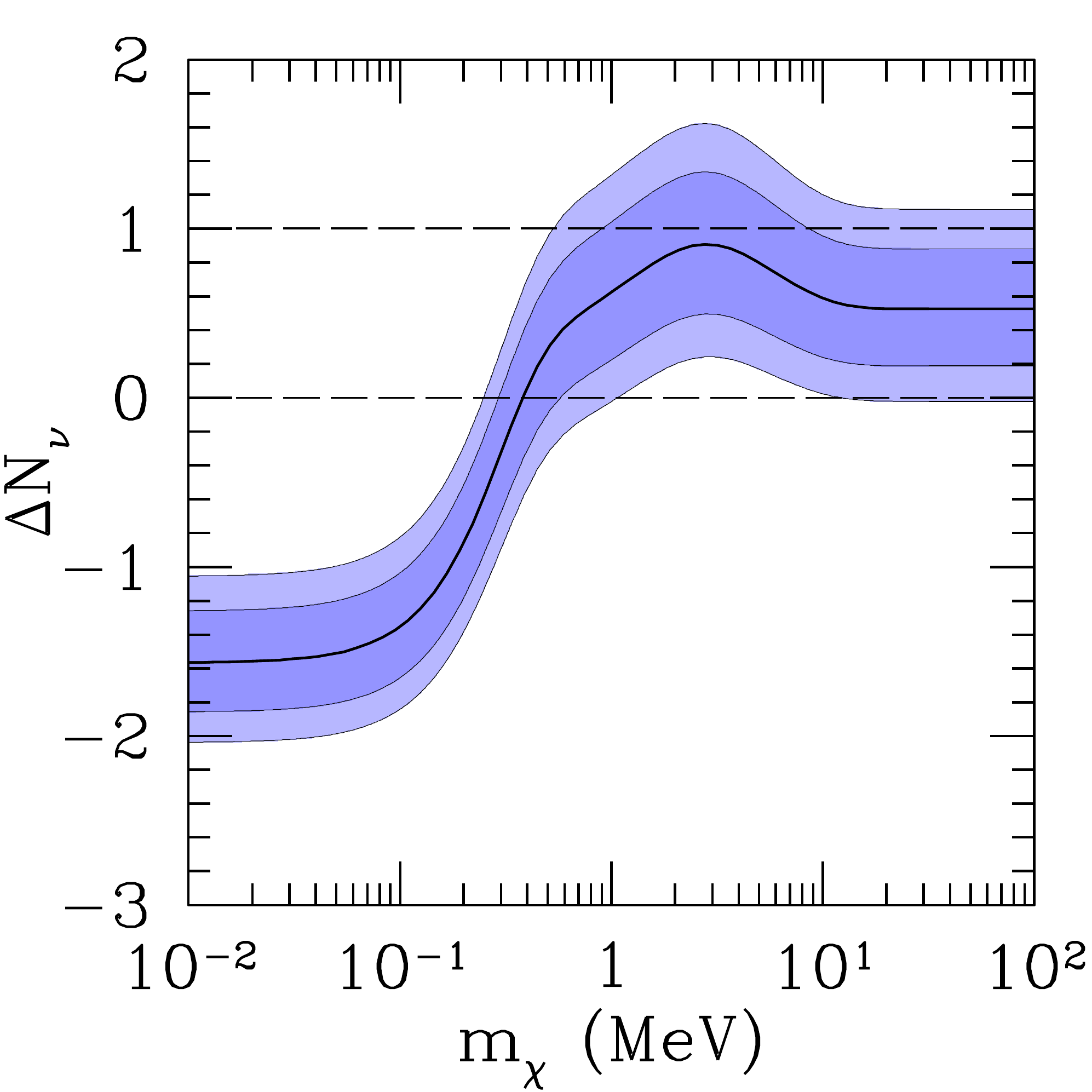}
\includegraphics[width=0.32\columnwidth]{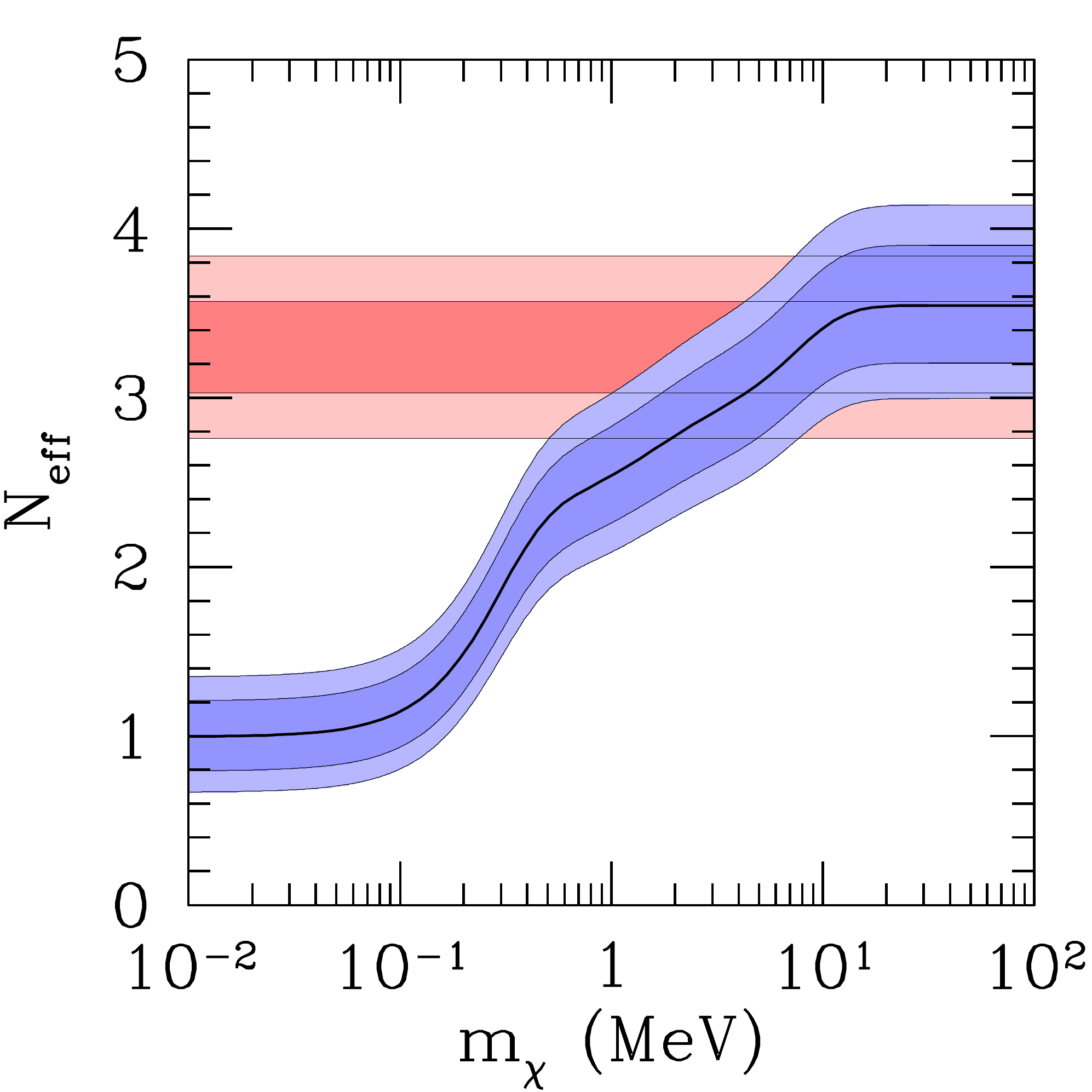}
\includegraphics[width=0.32\columnwidth]{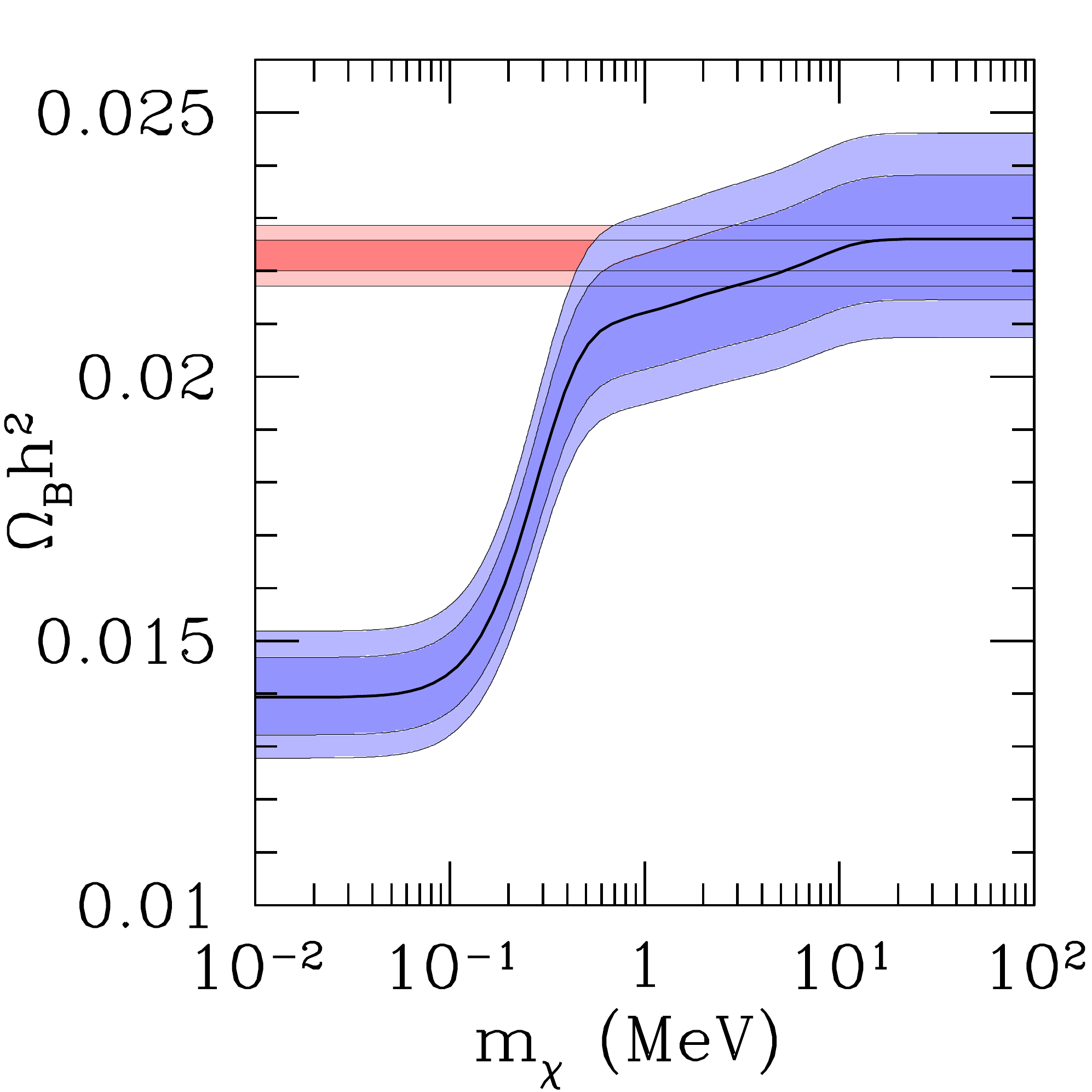}
\caption{(Color online) The various panels show \Deln~(left panel), \neff~(middle panel), and \omb~(right panel) as functions of the WIMP mass.  Darker and lighter blue (curved) contours show the 68\% and 95\% confidence-level regions of the joint likelihoods for each pair of parameters, using as constraints only D/H and Y$_{\rm P}$.  The thick dark curve running through the middle of these regions shows the best fit at each specified \mchi~value from 10 keV to 100 MeV.  In the left-hand panel the dashed lines show \Deln~= 0 and 1 as guides to the eye.  Underlain in pink horizontal bands in the middle and right panels are the 68\% and 95\% joint confidence-level regions corresponding to the $\Lambda\mathrm{CDM}+N_\mathrm{eff}$ fit from the Planck Collaboration, including both CMB and BAO data \cite{planck}, as discussed in the text.}
\label{fig:majorana-bbn-only}
\end{center}
\end{figure}

\begin{figure}[!t]
\begin{center}
\includegraphics[width=0.45\columnwidth]{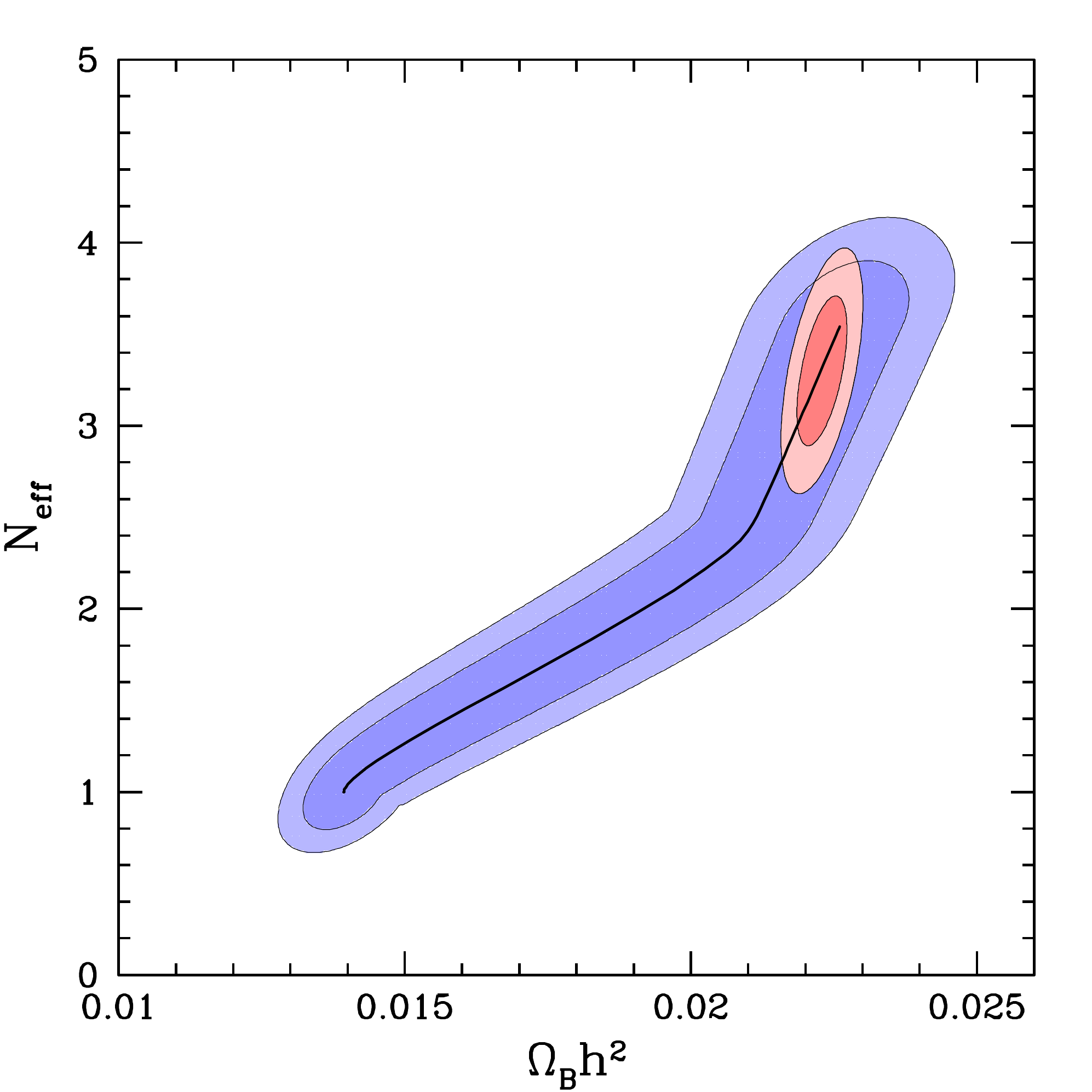}
\includegraphics[width=0.45\columnwidth]{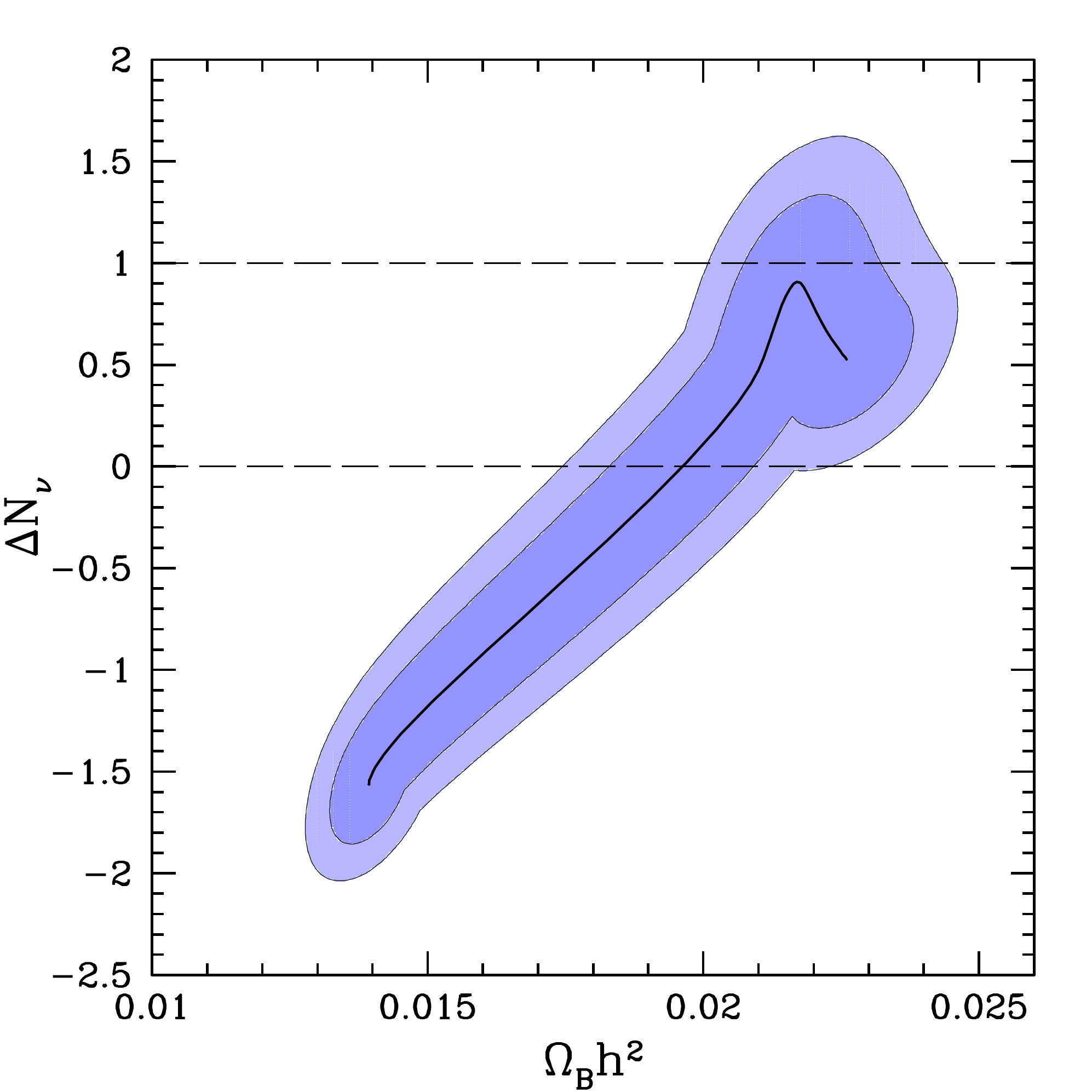}
\caption{(Color online) The left-hand panel shows the 68\% and 95\% contours for the
  BBN constraints in the \Deln~-- \omb~plane that follow from
  Fig.\,\ref{fig:majorana-bbn-only}.  The heavy curve is for perfect
  agreement with the adopted D and \4he abundances.  The right-hand
  panel shows the corresponding \neff~-- \omb~contours (and best BBN
  fit), again with guides to the eye at $\Deln = 0$ and 1.  Also
  shown, as pink ellipses, in the left-hand panel are the independent 68\% and
  95\% CMB+BAO contours.}
\label{fig:majorana-bbn-only-nnu-obh2}
\end{center}
\end{figure}

With two abundances (D and \4he) and three parameters (\omb, \Deln, and \mchi), the model parameters are not uniquely determined by BBN.  For every choice of \mchi, there is a pair of \{\omb,\,\Deln\} values that yield, precisely, the adopted abundances of D and \4he.  These fitted parameter values are shown as the solid curves in Figs.~\ref{fig:majorana-bbn-only} and \ref{fig:majorana-bbn-only-nnu-obh2}, with $m_\chi$ varying along the
curves.  Recall that N$^{0}_{\rm eff}$ is determined by \mchi, so that values of \mchi~and \Deln~fix the corresponding value of \neff.  In the following,  \Deln~is sometimes treated as a model parameter, and sometimes it is \neff; once $m_\chi$ is specified, either of these neutrino-counting parameters determines the other.  However, since it is possible that the largest value of \Deln~and the largest value of \neff~allowed by the data can occur at different values of \mchi, limits on the two parameters are not directly interchangeable without more information.  (The reader is reminded that by our definition,  \neff~specifies the post-BBN radiation density, which is measured by CMB observations.)

The reader's attention is called to the strong resemblance between the \Deln~vs.~\mchi~best-fit curve of the left-hand panel of Fig.~\ref{fig:majorana-bbn-only} and the Y$_{\rm P}$ vs.~\mchi~curves of Fig.~\ref{fig:Deln=0yields}.  This shows that the addition of a light WIMP with specified mass does not substantially change the roles of D/H and Y$_{\rm P}$ in constraining effective neutrinos: if there are only these two abundances and BBN is used to solve simultaneously for $\eta$ and \Deln, D/H mainly constrains $\eta$ while Y$_{\rm P}$ mainly constrains \Deln.

The results of constraining these parameters with the BBN abundances alone are shown in Figs.~\ref{fig:majorana-bbn-only} and \ref{fig:majorana-bbn-only-nnu-obh2} and in Table~\ref{tab:bounds}, where the best fit and 68\% and 95\% ranges of \Deln, \neff, and \omb~are shown as functions of the WIMP mass and are compared to the independent CMB constraints.  The contours in these figures show frequentist limits, corresponding to the ellipses in Fig.~\ref{fig:wimpyields-majorana}.  The ``projected'' one-dimensional intervals given in Table~\ref{tab:bounds} are profile likelihoods \cite{pdg2012,rolke}, treating all parameters but one as nuisance parameters.

\begin{figure}
\begin{center}
\includegraphics[width=0.32\columnwidth]{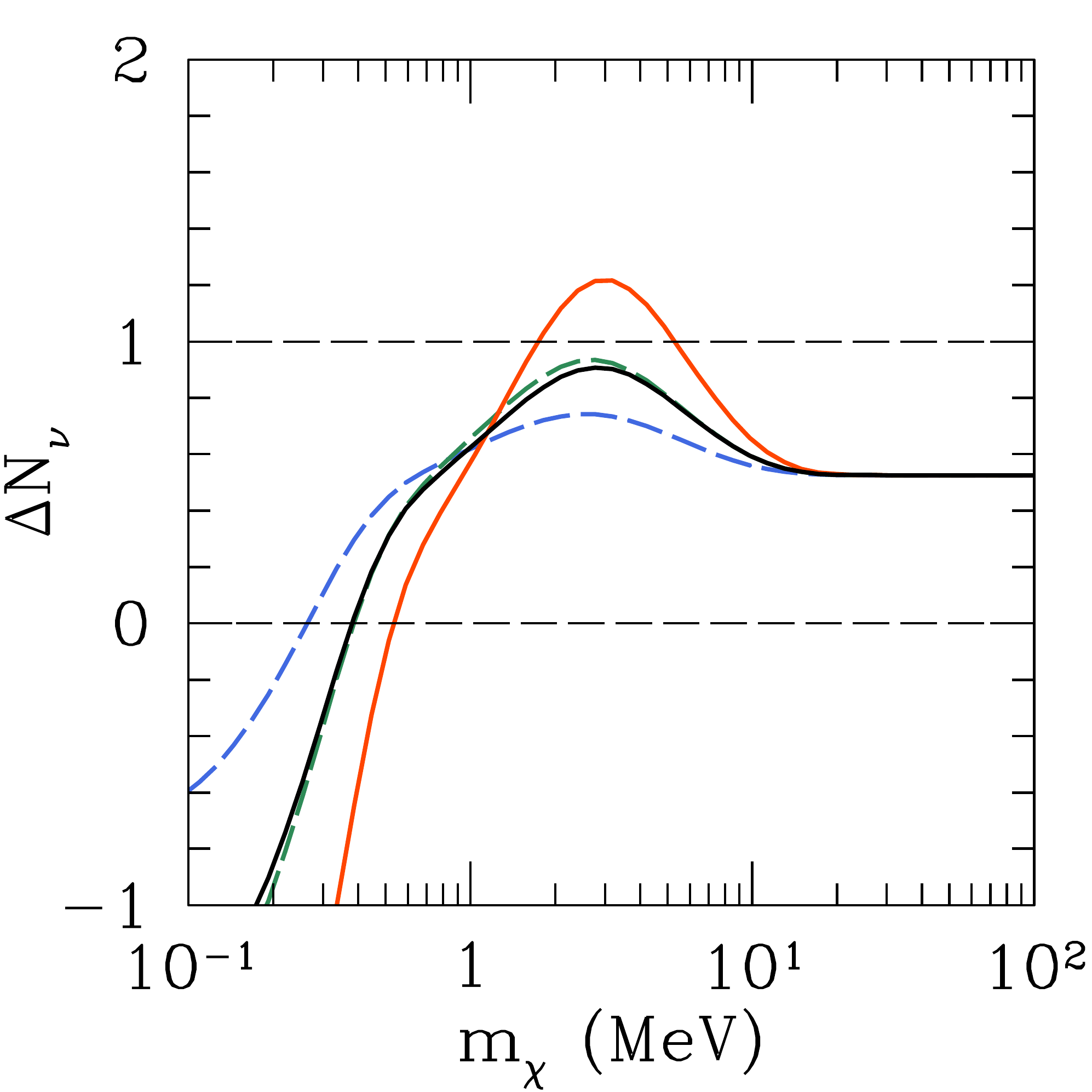}
\includegraphics[width=0.32\columnwidth]{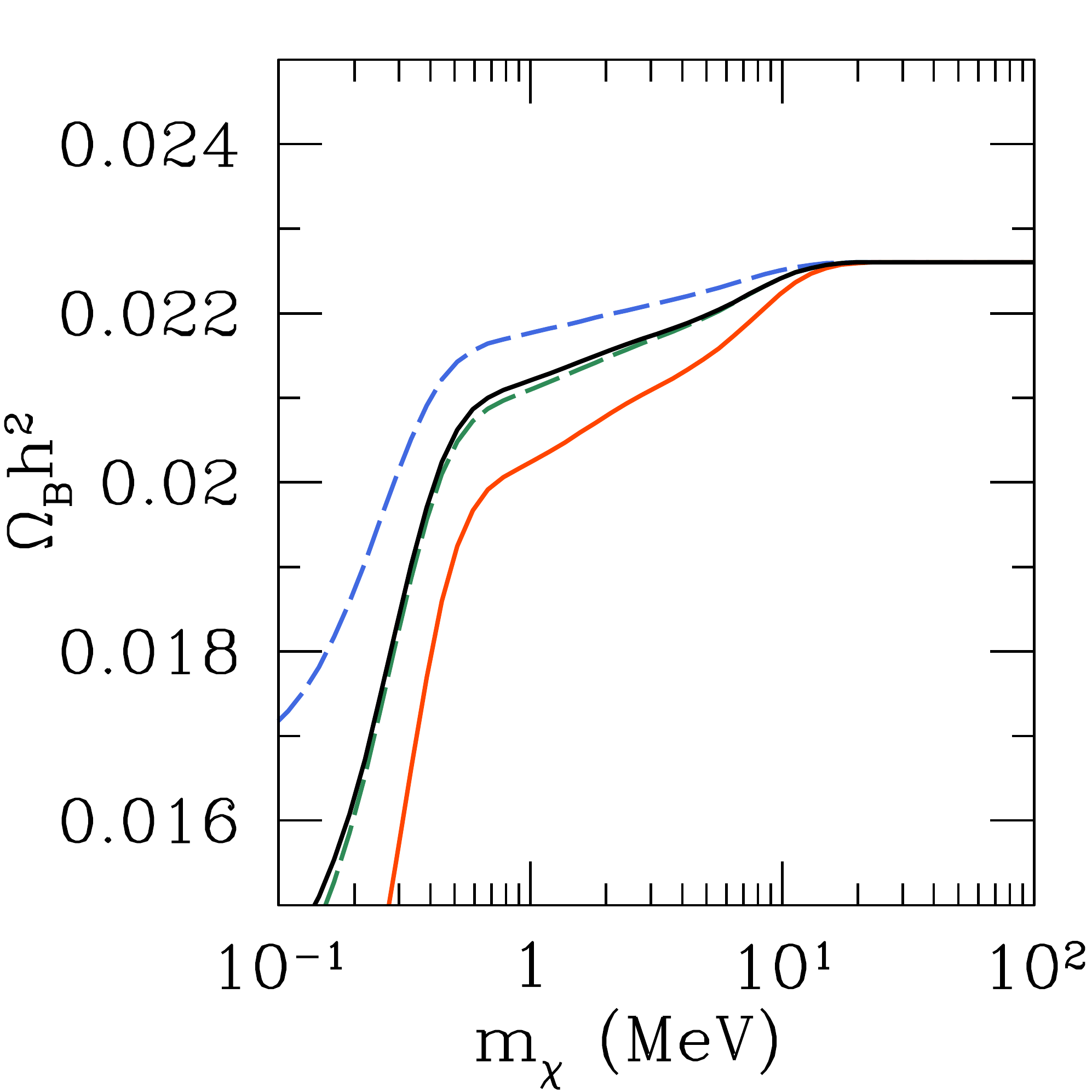}
\includegraphics[width=0.32\columnwidth]{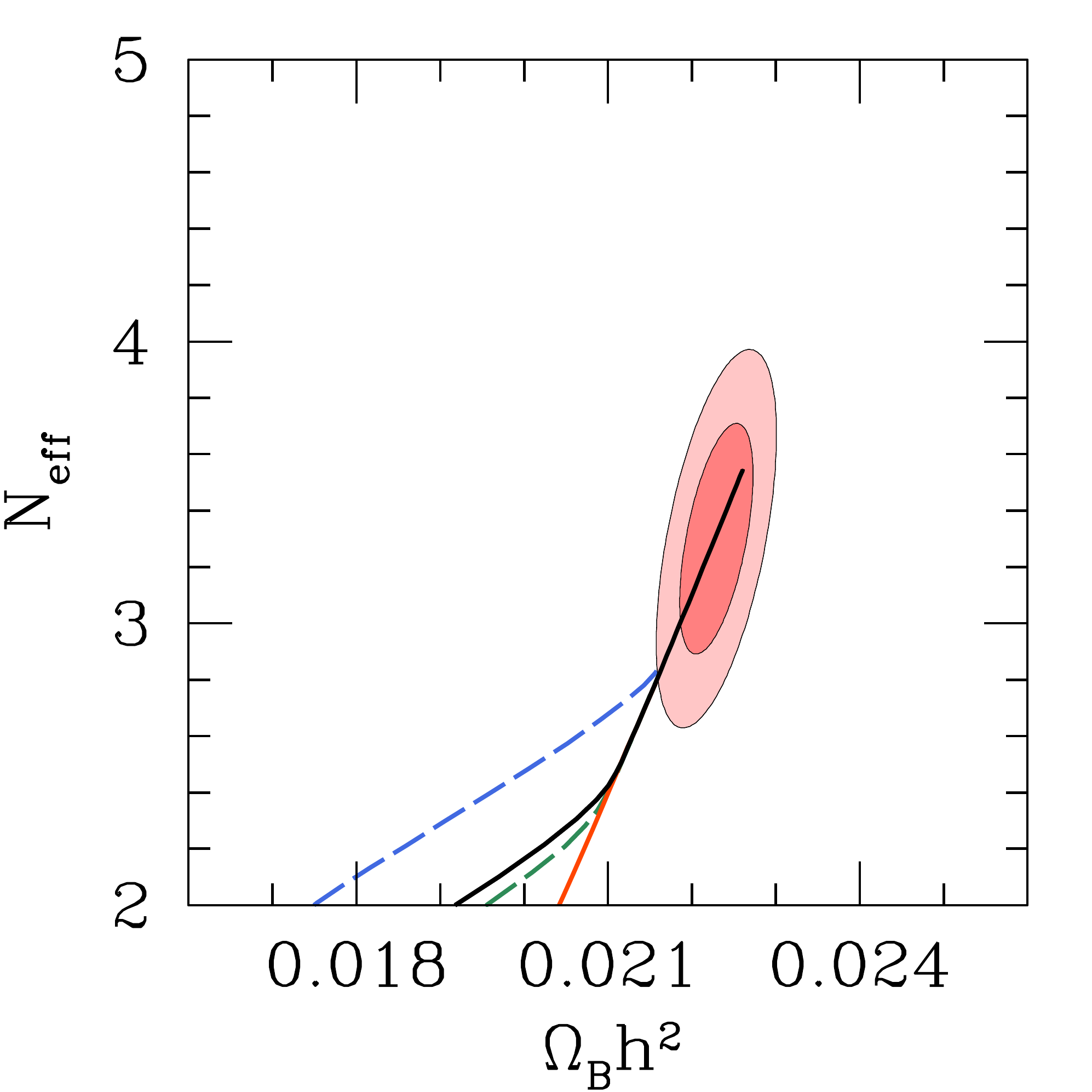}
\end{center}
\vskip -.2in
\caption{(Color online) Best-fit curves showing exact fits to the BBN
  data for the four types of WIMPs considered here, with color and
  line types as in Figs.\,\ref{fig:neffvsmem}
  and~\ref{fig:Deln=0yields} and appearing in the same order.  In the
  left and middle panels, these fits are shown as curves of \Deln~and
  \omb\ as functions of the WIMP mass.  The right-hand panel shows the
  corresponding relations between \neff~and \omb, along with 68\% and
  95\% contours from the Planck CMB+BAO fit\,\cite{planck}.}
  \label{fig:constraints-allkinds}
\end{figure}

It is clear from Fig.~\ref{fig:majorana-bbn-only} that BBN alone
provides no constraint on the WIMP mass without further information.
In fact, there is further information, without having to resort to the
CMB: it is known that there are three SM neutrino species, and it is
known from their mixing angles that they were well mixed and had a
single temperature both prior to and after decoupling of the electron
neutrinos from the plasma \cite{mangano,dolgov-osc,pastor}.  It is therefore
reasonable to assume that $\Deln \geq 0$.  From the left-hand panel of
Fig.~\ref{fig:majorana-bbn-only} and from Table~\ref{tab:bounds}, it
is clear that this constraint leads to a lower bound on \mchi.  The
strength of the lower bound depends on the nature of the WIMP, as
indicated in Fig.~\ref{fig:constraints-allkinds}.  At 95\% (one-sided)
confidence, a single-component scalar WIMP must have $\mchi > 150$
keV; a Dirac WIMP must have $\mchi > 420$ keV.  The lower limits for
two-component scalars and Majorana WIMPs fall between these two
values.  However, it should be noted that at $\Deln = 0$, the best
fit to BBN with light WIMPs allowed corresponds to \neff~in serious
conflict with CMB measurements.  This is shown in the first line of
Table \ref{tab:bounds} for each WIMP type.

\begingroup
\squeezetable
\scriptsize
\setlength{\tabcolsep}{-10pt}
\begin{table}
  \caption{Constraints on the model parameters are shown.  The BBN D and $^4$He abundances are always included and additional constraints are allowed to vary.  The 68\% confidence limits are shown first, then the 95\% confidence ranges.  Lower limits on $m_\chi$ are always 95\% one sided.  When there is a unique best fit but only a lower limit for $m_\chi$, the best fit is given in parentheses.  The input \omb~and \neff~constraints are always from the Planck $\Lambda\mathrm{CDM}+N_\mathrm{eff}$ fit including BAO [Eq.~(75) of Ref.~\cite{planck}], except where \Yp~is also given.  In this case, the constraints are from the Planck $\Lambda\mathrm{CDM}+N_\mathrm{eff}+{\rm Y}_{\rm P}$ fit including BAO [Eq.~(90) of Ref.~\cite{planck}], so that there is an additional constraint on \Yp~(beyond that from the direct observations).  The correlations between Planck parameters have been included as described in the text.  Where more significant figures are given than are merited by the error intervals, they are intended to keep the columns aligned and to show the effects of changing models or input data.}
  \label{tab:bounds}
\vskip 11pt
  \begin{tabular}{llldd}
{Constraints \hskip 1.6in} & {\hskip .2in $100\,\omb$ \hskip .8in} &  {\hskip .2in $N_\mathrm{eff}$} &\multicolumn{1}{c}{\hskip .6in $\Delta N_\nu$} & \multicolumn{1}{c}{$m_\chi \mathrm{(MeV)}$}\\\hline
\\
\multicolumn{5}{c}{Real scalar WIMP}\\
BBN only, $\Delta N_\nu = 0$ &
      $1.97\pm 0.10(0.19)$ & 2.41 & \multicolumn{1}{c}{\hskip .6in $\cdots$}  &
      0.26 ^{+0.09(0.30)}_{-0.07(0.12)}\\
BBN only, $\Delta N_\nu \geq 0$, 68\% &
      {\!\!(1.87,2.34)} & {\!\!(2.41,3.78)} & (0.0,1.00) &  
      > 0.15\\
      \hskip 1.35in 95\%  &
               {\!\!(1.78,2.42)} & {\!\!(2.41,4.02)} & (0.0,1.27) &  
                  > 0.15\\
BBN + CMB+BAO
 ($\eta$) &
      $2.23\pm 0.03(0.06)$  & $3.31^{+0.39(0.62)}_{-0.50(0.76)} $
                                           & 0.66^{+0.34(0.60)}_{-0.18(0.48)}  & 
      > 0.39  ~(5.4)\\
BBN + CMB+BAO
 ($N_\mathrm{eff}$) &
      $2.23\pm 0.08(^{+0.16}_{-0.15})$ & $3.30\pm 0.27(^{+0.47}_{-0.54})$
                                           &0.67^{+0.33(0.56)}_{-0.36(0.56)} & 
      > 0.43 ~ (5.2)\\ 
BBN + CMB+BAO
 ($\eta,N_\mathrm{eff}$) &
      $2.23\pm 0.03(0.06)$  & $3.30\pm 0.26(^{+0.46}_{-0.51})$  
                                       & 0.67^{+0.33(0.55)}_{-0.33(0.55)}  & 
      > 0.48 ~ (5.2)\\
BBN + CMB+BAO 
($\eta,N_\mathrm{eff},Y_P$)  &
      $2.23\pm 0.03(0.06)$  & $3.28^{+0.33(0.54)}_{-0.35(0.61)} $
                                       & 0.69^{+0.31(0.54)}_{-0.38(0.58)}  & 
      > 0.42 ~(4.7)\\
\\
\multicolumn{5}{c}{Majorana WIMP}\\
BBN only, $\Delta N_\nu = 0$ &
  $1.96\pm 0.09 (^{+0.18}_{-0.17})$ &  2.09 &\multicolumn{1}{c}{\hskip .6in $\cdots$} &
      0.38 ^{+0.10(0.37)}_{-0.06(0.11)}\\
BBN only, $\Delta N_\nu \geq 0$, 68\%  &
      {\!\!(1.88,2.34)}  & {\!\!(2.08,3.78)} & (0.0,1.19) &  
      > 0.28\\
      \hskip 1.35in 95\%  &
      {\!\!(1.79,2.42)}  & {\!\!(2.08,4.02)} & (0.0,1.48) &  
      > 0.28\\
BBN + CMB+BAO
 ($\eta$) &
      $2.23\pm 0.03(0.06)$  & $3.32^{+0.39(0.62)}_{-0.54(0.98)} $
                                             & 0.64^{+0.52(0.84)}_{-0.34(0.55)}  & 
      > 0.62  ~(8.1)\\
BBN + CMB+BAO
 ($N_\mathrm{eff}$) &
      $2.23\pm 0.08(^{+0.16}_{-0.15})$ & $3.30\pm 0.27(^{+0.47}_{-0.54}) $
                                             & 0.65^{+0.46(0.83)}_{-0.35(0.55)} & 
      > 1.66 ~ (7.9)\\ 
BBN + CMB+BAO
 ($\eta,N_\mathrm{eff}$) &
      $2.23\pm 0.03(0.06)$  & $3.30\pm 0.26(^{+0.46}_{-0.52}) $
                                             & 0.65^{+0.46(0.83)}_{-0.35(0.54)}  & 
      > 1.73~(7.9)\\
BBN + CMB+BAO
 ($\eta,N_\mathrm{eff},Y_P$)  &
      $2.23\pm 0.03(0.06)$  & $3.28^{+0.33(0.54)}_{-0.34(0.68)} $
                                             & 0.67^{+0.49(0.80)}_{-0.37(0.56)}  & 
      > 1.08 ~(7.5)\\
\\
\multicolumn{5}{c}{Complex scalar WIMP}\\
BBN only, $\Delta N_\nu = 0$ &
      $1.95^{+0.08(0.16)}_{-0.09(0.16)}$ & 2.00 & \multicolumn{1}{c}{\hskip .6in $\cdots$} &
      0.39 ^{+0.10(0.35)}_{-0.06(0.11)}\\
BBN only, $\Delta N_\nu \geq 0$, 68\% &
      {\!\!(1.87,2.34)} & {\!\!(1.99,3.78)} & (0.0,1.22) &  
      > 0.29\\
      \hskip 1.35in 95\%  &
      {\!\!(1.78,2.42)} & {\!\!(1.99,4.02)} & (0.0,1.51) &  
      > 0.29\\
BBN + CMB+BAO
 ($\eta$) &
      $2.23\pm 0.03(0.06)$  & $3.32^{+0.39(0.62)}_{-0.54(1.00)} $
                                              & 0.64^{+0.54(0.87)}_{-0.34(0.55)} & 
      > 0.72  ~(8.1)\\
BBN + CMB+BAO
 ($N_\mathrm{eff}$) &
      $2.23\pm 0.08(^{+0.16}_{-0.15})$ &$3.30\pm 0.27(^{+0.47}_{-0.54}) $
                                              &0.65^{+0.46(0.86)}_{-0.35(0.55)} & 
      > 1.94 ~ (7.9)\\ 
BBN + CMB+BAO
 ($\eta,N_\mathrm{eff}$) &
      $2.23\pm 0.03(0.06)$  & $3.30\pm 0.26(^{+0.46}_{-0.52}) $
                                              & 0.65^{+0.46(0.86)}_{-0.35(0.54)} & 
      > 2.01 ~ (7.9)\\
BBN + CMB+BAO
 ($\eta,N_\mathrm{eff},Y_P$)  &
      $2.23\pm 0.03(0.06)$  & $3.28^{+0.33(0.54)}_{-0.34(0.68)} $
                                              & 0.67^{+0.50(0.84)}_{-0.37(0.56)} & 
      > 1.32 ~(7.5)\\
\\
\multicolumn{5}{c}{Dirac WIMP}\\
BBN only, $\Delta N_\nu = 0$ &
      $1.94\pm 0.07(^{+0.14}_{-0.15})$ & 1.57 & \multicolumn{1}{c}{\hskip .6in $\cdots$} &
      0.54 ^{+0.15(0.50)}_{-0.08(0.13)}\\
BBN only, $\Delta N_\nu \geq 0$, 68\%   &
      {\!\!(1.87,2.34)} & {\!\!(1.56,3.78)} & (0.0,1.54) &  
      > 0.42\\%
      \hskip 1.35in 95\%  &
                  {\!\!(1.79,2.42)} & {\!\!(1.56,4.02)} & (0.0,1.82) &  
                  > 0.42\\
BBN + CMB+BAO
 ($\eta$) &
      $2.23\pm 0.03(0.06)$  & $3.32^{+0.39(0.61)}_{-0.54(1.04)} $
                                            & 0.63^{+0.56(1.18)}_{-0.33(0.54)}  & 
      > 2.36  ~(10.5)\\%
BBN + CMB+BAO
 ($N_\mathrm{eff}$) &
      $2.23\pm 0.08(^{+0.16}_{-0.15})$ &$3.30\pm 0.27(^{+0.47}_{-0.54}) $
                                            &0.64^{+0.42(0.98)}_{-0.34(0.54)} & 
      > 4.84 ~ (10.2)\\ %
BBN + CMB+BAO
 ($\eta,N_\mathrm{eff}$) &
      $2.23\pm 0.03(0.06)$  & $3.30\pm 0.26(^{+0.46}_{-0.52}) $
                                            &0.64^{+0.42(0.97)}_{-0.34(0.53)}  & 
      > 4.95 ~ (10.2)\\%
BBN + CMB+BAO
 ($\eta,N_\mathrm{eff},Y_P$)  &
      $2.23\pm 0.03(0.06)$  & $3.27\pm 0.34(^{+0.55}_{-0.68}) $
                                            &0.66^{+0.47(1.09)}_{-0.36(0.56)}  & 
      > 3.70 ~(10.2)\\%
  \end{tabular}
\end{table}
\endgroup

Limits can also be set on \omb~purely from BBN in the context of this
model.  Figure \ref{fig:majorana-bbn-only} shows how this is possible.
At $\mchi\gtrsim 20$ MeV, the WIMP no longer has any effect on BBN or
\neff, and the familiar model of BBN plus equivalent neutrinos appears
as the limiting case at the high-\omb~side of the graph.  The upper
limit to \omb~provided by BBN abundances in a model with no WIMPs is
then the upper limit to \omb~in the model with a light WIMP.  Lower
values of \omb~can be achieved by lowering \mchi, but eventually, the
requirement that $\Deln \geq 0$ limits how far we can go in this
direction.  As a result,
\begin{equation}
0.018 < \omb < 0.024~~(95\%~\mathrm{C.L.}).
\end{equation}
The lower limit is nearly independent of the nature of the WIMP (real
or complex scalar, Majorana or Dirac fermion), as can be understood by
comparing the left and middle panels of
Fig.~\ref{fig:constraints-allkinds}.  There is an analogous BBN-only
upper limit to \Deln, corresponding to the highest value of \Deln~at
which the curves shown in Figs.~\ref{fig:wimpyields-majorana} and
\ref{fig:wimpyields-allkinds} intersect the observations at any value
of \omb.  \Deln~values higher than this maximum produce too much 
$^4$He under all circumstances.  As expected, the upper limit on
\Deln~depends on the nature of the WIMP, ranging from $(\Delta{\rm
  N}_{\nu})_{\rm max} \lsim 1.3$ (real scalars) to $(\Delta{\rm
  N}_{\nu})_{\rm max} \lsim 1.8$ (Dirac fermions) at 95\%
C.L.~(two sided).

Figure~\ref{fig:majorana-bbn-only} also shows, for comparison with the BBN constraints, the recently inferred values of \omb~and~\neff~from the CMB and baryon acoustic oscillation (BAO) data [Eq.\,(75) of Ref.\,\cite{planck}].\footnote{This is from a fit to the six $\Lambda$CDM parameters plus \neff~and is for the fit and data combination denoted: base\_nnu\_planck\_lowl\_lowLike\_highL\_BAO.  The constraint on \omb~and its correlation with \neff~were taken from files published online in the Planck Legacy Archive \cite{planck-web}.  This fit is not, strictly speaking, compatible with the light-WIMP scenario, because the BBN relation among \omb, \neff, and Y$_{\rm P}$ with no light WIMP was used to fix the composition of the baryon-photon fluid.  However, the sensitivity of the CMB fits to Y$_{\rm P}$ is weak (see below).}  Since these non-BBN constraints have only limited overlap with the BBN bounds on the model parameters, combining them with the BBN data will produce tighter constraints than from either data set alone.  In particular, the lower bound to the WIMP mass found above can be increased by a few tenths of an MeV.  This BBN + CMB combination is explored next, in Sec.\,\ref{sec:constraints}.

\section{BBN And CMB Constraints On \mchi, \neff, \Deln, and \omb}
\label{sec:constraints}

\subsection{Simultaneous light WIMPs and equivalent neutrinos}

\begin{figure}[!t]
\begin{center}
\includegraphics[width=0.32\columnwidth]{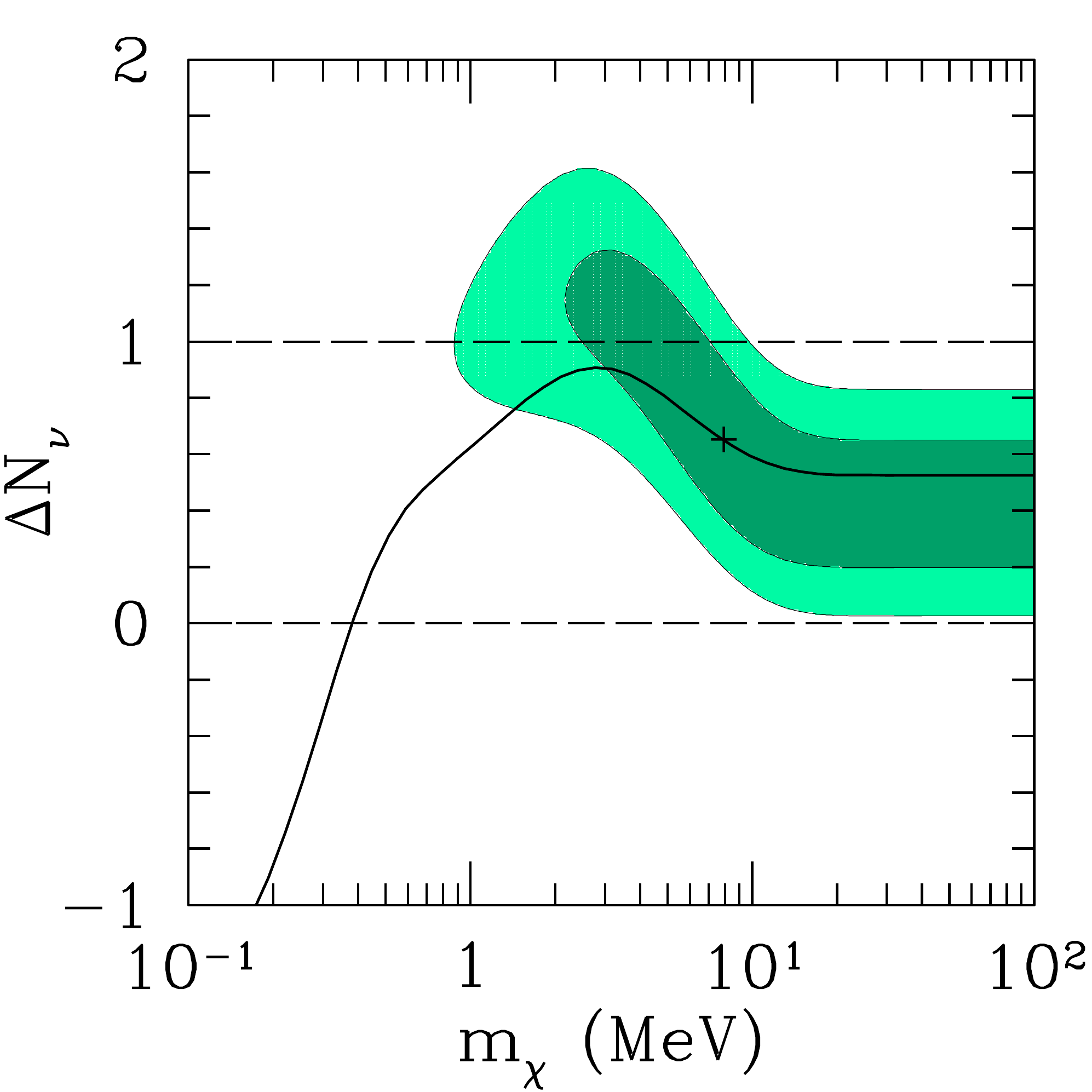}
\includegraphics[width=0.32\columnwidth]{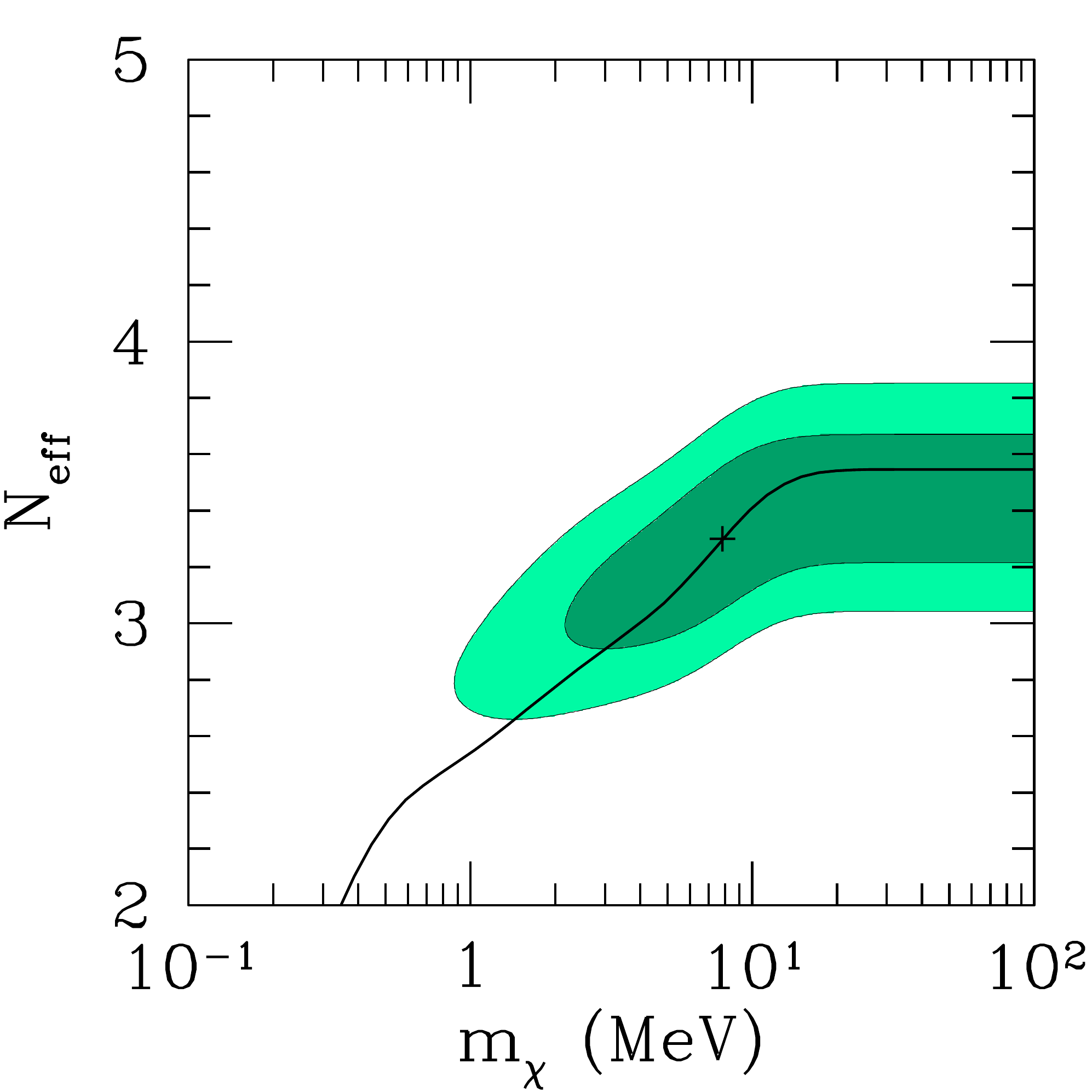}
\includegraphics[width=0.32\columnwidth]{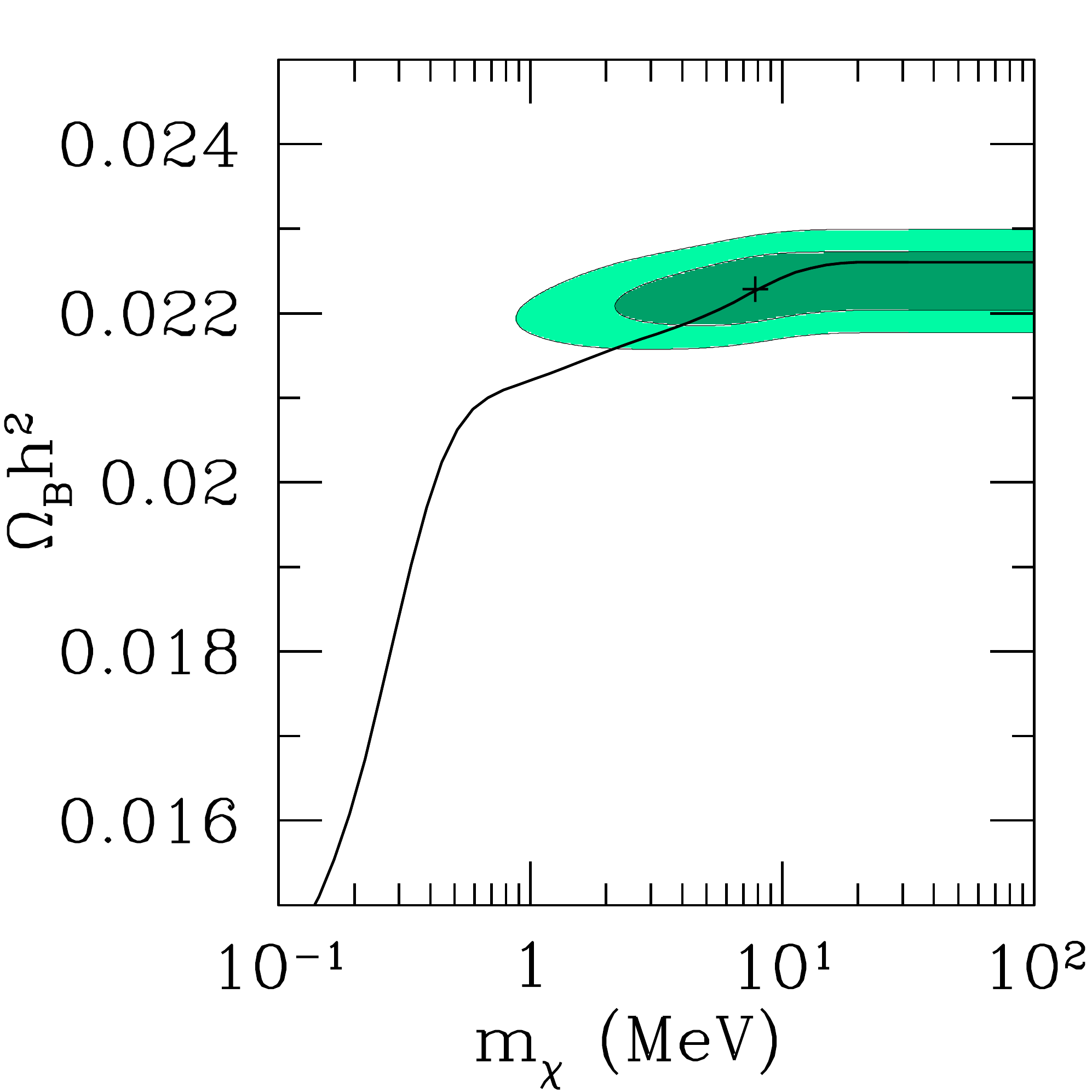}
\caption{(Color online) The 68\% and 95\% contours, darker and lighter
  green respectively, from the joint BBN + CMB fits.  In the left
  panel, the \Deln~-- \mchi~contours are shown, along with the exact
  BBN fit curve.  Also shown in the left panel (dashed lines), as
  guides to the eye, are \Deln~= 0 and 1.  The middle panel shows the
  corresponding contours for \neff~vs.~\mchi, along with the
  corresponding exact BBN fit curve.  The right-hand panel shows the
  joint contours in the \neff~-- \omb~plane, along with the
  corresponding exact BBN fit curve. The cross in each figure
  indicates the best-fit model.}
\label{fig:majorana-bbn+cmb}
\end{center}
\end{figure}

Neither BBN nor the CMB, alone, can constrain the mass of a light WIMP
unless some additional assumption is made about \Deln~(\eg,
\Deln~$\geq 0$), and even then the constraint is rather weak.  Indeed,
the CMB data for \neff~and \omb~are agnostic to the presence of a
light WIMP.  It was seen above that in addition to the dependence of
the BBN yields on the baryon density, \omb, and on the number of
equivalent neutrinos, \Deln, they also depend on \mchi.  This ensures
that a comparison of the BBN-predicted yields with the observationally
inferred primordial abundances for only two nuclides, D and \4he, is
insufficient to constrain all three parameters.  Indeed, as seen in
Figs.\,\ref{fig:majorana-bbn-only}\,and\,\ref{fig:majorana-bbn-only-nnu-obh2},
for every choice of \mchi, a combination of \{\omb,\,\Deln\} (or, of
\{\omb,\,\neff\}) can be found so that the BBN yields agree, exactly,
with the adopted values of \Yp~and \yd.  By combining the
WIMP-mass-dependent ranges of \neff~and \omb~allowed by BBN with the
WIMP-mass-independent ranges identified by the CMB, the WIMP mass can
be constrained and the joint BBN + CMB best-fit values identified,
along with their 68\% and 95\% ranges.  In
Fig.~\ref{fig:majorana-bbn+cmb}, the BBN + CMB-allowed regions for
\mchi~with \neff, \Deln, and \omb~are shown for the case of a Majorana
WIMP.  The numerical results for Majorana and Dirac WIMPs and for real
and complex scalar WIMPs are given in Table \ref{tab:bounds}.  The
results in the graphs and in most of the table are based on the Planck
Collaboration fit of CMB+BAO data to the six $\Lambda$CDM parameters
plus \neff~\cite{planck,planck-web}.  This Planck fit provides
relatively tight constraints based on consistent data.

It is found here that the 95\% confidence lower bounds to the WIMP
mass range from $\mchi \gsim 0.5\,{\rm MeV}$ for self-conjugate
scalars to \mbox{$\mchi \gsim 5$ MeV} for Dirac fermions.  Although
the absence of a light WIMP (\mchi~$\gsim 20\,{\rm MeV}$) is
consistent with the BBN and CMB data well within the 68\% confidence
interval, the combined BBN + CMB data do have a slight preference for
a light WIMP with a mass of \mchi~$\sim 5 - 10\,{\rm MeV}$, depending
on the nature of the WIMP.  In all cases, \neff~$\approx 3.3 \pm 0.3$
($1\sigma$) and $\eta_{10} \approx 6.1 \pm 0.1$, corresponding to
\omb~$\approx 0.022$.  As our analysis shows, the tight and robust CMB
constraint on \omb~dominates that from the deuterium abundance, while
the precise, new \4he~abundance determination \cite{izotov} ensures
that for constraining \neff~(\Deln), BBN and the CMB are competitive.

\begin{figure}[!t]
\begin{center}
\includegraphics[width=0.45\columnwidth]{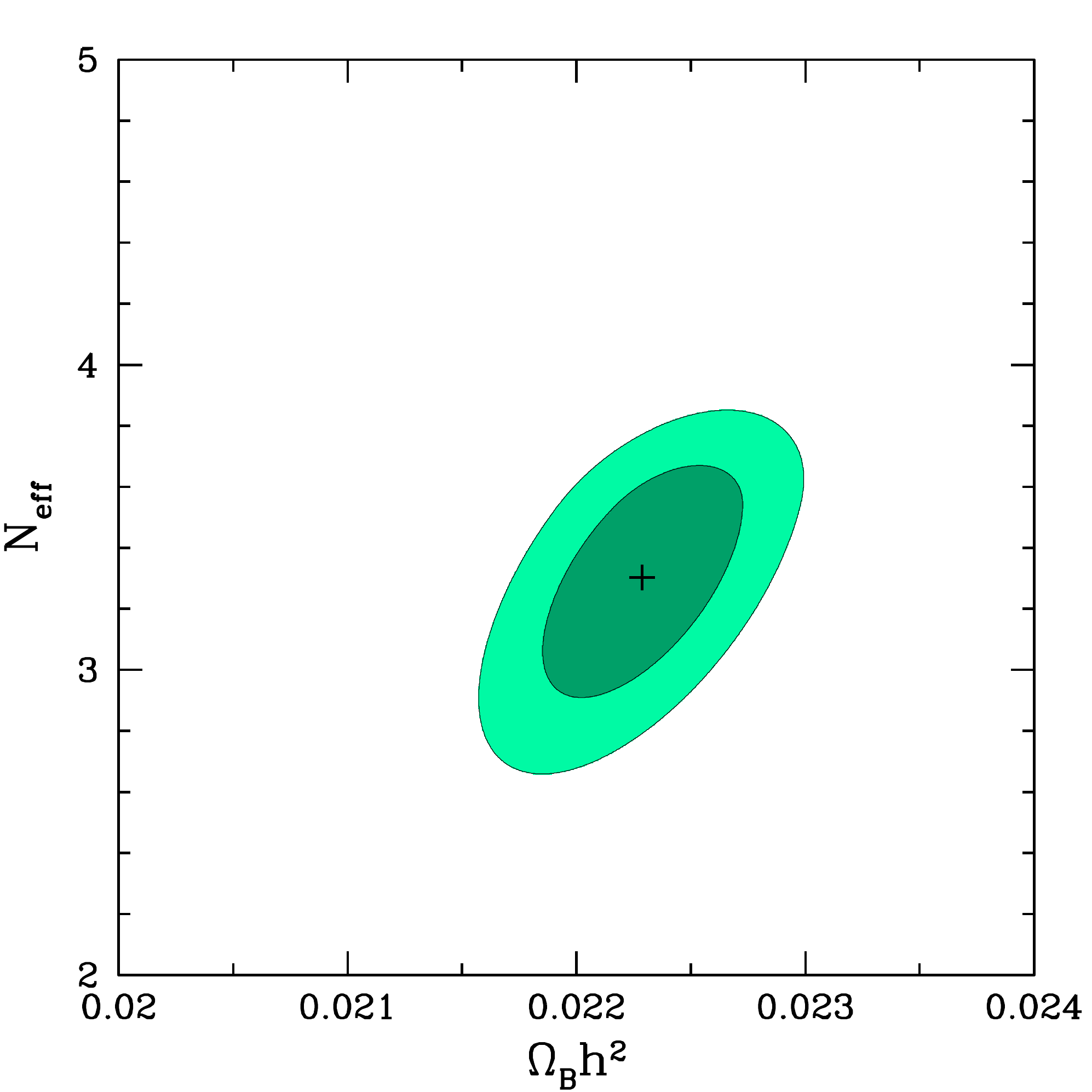}
\includegraphics[width=0.45\columnwidth]{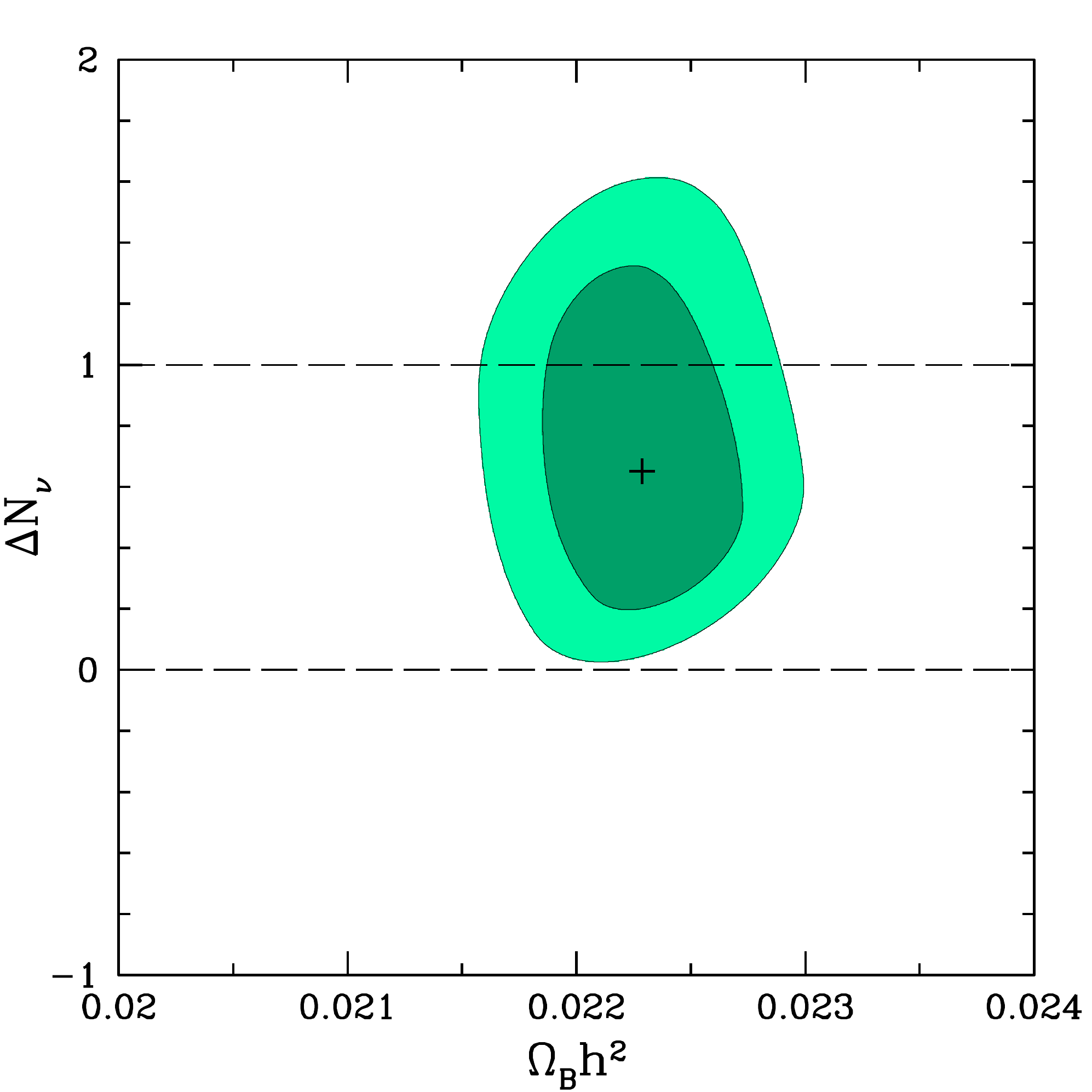}
\caption{(Color online) The joint BBN+CMB 68\% and 95\% contours in
  the \neff~-- \omb~plane (left panel) and in the \Deln~-- \omb~plane
  (right panel) for a Majorana WIMP.  The cross in each panel is at
  the best-fit point, corresponding to \mchi~= 7.9 MeV.}
\label{fig:majorana-bbn+cmb-nnu-obh2}
\end{center}
\end{figure}

With or without a light WIMP, the combined BBN and CMB data have a
preference (at $\sim 95\,\%$ confidence) for \Deln~$> 0$ (see
Figs.~\ref{fig:majorana-bbn+cmb} and
\ref{fig:majorana-bbn+cmb-nnu-obh2} and Table \ref{tab:bounds}),
suggesting the presence of dark radiation.  However, without a light
WIMP, a sterile neutrino with $\Deln = 1$ is disfavored (at $\sim
99\,\%$ confidence).  These results are driven by the primordial
helium abundance adopted in our analysis, both its central value and
its relatively small uncertainty \cite{izotov}.  Note that the error
adopted for the observationally inferred value of \Yp~ensures that the
errors for \Deln~(or \neff) from BBN and the CMB are comparable ($\sim
0.2 - 0.3$).  Further reduction in the CMB error on \neff, anticipated
from the Planck polarization data~\cite{bashinsky}, will have the
potential to support (or weaken) these parameter constraints,
depending on whether or not the central value for \neff~from the CMB
changes.

In the context of a light-WIMP model, the BBN and CMB data adopted
here are remarkably consistent.  The important results in Table
\ref{tab:bounds} are those for BBN only (lines 1--3 for each WIMP type)
and for BBN data combined with the CMB+BAO fit to the $\Lambda$CDM
parameters plus \neff~(line 6).  Three more cases are included in the
table in order to examine separately the roles of the CMB+BAO
constraints on \omb~and on \neff.  In lines 4 and 5 for each WIMP
type, are shown combinations of CMB+BAO constraints with BBN data,
first with the CMB+BAO constraint on \neff~removed and then with the
constraint on \omb~removed.  In both cases, the remaining parameter is
chosen to have its value from the $\Lambda$CDM+\neff~fit; these are
intended to reveal the role of the omitted parameter, not to represent
a realistic fit.  What is seen is that if the \neff~constraint is
removed (line 4), the \omb~constraint suffices to keep the combined
fit centered on the CMB+BAO value of \neff.  Similarly, if the
\omb~constraint is removed (line 5), the \neff~constraint suffices to
keep the combined fit centered on the CMB+BAO value of \omb~(though
with a larger allowed interval).  The consistency may be seen in two
other ways: 1) the best-fit point of the combined fit (the ``$+$'' in
Figs.~\ref{fig:majorana-bbn+cmb}-\ref{fig:majorana-bbn+cmb-nnu-obh2})
lies on the BBN-only best-fit curve; or 2) $\chi^2 \ll 1$ at the best
fit.\footnote{Relatively late in the preparation of this work we
  learned of a new observational determination of \yd~\cite{cooke}
  whose central value is consistent with the value adopted here but
  whose error is smaller by a factor of three.  If this D abundance
  were adopted it would slightly increase the preferred value of
  \omb~(by $\sim 0.0003$) and reduce the favored value of \Deln~(by
  $\sim 0.01$), introducing some tension between BBN and the CMB
  which, nonetheless, would remain consistent within 68\% confidence.
  These small shifts are well within the BBN and CMB errors (and, even
  within the round off errors).}

\begin{figure}
\begin{center}
\includegraphics[width=0.5\columnwidth]{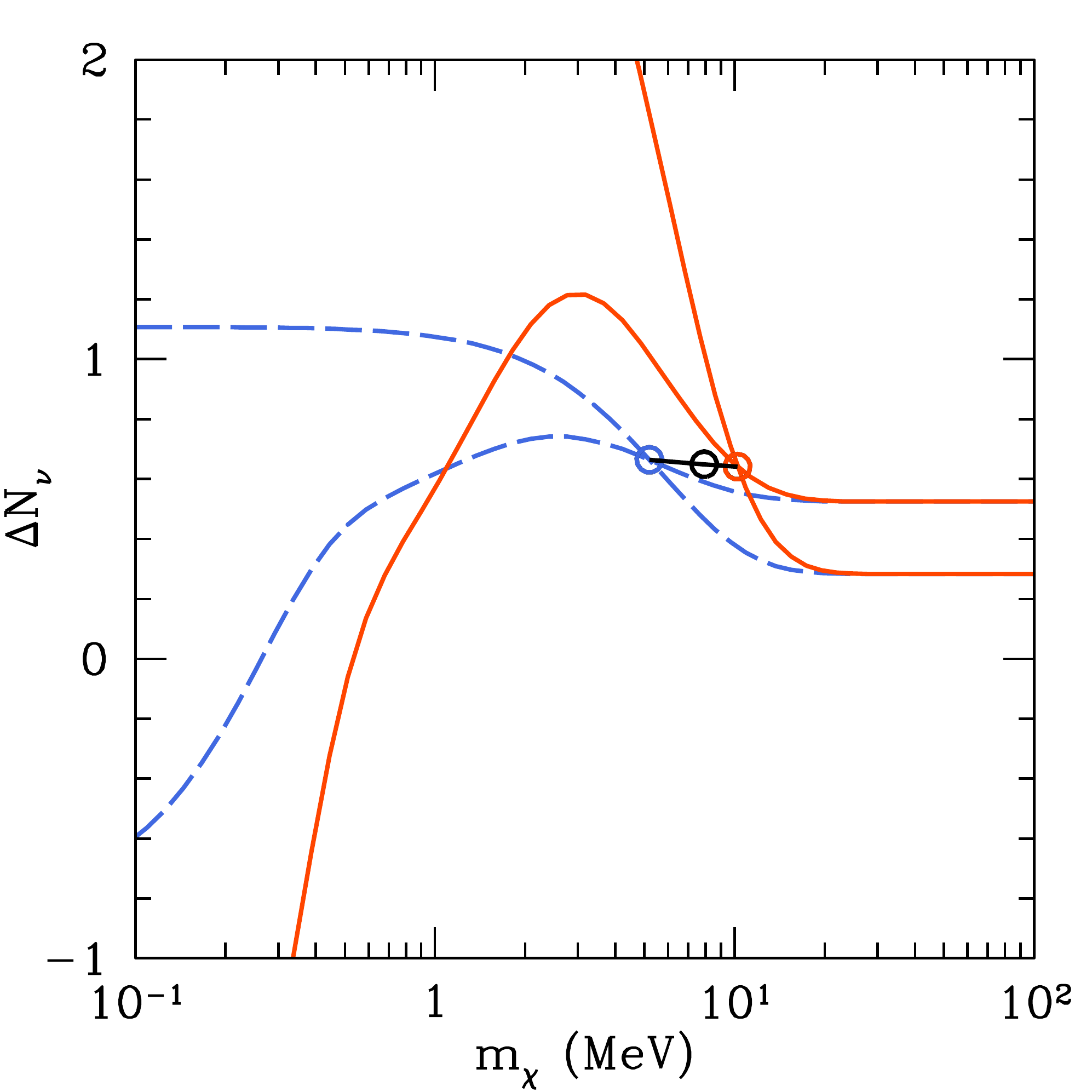}
\end{center}
\vskip -.2in
  \caption{(Color online) Demonstration of the relatively weak
    dependence of the best-fit \Deln~values on the WIMP type.  The
    dashed blue curves show the best fits of a real scalar WIMP to BBN
    and to CMB data separately (cf.~Figs.~\ref{fig:neffvsmnu} and
    \ref{fig:constraints-allkinds}).  The best (and coincidentally
    exact) simultaneous fit to BBN and CMB data for a real scalar
    light WIMP lies at the intersection of these curves, which is
    circled.  The two analogous curves for a Dirac light WIMP are also
    shown (solid orange), with their intersection point circled.  Fit
    curves are not shown for other WIMP types, but the joint-fit
    points for those curves (Majorana and complex scalar) are also
    circled (and indistinguishable at this scale).  The nearly
    horizontal straight lines joining the simultaneous-fit points show
    the near invariance of the best fit for \Deln, also evident in
    Table \ref{tab:bounds}.}
  \label{fig:delta-nnu-invariance}
\end{figure}

The invariance of the fitted parameters with WIMP type is remarkable.
It is less surprising for \omb~than for other parameters, since this
parameter is constrained very directly by the CMB data and by \yd.
Likewise, \neff~is connected directly to physics in the CMB and (for
the relatively large best-fit \mchi) at late times in BBN.  But why is
it that in simultaneous fitting, \mchi~seems to be chosen in each case
to give very nearly the same \Deln?  Exact fits to CMB+BAO and to BBN
are shown as functions of \mchi~in Fig.~\ref{fig:delta-nnu-invariance}
for two types of WIMPs.  The crossings of the BBN and CBM+BAO curves
always occur very close to $\Deln=0.65$, regardless of the type of
light WIMP.

\begin{figure}
\includegraphics[width=0.45\columnwidth]{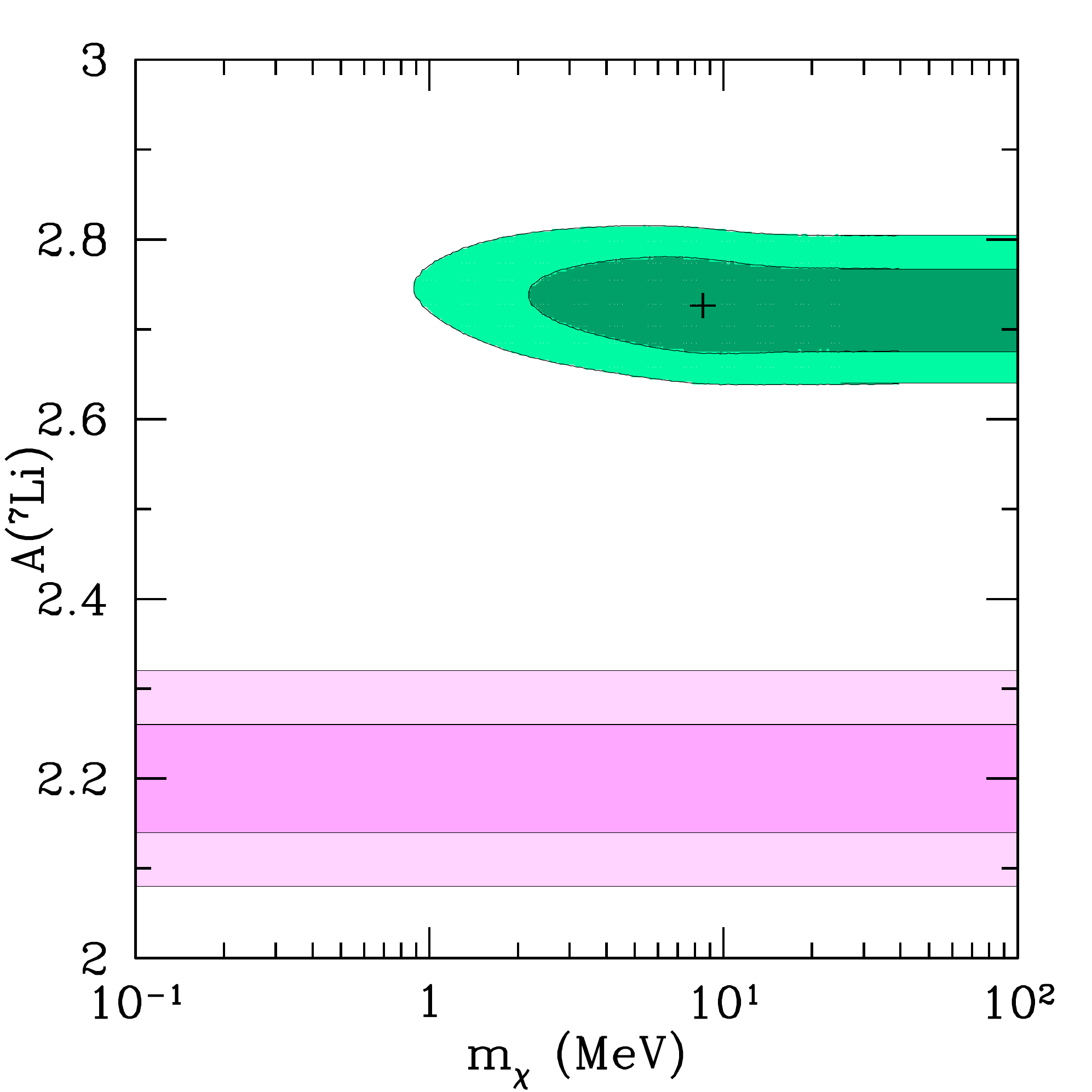}
  \caption{(Color online) Joint BBN and CMB+BAO constraints at 68\%
    and 95\% confidence on a Majorana light WIMP from
    Figs.~\ref{fig:majorana-bbn+cmb} and
    \ref{fig:majorana-bbn+cmb-nnu-obh2}, shown in terms of lithium
    yields and \mchi, with nuclear errors on BBN yields included and
    with $A(\mathrm{Li})\equiv 12+\log (\mathrm{Li/H})$ (curved green
    contours).  The ``$+$" corresponds to the best-fit point shown in
    Figs.~\ref{fig:majorana-bbn+cmb} and
    \ref{fig:majorana-bbn+cmb-nnu-obh2}.  The width in the Li/H
    direction is dominated by the nuclear uncertainty on the BBN
    lithium yield.  The straight horizontal bands show 68\% and 95\%
    confidence intervals from observations of lithium in
    low-metallicity stars presented in Ref.~\cite{spite},
    demonstrating that the light-WIMP scenario cannot alleviate the
    well-known ``BBN lithium problem.''}
  \label{fig:lithium}
\end{figure}

Another invariance of the light-WIMP model with respect to WIMP type
occurs in the yields of \3he and \7li.  This has already been
commented upon above.  In fitting BBN models to data without further
constraints, \yd~mainly determines \omb.  When a CMB constraint on
\omb~is added, \yd~constrains expansion time scales late in BBN.  (In
fact, the \omb~constraint from \yd~traces back to nuclear time scales
late in BBN~\cite{esmailzadeh}.)  With conditions late in BBN
constrained by deuterium and the CMB, even an extended model with
light WIMPs and equivalent neutrinos has little room for variation of
\3he or \7li away from whatever abundances occur where SBBN agrees
with \yd.  This is shown in Fig.~\ref{fig:lithium} where, in fact,
there would be an even narrower band of predicted \7li/H were it not
for the nuclear uncertainty on the \7li~yield.  With \omb~fixed by the
CMB and conditions during the final nuclear-burning phase of BBN fixed
by the deuterium observations, the lithium abundance is nearly
uniquely determined.  The pinning of conditions late in BBN on \yd~has
the consequence that in a model that differs from SBBN mainly in
time scales, the well-known ``BBN lithium problem'' persists.  We find
$A(\mathrm{Li}) \equiv 12+\log_{10}(\mathrm{Li/H}) = 2.73\pm 0.04$,
very nearly the same as in SBBN and essentially independent of the
type of light WIMP.  BBN and the CMB with light WIMPs in fact favor a
very slightly higher $A(\mathrm{Li})$ with a very slightly higher
upper limit than in SBBN.  With equivalent neutrinos but not a light
WIMP, the simultaneous fit of \Yp, \yd, and the CMB gives
$A(\mathrm{Li}) = 2.72\pm 0.03$.

Since the helium abundance affects the composition of the oscillating
fluid~\cite{hou}, the acoustic oscillations probed by the CMB, and
particularly their damping at small wavelengths, depend on \Yp.  This
circumstance forces some value of \Yp~to be assumed in any analysis of
the CMB data.  The usual assumption made in recent years is that
\Yp~depends on \omb~as predicted by SBBN, or on \omb~and \neff~jointly
with $\neff= 3.05 + \Deln$ and no light WIMP.  Even in the standard,
$\Deln = 0$, $\mchi\rightarrow\infty$ model, this introduces some
small inconsistencies with the analysis here.  For example, the Planck
analysis \cite{planck} was based on an older recommended value of the
neutron lifetime than used here, and on a BBN code with small
($\Delta{\rm Y}_{\rm P} \sim 0.0005$) systematic differences from the
code used here.  Perhaps the most consistent combination of a more
elaborate BBN model (like the light-WIMP scenario) with the CMB data
would use fits from the Planck Collaboration in which \Yp~is fitted as
an independent parameter along with \neff~and the base $\Lambda$CDM
parameters.  Indeed, such fits were considered by the Planck
Collaboration, \eg, Eq.\,(90) of Ref.\,\cite{planck}.  Using the details
of this fit provided at the Planck Web site \cite{planck-web}
(including the correlations among \neff, \Yp, and \omb), we repeated
our analysis of the joint BBN and CMB+BAO fit.  In the Planck fit
alone, \Yp~is not significantly constrained, with \Yp~$=
0.260^{+0.034}_{-0.029}$ at a 68\% C.L.  The best-fit values of \omb~and
\neff~are not changed significantly in the Planck fit from their
values when \Yp~was fixed to a BBN relation.  However, the current CMB
data contain a strong degeneracy between \Yp~and \neff, connected to
the physics of Silk damping~\cite{hou}.  This degeneracy greatly
weakens the constraint on \neff~once \Yp~is allowed as an additional
parameter, so that the 68\% limit on the error for \neff~grows from
$0.27$ in the $\Lambda$CDM+\neff~fit to $\sim 0.5$ in the
$\Lambda$CDM+\neff+\Yp~fit, without the best-fit value of
\neff~changing significantly.  Using the model with fitted \Yp~in the
joint BBN and CMB+BAO analysis (line 7 for each WIMP type in Table
\ref{tab:bounds}) has roughly the same effect as removing the CMB
\neff~from the fitted constraints (line 4).  This is because
\omb~changes very little among alternate fits of CMB data, while the
CMB+BAO values of \omb~and \neff~constrain the light-WIMP model in
similar ways once BBN is included.  The very large likelihood interval
for \Yp~in the CMB fit (compared with the \hii~region observations)
keeps the correlation of \Yp~with other parameters from being
important in the joint fit.  The value of \Yp~in the fit including
both BBN and CMB+BAO is still essentially that from the \hii~regions
[where $\sigma({\rm Y}_{\rm P}) = 0.003$] because the CMB+BAO
constraint on this parameter is so weak.  There is very little
difference between this case and the Planck $\Lambda$CDM+\neff~fit
shown in Figs.~\ref{fig:majorana-bbn+cmb} and
\ref{fig:majorana-bbn+cmb-nnu-obh2}.

The Planck Collaboration has also published results of fits with
additional parameters.  The central value of \omb\ in these fits is
always within 68\% confidence limits of the $\Lambda$CDM value, and
degeneracy with the added parameters does not change the bounds on
\omb~significantly.  Constraints on \neff\ are influenced
significantly in two published choices of additional parameters and
observational constraints.  Adding the sum of neutrino masses as an
independent parameter widens the confidence intervals on \neff.
Introducing the Hubble constant as measured by the Hubble Space
Telescope key project into the Planck fits increases the best-fit
\neff\ by about 0.3, but it also nearly doubles the error.  For the
analysis here using the $\Lambda$CDM+\neff~fit, \neff\ has constraints
of roughly equal strength from the CMB and from BBN.  These alternate
CMB fits would thus tip the scale in favor of the BBN constraint in
the parts of parameter space where there is tension and increase the
errors on our fitted parameters slightly.  The Hubble constant
constraint improves the agreement between Planck and SBBN, so that
slightly higher \mchi\ and lower \Deln\ would be favored if we
considered the corresponding Planck fit
(cf.~Figs.~\ref{fig:majorana-bbn-only-nnu-obh2} and
\ref{fig:majorana-bbn+cmb}).

A final note is in order concerning the CMB constraints used.  In the
analysis here the CMB constraints have been modeled as Gaussian
likelihood functions with the correlations among the parameters given
by their covariances as found in the Markov-chain Monte Carlo (MCMC)
analyses by the Planck Collaboration \cite{planck}.  Although this
Gaussian approximation is not exact, it is useful for the present
analysis.  The confidence limits presented are frequentist, a choice
driven partly to avoid setting priors on \mchi~when our model reaches
an asymptotic limit for all values of $\mchi \gtrsim 20$ MeV.  In the
opinion of the authors, this makes profile likelihoods simpler than a
Bayesian analysis by MCMC.  The parameters used here as Gaussian
approximations to the Planck Collaboration fits are given in Table
\ref{tab:cmb-fits}.  The differences between the Planck MCMC
likelihood distributions and the Gaussian approximations are small and
are more important for the 95\% than for the 68\% intervals.  Where the
original likelihoods were asymmetric, the symmetric approximations
were chosen to be more accurate on the low-\neff~side of the
distributions, since this side is more interesting for setting limits
on \mchi~(cf.~Fig.~\ref{fig:majorana-bbn-only}).  Planck fits that
include the BAO constraint were used, since this constraint appears to
be consistent with the CMB data and therefore allows a slightly
tighter constraint on parameters than the CMB data alone \cite{planck}.

\begin{table}
  \caption{Planck Collaboration fits to cosmological parameters and the parameters of Gaussian approximations to them (68\% C.L.~errors) used in this work.  The first of these [Eq.\,(75) of Ref.\,\cite{planck}] is denoted ``base\_nnu\_planck\_lowl\_lowLike\_highL\_BAO'' by the Planck Collaboration, and the second [Eq.\,(90) of Ref.\,\cite{planck}] is denoted ``base\_nnu\_yhe\_planck\_lowl\_lowLike\_highL\_post\_BAO.''}
  \label{tab:cmb-fits}
    \begin{tabular}{lllll}
Data/Parameters fit & \multicolumn{1}{c}{$\omb$} &
\multicolumn{1}{c}{Approx.~$\sigma(\omb)$} &
\multicolumn{1}{c}{$N_\mathrm{eff}$} &
\multicolumn{1}{c}{Approx.~$\sigma(\neff)$}
\\\hline CMB+BAO
($\eta,N_\mathrm{eff})$ & $0.02229\pm 0.00029$ & {\hskip .25in $0.00029$} & $3.30\pm 0.27$ & {\hskip .25in $0.27$}\\ 
CMB+BAO ($\eta,N_\mathrm{eff},Y_P)$ & $0.02233\pm 0.00031$ & {\hskip 0.25in $0.00031$} & 
$3.19^{+0.43}_{-0.54}$ & {\hskip .25in $0.54$} \\
  \end{tabular}
\end{table}

\subsection{BBN and CMB constraints without a light WIMP}
\label{nowimp}

\begin{figure}[!t]
\begin{center}
\includegraphics[width=0.45\columnwidth]{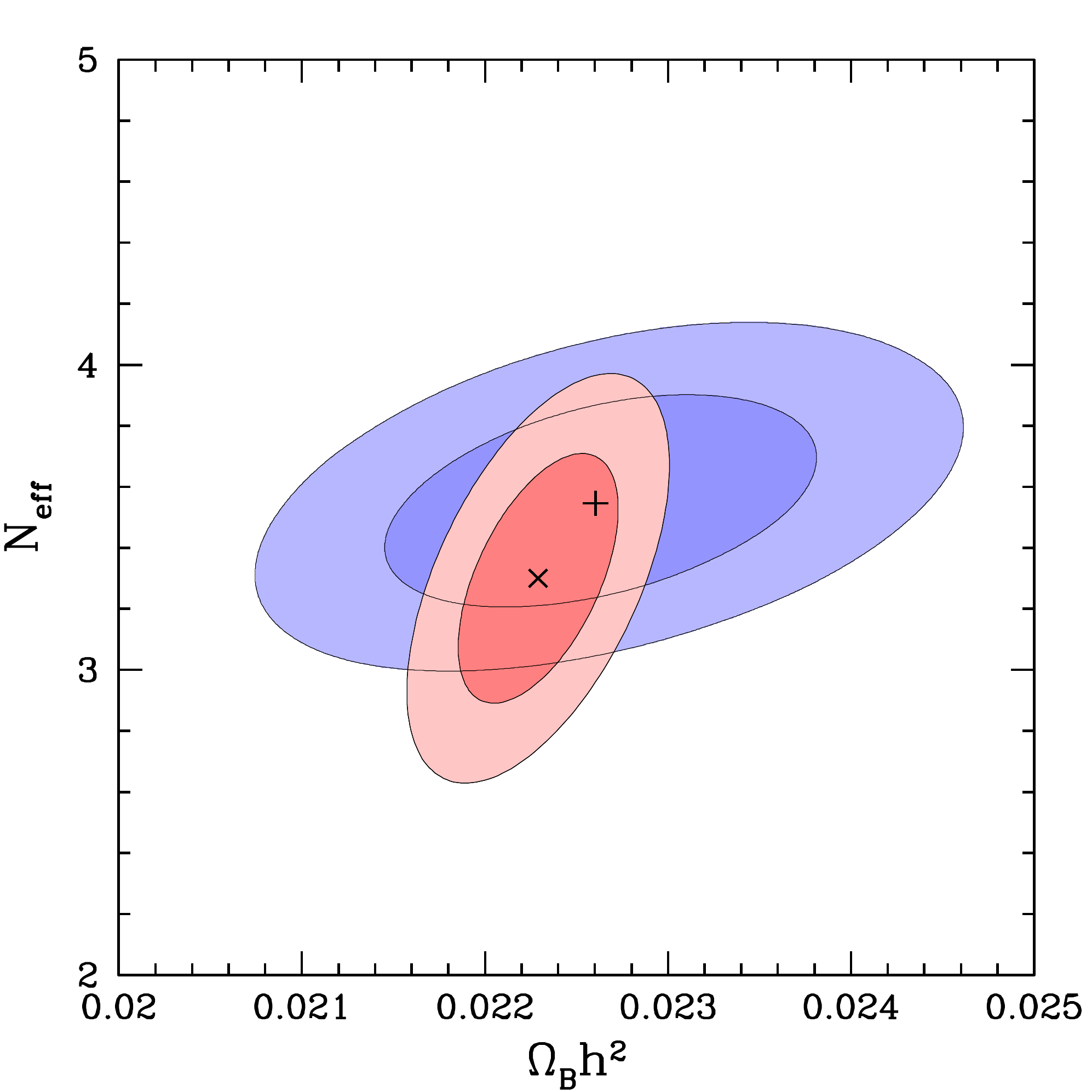}
\includegraphics[width=0.45\columnwidth]{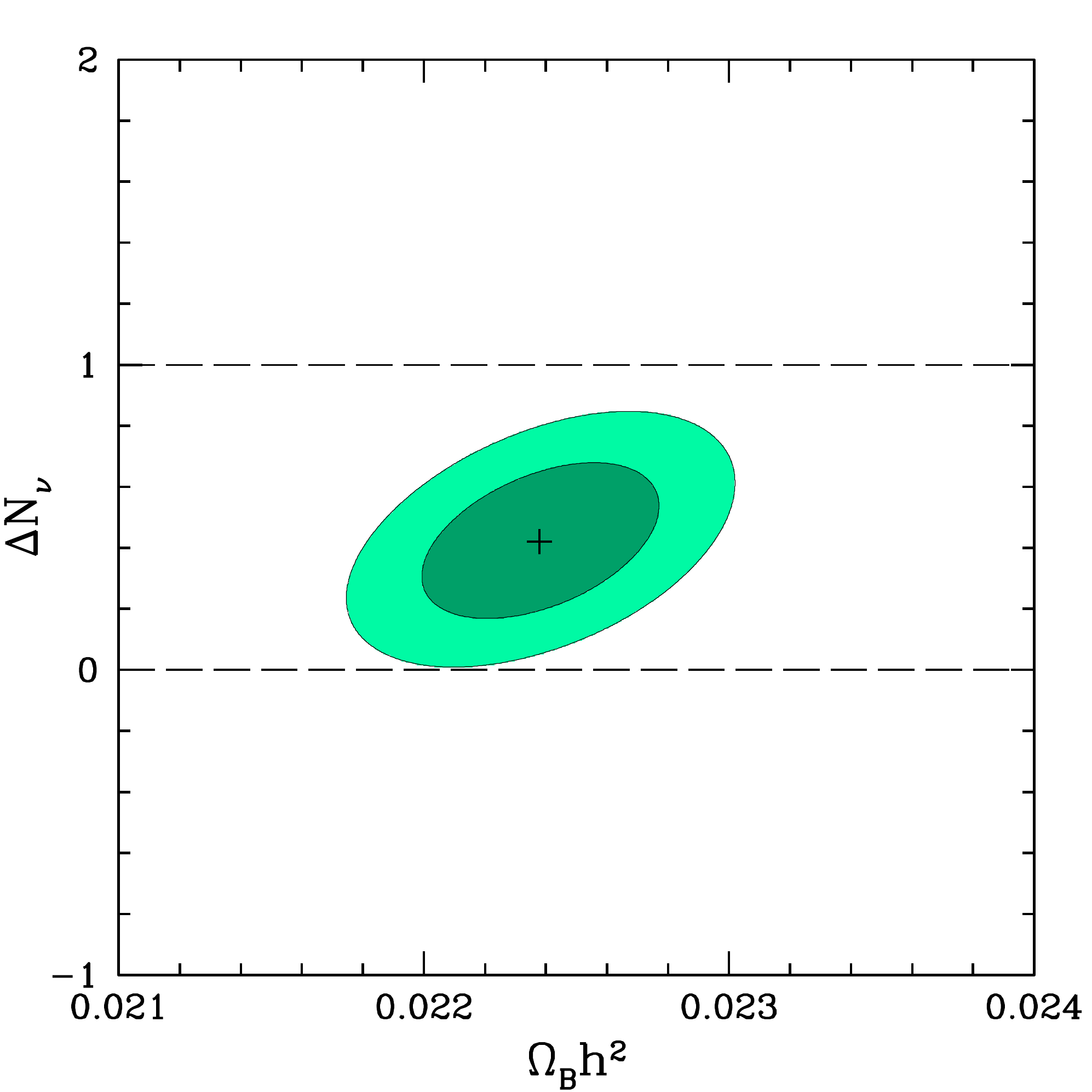}
\\\vskip 0.1in
\caption{(Color online). The left panel compares the 68\% (darker) and
  95\% (lighter) contours in the \neff~--~\omb~plane derived
  separately from BBN (blue, higher, and shallowly tilted) and the CMB
  (pink, lower, and steeply tilted) in the absence of a light WIMP.
  The ``$\times$'' symbol marks the best-fit CMB point, and the ``$+$''
  is the best-fit BBN point.  The right panel shows the joint BBN+CMB
  best-fit 68\% (darker) and 95\% (lighter) contours for
  \Deln~--~\omb.  The ``$+$'' symbol marks the best-fit point.  The
  full correction of Ref.~\cite{mangano} for neutrino heating is not
  included here, though it is in Eq.~(\ref{eq:fitted-params-no-wimp})
  and just below.}
\label{fig:neff}
\end{center}
\end{figure}

Given that the analysis here has used very recently revised abundances
for D and \4he, along with updated nuclear rates for some of the key
reactions in the BBN analysis, and the BBN results have been combined
with independent data from the CMB, it is interesting to explore the
cosmological parameter estimates in the absence of a light WIMP
(\mchi~$\gsim 20\,{\rm MeV}$).  In this case the BBN-predicted yields
depend on only two parameters, \Deln~(or, $\neff = 3.05 +
\Deln$~\cite{mangano}) and the baryon-to-photon ratio, $\eta_{10}$
(or, the baryon density parameter \omb).  Two observationally inferred
primordial abundances, \eg, of D and \4he, are sufficient to determine
\Deln~and $\eta_{10}$.  For BBN without light WIMPs, it is found
that,\footnote{BBN calculations with no light WIMP have been carried
  out by other authors, abandoning the instantaneous decoupling
  approximation and taking full account of the entropy from $e^\pm$
  annihilation that heats the neutrinos, with neutrino mixing included
  \cite{dolgov,hannestad,mangano,dolgov-osc,pastor}.  In this case,
  N$^{0}_{\rm eff} = 3.05$, and \Yp~is increased by 0.0002 compared to
  the result when all $e^\pm$ annihilation heating of the neutrinos is
  ignored \cite{mangano}.  Fitting fixed BBN abundances, this would
  slightly decrease \Deln~and slightly increase \neff~relative to the
  $\mchi \rightarrow \infty$ limit of the present calculations.  The
  results given in this section include these corrections for
  incomplete decoupling.  These very small differences are well within
  the statistical errors, as well as simple round off errors: without
  the correction, the limits are $\Deln = 0.53\pm 0.23$ and $\neff =
  3.55\pm 0.23$ for BBN only and \neff~$= 3.44 \pm 0.17$ and \Deln~$= 0.42 \pm
  0.17$ for BBN+CMB.} at 68\% C.L.,
\begin{equation}
 \eta_{10} = 6.19 \pm 0.21\,, \ \ \Deln = 0.51 \pm 0.23\,,
\label{eq:fitted-params-no-wimp}
 \end{equation}
 corresponding to \omb$= 0.0226 \pm 0.0008$ and \neff~$= 3.56 \pm
 0.23$.  These BBN values are in excellent agreement with the
 independent determinations from the CMB (\neff~$= 3.30 \pm 0.27$,
 \omb~$= 0.0223 \pm 0.0003$).  These two results are compared in the
 left panel of Fig.~\ref{fig:neff}, where correlations between fitted
 parameters are visible.

The right panel of Fig.\,\ref{fig:neff} shows the joint BBN + CMB fit
with no light WIMP, which gives \neff~$= 3.45 \pm 0.17$ (\Deln~$= 0.41
\pm 0.17$) and $\eta_{10} = 6.13 \pm 0.07$ (\omb~$= 0.0224 \pm
0.0003$).  Recall that for SBBN it is assumed that there are no
equivalent neutrinos (\Deln~= 0).  It can be seen in
Fig.~\ref{fig:neff} that SBBN just misses consistency with
light-element abundances and CMB observations at a bit more than the
95\% confidence level, as discussed below.  The presence of one
sterile neutrino (\Deln~= 1) is even more disfavored, at $\gsim 99\%$
confidence.
 
\subsection{BBN with no light WIMP and $\Deln = 0$ or 1}

\begin{figure}[!t]
\begin{center}
\includegraphics[width=0.45\columnwidth]{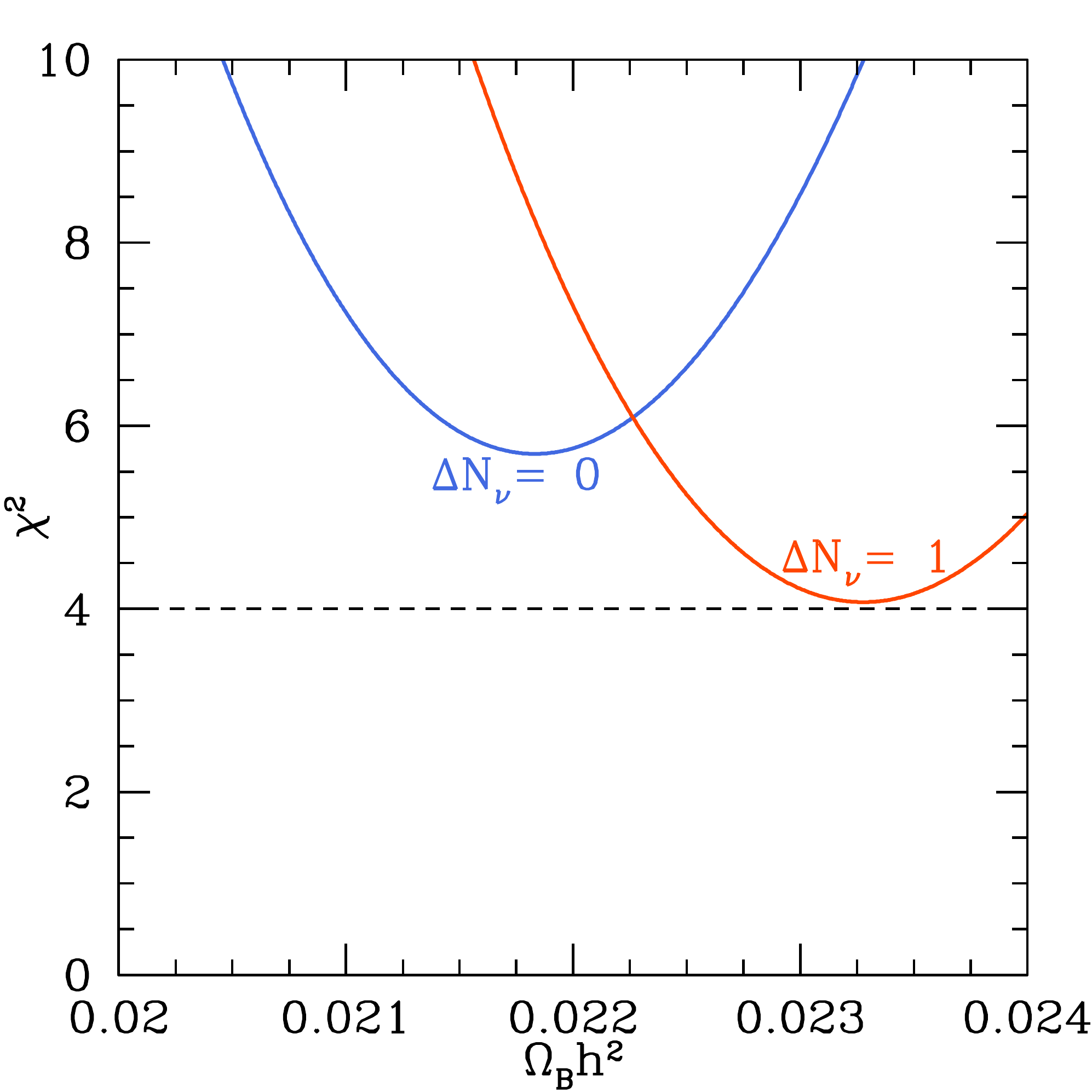}
\caption{(Color online) $\chi^{2}$ for the one-parameter ($\eta_{10}$) BBN fits to D and \4he in the absence of a light WIMP for \Deln~= 0 (no dark radiation) and \Deln~= 1 (sterile neutrino).}
\label{fig:chisq}
\end{center}
\end{figure}

Standard big bang nucleosynthesis (SBBN) is a subset of the
no-light-WIMP case considered above, in which there are also no
equivalent neutrinos (\Deln~= 0).  This leaves only one
overconstrained parameter (\omb~or $\eta_{10}$) to fit to the two
adopted primordial abundances (\yd, \Yp).  The data for D and \4he
could, in principle, be in conflict; i.e., a fit of the model to the
observations could have a poor $\chi^{2}$.  In fact, there is some
tension between the SBBN predictions and our adopted primordial
abundances, as shown in Fig.~\ref{fig:chisq} and indicated already in
Fig.~\ref{fig:neff}.  For SBBN alone, it is found that $\eta_{10} =
6.0 \pm 0.2$ ($\omb = 0.0219 \pm 0.0007$) and $\chi^{2}_\mathrm{min} =
5.7$, suggesting a poor fit ($p$ value 1.7\%).  The baryon density
fitted in this case is in good agreement with the CMB.  However, the
current D and \4he abundances by themselves suggest that \Deln~$\neq
0$, disfavoring SBBN.

Possibly the next best motivated model of BBN after SBBN includes a
sterile neutrino kept in thermal contact with the others through the
BBN era by mixing.  Except at special values of the mixing parameters,
this model has $\Deln = 1$, which was shown in Fig.~\ref{fig:neff} to
be even more disfavored than $\Deln = 0$ in a joint fit of BBN and
CMB+BAO data.  Figure~\ref{fig:chisq} shows that when \omb~is fitted to
D and \Yp~data with $\Deln = 1$ enforced, the one-parameter fit is
better than for $\Deln = 0$ but is still excluded at about 95\%
C.L.~with $\chi^2_\mathrm{min} = 4.1$.  This fit is still within
$1.5\sigma$ of the CMB value of \omb, since it has $\eta_{10} = 6.4
\pm 0.2$ or $\omb = 0.0234\pm 0.0007$.  It appears that if there are
equivalent neutrinos (and no light WIMPs), no subset of the present
data favors their being fermions with the same temperature as the
standard-model neutrinos.  Since experimental anomalies interpretable
as strong mixing with sterile neutrinos are a primary motivation for
cosmological models with extra neutrinos, BBN seems to present serious
difficulties for such models unless there is a light WIMP.

\section{Summary And Conclusions}
\label{summary}

There are now observations constraining the relativistic energy
density of the Universe at the very different times of BBN and just
before recombination.  Traditionally, these lines of evidence have
been presented and interpreted as providing counts of thermalized
neutrino-like species.  It was stressed in Ref.~\cite{chimera} that
the simplest ``counting'' interpretation of these data hinges on the
assumption that the temperatures of any additional species have the
same ratio to the photon temperature as neutrino temperatures in the
standard model, $(T_{\nu}/T_{\gamma})^{3}_{0} = 4/11$.  An additional
species could have a lower temperature if it decoupled from the
electromagnetic plasma earlier than the standard-model neutrinos,
leading to less than one ``equivalent neutrino'' species.  Moreover,
it was shown in Ref.~\cite{chimera} that even a species at the same
temperature as the standard-model neutrinos could be ``hidden'' from
CMB constraints if in addition to equivalent neutrinos there were also
light ($m_\chi \lesssim 10$ MeV) WIMPs that remained coupled to the
electromagnetic plasma after neutrino decoupling.

In this report, the consequences of simultaneous light WIMPs and
equivalent neutrinos during BBN were computed.  These results were
then combined with observed light-nuclide abundances and cosmological
parameters from the Planck Collaboration to constrain the degeneracy
between light WIMPs and equivalent neutrinos.  It was found that a
lower limit on the possible WIMP mass \mchi~results, below which there
is no simultaneous fit to all of the data for any (possibly
noninteger) number \Deln~of equivalent neutrino species.  This lower
limit ranges from $\sim 500$ keV to $\sim 5$ MeV, depending on the
nature of the WIMP.  Good simultaneous fits to all data are found for
all types of light WIMP (real and complex scalar, Majorana and Dirac
fermion), so the model is consistent with the data.  However, the best
fit with finite \mchi~is not significantly better than the best fit
with no light WIMPs.

The original motivation for exploring light WIMPs was that they might
significantly loosen the constraints on \Deln, which are naively
$\Deln \simeq 0.4 \pm 0.2$ at a 68\% C.L.  Allowing light WIMPs shifts
the best-fit \Deln~to $\sim 0.65$, regardless of WIMP type, with the
tradeoff that the best fit requires a WIMP of mass 5 to 10 MeV.  The
standard-model value $\Deln = 0$ falls just beyond the 95\% C.L.~of
the combined data whether or not a light WIMP is allowed, and this
is true of any fit using the chosen values of the observed BBN
abundances.  An additional neutrino species at the same temperature as
the standard-model neutrinos, $\Deln = 1$, is allowed at better than
68\% C.L., but only if there are light WIMPs.  The available data do
not allow $\Deln = 2$.

The data constraining equivalent neutrinos will continue to improve in
the near future.  The precision of the BBN data appears to be
improving.  A particularly dramatic improvement can be expected in the
CMB data, where polarization measurements at small angular scale will
include a signature of relativistic energy density that does not
suffer from degeneracies with other effects \cite{bashinsky}.  When
these data are interpreted to provide limits on particle properties,
it should be kept in mind that limits on parameters like \Deln~depend
on the model assumed.  While not strongly motivated by particle
physics, the scenario examined here includes the standard model as a
special case and represents an example of the kind of relaxed
assumptions that might be useful for conservative parameter
constraints.

\acknowledgments

We would like to thank the anonymous referee for several very useful
suggestions that improved this paper.  We are grateful to the Ohio
State University Center for Cosmology and Astro-Particle Physics for
hosting K.\,M.\,N.'s visit during which most of the work described
here was done.  K.\,M.\,N. is pleased to acknowledge support from the
Institute for Nuclear and Particle Physics at Ohio
University. G. S. is grateful for the hospitality provided by the
Departamento de Astronomia of the Instituto Astron\^{o}mico e
Geof\'{i}sico of the Universidade de S\~{a}o Paulo, where much of the
writing was done.  The research of G. S. is supported at OSU by 
U.S.~DOE Grant No.~DE-FG02-91ER40690.

\end{document}